\documentclass[useAMS,usenatbib]{mn2e} 

\usepackage[T1]{fontenc}
\usepackage{ae,aecompl}

\usepackage{graphicx}	
\usepackage{amsmath}	
\usepackage{amssymb}	
\usepackage{epsfig}     
\usepackage{pifont}     
\usepackage{grffile}    

\def\spose#1{\hbox to 0pt{#1\hss}}
\def\ltsimm{\mathrel{\spose{\lower 3pt\hbox{$\sim$}}
        \raise 2.0pt\hbox{$<$}}}
\def\gtsimm{\mathrel{\spose{\lower 3pt\hbox{$\sim$}}
        \raise 2.0pt\hbox{$>$}}}

\title[Shock-filament simulations]{Numerical Simulations of a
  Shock-Filament Interaction}

\author[J.~M.~Pittard \& K.~J.~A.~Goldsmith]
{J.~M.~Pittard\thanks{E-mail: jmp@ast.leeds.ac.uk (JMP)} \& K.~J.~A.~Goldsmith\\
School of Physics and Astronomy, University of
       Leeds, Woodhouse Lane, Leeds LS2 9JT, UK
}

\date{Accepted 11$^{th}$ February 2016. Received 11$^{th}$ February 2016; in original form 18$^{th}$ December 2015}

\pubyear{2016}

\begin{document}
\label{firstpage}
\pagerange{\pageref{firstpage}--\pageref{lastpage}}
\maketitle

\begin{abstract}
  We present 3D hydrodynamic adiabatic simulations of a shock
  interacting with a dense, elongated cloud. We compare how the nature
  of the interaction changes with the filament's length and its
  orientation to the shock, and with the shock Mach number and the
  density contrast of the filament. We then examine the
  differences with respect to 3D spherical-cloud calculations. We find
  significant differences in the morphology of the interaction when
  $M=10$ and $\chi=10^{2}$: in many cases 3 parallel rolls are
  formed, and spread further apart with time, and periodic vortex shedding
  can occur off the ends of oblique filaments. Sideways-on filaments
  are accelerated more quickly, and initially lose mass more quickly
  than spherical clouds due to their greater surface area to volume
  ratio. However, at late stages they lose mass more slowly, 
  due to the reduced relative speed between the filament and the
  postshock flow. The acceleration and mixing timescales can vary by a
  factor of 2 as the filament orientation changes. Oblique filaments can
  achieve transverse velocities up to 10\% of the shock speed. Some
  aspects of our simulations are compared against experimental and
  numerical work on rigid cylinders. 
\end{abstract}




\section{Introduction}
\label{sec:intro}
In recent years, filaments have gained increasing significance in many
areas of astrophysics. 
\emph{Spitzer} has revealed filamentary structures throughout the
galactic disk \citep[e.g.,][]{Benjamin:2003,Carey:2009,Churchwell:2009}, while
\emph{Herschel} images have revealed a complex web of filamentary
structures in every interstellar cloud
\citep[e.g.][]{Andre:2010,Menshchikov:2010,Molinari:2010,Henning:2010,
  Motte:2010}, whether or not stars have formed in
them. \citet{Goodman:2014} argue that filamentary molecular structures
can extend many hundreds of parsecs in length, and make up the
``bones'' of the Milky Way.
The filaments are thought to form out of compressed gas due to
converging flows, which are driven by interstellar turbulence, by
large scale spiral arms, or by gravitational disk instabilities. In
support of these observations, simulations of molecular clouds
consistently show the formation of filamentary structure
\citep[e.g.,][]{Klessen:2000,Ballesteros-Paredes:2002,Bate:2005,Padoan:2006,Vazquez-Semadeni:2006,Heitsch:2008,Federrath:2010,Krumholz:2011,Glover:2012,Bonnell:2013,Hennebelle:2013,Gomez:2014,Smith:2014,Kirk:2015,Moeckel:2015}.

Some molecular cloud filaments are likely to be transitory objects,
but others collapse further under gravity and fragment into
star-forming cores. The resulting stars help to disperse the remaining
parts of the filament through their ionizing radiation and
winds/outflows. However, the efficiency of this process remains
contentious: \citet{Dale:2011} find that the ionizing flux from
massive clusters forming at the base of strong filamentary accretion
flows has little effect on the filaments, while \citet{Colin:2013}
find that massive stars are able to completely evacuate dense gas to
10\,pc distances.  Filaments located near regions of massive star
formation can also be strongly affected by high speed shocks and
winds. For example, the B59 filament in the Pipe Nebula is clearly
being shaped by a wind from the nearby Sco OB2 association
\citep{Peretto:2012}. Similarly, the winds and radiation from the OB
stars in the Cyg~OB2 and Cyg~OB9 associations appear to be affecting
the molecular clouds of the Cyg~X complex \citep{Schneider:2006},
while \citet{Wilson:2005} suggest that stellar wind driven compression
is consistent with the kinematics and structure of the CO gas in
Orion~A and Orion~B. Winds may also play a part in disrupting
molecular clouds, such as the Rosette nebula \citep{Bruhweiler:2010}.
\citet{Dent:2009} note that the near constant velocity gradient over
an extended clump in the Rosette nebula is difficult to explain with
radiatively driven models of clump acceleration. The lifetime of
filamentary features in a molecular clump exposed to stellar winds and
supernova explosions was investigated by \citet{Rogers:2013}.

Filaments are also observed in young protoclusters
\citep[e.g.,][]{Vig:2007}, in galactic outflows
\citep[e.g.,][]{Shopbell:1998,Ohyama:2002,Martin:2002,Strickland:2004,Hoopes:2005,Veilleux:2005,Engelbracht:2006,Westmoquette:2011,Heeson:2011,Bolatto:2013}
and in clusters of galaxies
\citep[e.g.,][]{Conselice:2001,Crawford:2005,Forman:2007,McDonald:2010,Canning:2011}
and are reproduced in simulations of these sources
\citep[e.g.,][]{Rodriguez-Gonzalez:2008,Strickland:2000,Cooper:2008,Fujita:2009,Gaspari:2012}.
The emission from galactic winds is dominated by shock excitation
\citep[e.g.,][]{Veilleux:2002,Sharp:2010,Rich:2010}, indicating that
these filamentary regions are inundated by shocks.

As a first step towards understanding the effect of mechanical
feedback on a filament, we perform idealized numerical simulations of
an adiabatic shock striking a non-magnetized filament. This represents
the most basic setup for a shock-filament interaction, and is
appropriate for our initial study. The assumption of adiabaticity
implies that our work is relevant in the small cloud and/or low
density limit \citep[see, e.g.,][]{Klein:1994}. It requires that the
evolutionary timescale of the cloud is shorter than the cooling
timescale. In the interstellar medium (ISM), diffuse filaments in an
ionized or atomic phase may satisfy this criteria.  In future work we
will examine the effects of winds, magnetic fields and radiative
cooling.  In Sec.~\ref{sec:definition} we introduce the shock-filament
problem and review the current literature. In Sec.~\ref{sec:numerics}
we describe the numerical methods, the initial and boundary
conditions, and the time-scales and statistics used in our study. In
Sec.~\ref{sec:results} we present our results and in
Sec.~\ref{sec:conclusions} we summarize our conclusions. A resolution
test is presented in an Appendix.

\section{Problem Definition}
\label{sec:definition}
An adiabatic shock striking a non-magnetized filament represents the
most basic idealized scenario of a shock-filament interaction. The
interaction depends on five parameters: 1) the Mach number of the
shock, $M$; 2) the density contrast of the filament, $\chi$; 3) the
filament aspect ratio; 4) the orientation of the filament; 5) the
ratio of specific heats, $\gamma$. As no physical scales are imposed
on the problem, our simulations can be scaled to any desired length-
or time-scales. In the strong shock limit ($M >> 1$), we can expect
Mach-scaling to also hold \citep[e.g.,][]{Klein:1994}, so our high Mach
number calculations will be representative of any other high Mach
number simulation.

Our idealized filament is assumed to have a cylindrical centre, of
length $l\,r_{\rm c}$, where $r_{\rm c}$ is the filament radius, which
is capped by a hemisphere at each end (see, e.g.,
Fig.~\ref{fig:m10c2l8s_XY_XZ_montage}). We choose this definition so
that spherical clouds have $l=0$. The total length of the filament is
$(l + 2)\,r_{\rm c}$, and the ratio of the lengths of its major and
minor axes is $(l+2)/2$. The filament's orientation is specified by
the angle $\theta$, which is the angle between its major-axis and the
shock \emph{surface}.  Our setup is clearly idealized, not least
because observations of actual filaments in molecular clouds show a
great deal of sub-structure
\citep[e.g.,][]{Hacar:2013,Fernandez-Lopez:2014,Henshaw:2014}, and may
be composed of multiple strands of gas.
 
Nevertheless, this description is adequate for our
purposes and allows us to investigate how the interaction changes as
$M$, $\chi$, $l$, and $\theta$ are varied.  The filament is
assumed to have soft edges over about 10 per cent of its radius
\citep[we adopt $p_{1}=10$ in keeping with our earlier work,
cf.][]{Pittard:2009,Pittard:2010,Pittard:2016}.  Filaments in molecular clouds with
a Plummer-like density profile are like soft-edged clouds and we can
expect that instabilities will take relatively longer to develop
\citep[see][]{Nakamura:2006}.  The filament is also in pressure
equilibrium with its surroundings which have a mass density of
$\rho_{\rm amb}$.

\subsection{Previous numerical studies of shock-cloud interactions}
The first numerical studies of a planar adiabatic shock striking a
single isolated spherical cloud were published in the 1970s. The
nature of the interaction changes with the specific values of $M$ and
$\chi$, and the cloud's density profile, but in all cases the cloud is
initially compressed and then re-expands. Vorticity is deposited at
the surface of the cloud, and the cloud is subject to numerous
dynamical instabilities, including Kelvin-Helmholtz (KH),
Rayleigh-Taylor (RT) and Richtmyer-Meshkov (RM)
\citep*[e.g.,][]{Klein:1994}. The vorticity deposition and growth of
KH instabilities is strongest when the cloud has sharp edges. The
interaction is more gentle at lower Mach numbers and for clouds with
soft edges \citep[e.g.,][]{Nakamura:2006,Pittard:2010,Pittard:2016}.
The cloud is more rigid, and is better able to resist the passage of
the shock, as $\chi$ increases \citep[see, e.g.,][]{Pittard:2010}. For
all but weak shocks, the interaction is characterized by a series of
shocks and rarefaction waves moving through the cloud which cause the
cloud to reverberate - at some moments in time the cloud is pancaked,
while at other moments the cloud becomes hollow and ``voided''.
Shocks leaving the cloud introduce reflected rarefaction waves into
the cloud, and shocks into the external medium. Rarefaction waves
within the cloud which reach the cloud boundary introduce a
transmitted rarefaction wave into the external medium and a reflected
shock which moves back into the cloud.

For relatively sharp-edged clouds, the cloud lifetime, $t_{\rm life}$,
defined as when material from the core of the cloud is well mixed with
the ambient material, is $ \sim 6\,t_{\rm KHD}$, where $t_{\rm KHD}$
is the growth-timescale for the most disruptive, long-wavelength, KH
instabilities \citep{Pittard:2010,Pittard:2016}. At high Mach numbers
and for $\gamma=5/3$, this is equivalent to $t_{\rm life}\sim8\,t_{\rm
  cc}$. In 3D simulations, non-axisymmetric instabilities develop
\citep[][]{Stone:1992,Klein:2003,Niederhaus:2008}, but for sharp-edged
and reasonably sharp-edged clouds these do not significantly affect
either the cloud lifetime or its acceleration \citep{Pittard:2016}.
\citet{Klein:2003} argued that the break-up of the vortex ring in 3D
experiments and simulations led to significant differences in the
hollowing or ``voiding'' of the cloud compared to 2D
simulations. However, \citet{Pittard:2016} showed that 2D axisymmetric
and fully 3D simulations evolve almost identically until the late
stages of the interaction, and put this contradiction down to the
different numerical codes and initial conditions used by
\citet{Klein:2003}.

While the hydrodynamical nature of the interaction of a shock with a
spherical cloud is now well understood, very little work exists in the
current literature on how the interaction changes when the cloud is
not spherical.  \citet{Klein:1994} were amongst the first to look into
this when they investigated the interaction of a shock with a cloud with
an aspect ratio of 3:1 (see their~Sec 9.1). Their calculation was in 2D
axisymmetry with the long-axis of the cloud aligned with the
direction of propagation of the shock (so the ``filament'' was hit from
one end). They introduced a modified cloud crushing timescale ($t_{\rm
  cc}'$, their Eq.~9.2) and found reasonably good agreement with the
spherical case when comparing as a function of $t_{\rm
  cc}'$. Therefore, \citet{Klein:1994} noted that their main
conclusions were unaffected by modest changes in the initial shape of
the cloud.

\citet{Xu:1995} conducted some of the first 3D shock-cloud simulations
in order to investigate how the interaction depends on the shape of
the cloud. Their use of 3D allowed arbitrary orientations of the cloud
relative to the shock, thus avoiding the restrictions in
\citet{Klein:1994}'s work. They instead noted that the nature of the
interaction \emph{is} quite different for spherical and prolate
clouds: an aligned prolate cloud does not form a vortex ring, and
accelerates faster than a spherical cloud, whereas an inclined prolate
cloud is ``spun around'' and instabilities develop differently. They
further found that the mixing rate of a prolate cloud is much faster
than a spherical cloud of the same mass as a result of the higher
surface-to-volume ratio of the cloud.  In both of their calculations
an axial ratio of 2:1 was adopted.

To our knowledge, there exists no other dedicated numerical study of
shocks striking prolate or filamentary clouds. The current work therefore
extends the investigations of \citet{Klein:1994} and \citet{Xu:1995} to
interactions where the aspect ratio of the cloud and its density
contrast are greater. Our work builds on recent spherical-cloud
simulations where the dependence of the interaction with a shock was
explored for a variety of density contrasts and shock Mach numbers
\citep{Pittard:2010,Pittard:2016}.

\section{The Numerical Setup}
\label{sec:numerics}
The calculations were performed on a 3D XYZ cartesian grid using the
{\sc MG} Eulerian adaptive mesh refinement (AMR) hydrodynamic code \citep{Falle:1991}.
The normal Eulerian equations are solved for conservation of mass,
momentum and energy:
\begin{equation}
\label{eq:mass}
\frac{\partial \rho}{\partial t} + \nabla \cdot (\rho {\bf u}) = 0,  
\end{equation} 
\begin{equation}
\label{eq:mtm}
\frac{\partial \rho {\bf u}}{\partial t} + \nabla \cdot (\rho {\bf
  uu}) + \nabla P = 0, 
\end{equation} 
\begin{equation}
\label{eq:energy}
\frac{\partial E}{\partial t} + \nabla \cdot [(E + P){\bf u}] = 0, 
\end{equation} 
where $\rho$ is the mass density, ${\bf u}$ is the velocity, 
\begin{equation}
E = \frac{P}{\gamma - 1} + \frac{1}{2}\rho{\bf u}^2
\end{equation} 
is the total energy density, and $P$ is the thermal pressure.

{\sc MG} uses piece-wise linear cell interpolation and solves the
Riemann problem at cell interfaces to obtain the conserved fluxes for
the time update. A linear solver is used for most cases, with the code
switching to an exact solver when there is a large difference between
the two states. The flux update occurs for all
directions simultaneously. The time integration proceeds first with a
half time-step to obtain fluxes at this point. The conserved variables
are then updated over the full time-step. The code is thus 2$^{\rm
  nd}$-order accurate in space and time. 

The two coarsest levels ($G^0$ and $G^1$) cover the whole
computational domain, and finer cells are added or removed as
needed. Steep gradients in the fluid variables, such as at shocks or
contact discontinuities, cause the mesh to be more refined than in
more uniform regions. Each level is generated from its predecessor by
doubling the number of computational grid cells in each spatial
direction. This technique enables the generation of fine grids in
regions of high spatial and temporal variation, and conversely,
relatively coarse grids where the flow field is numerically
smooth. Values for newly created cells are obtained by linear
interpolation from the coarser level. To maintain accuracy, the
solution in finer grid cells overwrites that in coarser cells, and
flux corrections are applied at coarse-fine boundaries. Double
refinements are not allowed.

Refinement is performed on a cell-by-cell basis and is controlled by
the difference in the solutions on the coarser grids. In this work
only the difference in density is compared, and cells are marked for
refinement if $|\rho_{\rm f} - \rho_{\rm ic}| > 0.01\,\rho_{\rm f}$,
where $\rho_{\rm f}$ is the density in the fine cell and $\rho_{\rm
  ic}$ is the interpolated density from the coarser cell,
\emph{before} the solution in the coarse cell has been overwritten by
that in the finer cells. In this way the refinement algorithm really
is comparing an error. In all of our calculations the $G^0$ grid has a
cell width of $2\,r_{\rm c}$, so that calculations with 32 cells
across the filament semi-minor axis on the finest grid (henceforth
$R_{32}$) use 7 grid levels in total ($G^0$ to $G^6$).

In this work we conduct a purely hydrodynamic study, and ignore the
effects of magnetic fields, thermal conduction, cooling and
self-gravity.  All calculations were perfomed for an ideal gas with
$\gamma=5/3$ and are adiabatic. Our calculations are thus scale-free
and can be easily converted to any desired physical scales.

An advected scalar is used to distinguish between cloud and ambient
material, and thus to track the ablation and mixing of the cloud, and
the cloud's acceleration by the passage of the shock and subsequent
exposure to the post-shock flow. The filament is initially centered at
the grid origin $(X,Y,Z)=(0,0,0)$ while the shock is located at
$X=-10$.  The grid has zero gradient conditions on each boundary and
is set large enough so that the filament is well-dispersed and mixed
into the post-shock flow before the shock reaches the downstream
boundary. The grid extent is dependent on $M$, $\chi$, and the filament
orientation and is noted in Table~\ref{tab:gridextent}. Note that we
do not impose any symmetry constraints on the interaction, unlike the
earlier work of \citet{Xu:1995}.  We define motion in the direction of
shock propagation as ``axial'' (the shock propagates along the
X-axis), while motion perpendicular to this (in the Y and Z
directions) is termed ``radial'' or ``transverse''.

\begin{table}
\centering
\caption[]{The grid extent as a function of $M$ and $\chi$. The unit of
  length is the filament radius, $r_{c}$. Note that simulations
  \emph{m10c2l8o30}, \emph{m10c2l8o60} and \emph{m10c2l8o85} actually
  had $Y$ and $Z$ extents of $-16 < Y < 16$, and $-33 < Z < 15$, $-25
  < Z < 15$, and $-20 < Z < 20$, respectively. In all cases the
  filament is centered at the grid origin, while the shock is
  initialized at $X=-10$ and propagates in the $+X$ direction.}
\label{tab:gridextent}
\begin{tabular}{llccc}
\hline
$M$ & $\chi$ & $X$ & $Y$ & $Z$ \\
\hline
10 & 10 & $-20 < X < 160$ & $-24 < Y < 24$ & $-16 < Z < 16$ \\ 
10 & $10^{2}$ & $-20 < X < 160$ & $-24 < Y < 24$ & $-16 < Z < 16$ \\ 
10 & $10^{3}$ & $-20 < X < 460$ & $-24 < Y < 24$ & $-16 < Z < 16$ \\ 
3 & 10,$10^{2}$ & $-20 < X < 160$ & $-24 < Y < 24$ & $-16 < Z < 16$ \\ 
1.5 & 10,$10^{2}$ & $-120 < X < 160$ & $-24 < Y < 24$ & $-24 < Z < 24$ \\
\hline
\end{tabular}
\end{table}

The filament evolution is studied through various integrated
quantities
\citep[see][]{Klein:1994,Nakamura:2006,Pittard:2009,Pittard:2016}. Averaged
(mass-weighted) quantities $\langle f \rangle$, are constructed by
\begin{equation}
\langle f\rangle = \frac{1}{m_{\beta}}\int_{\kappa \geq \beta} \kappa \rho f \;dV,
\end{equation}
where the mass identified as being part of the filament is
\begin{equation}
m_{\beta} = \int_{\kappa \geq \beta} \kappa \rho \;dV.
\end{equation}
$\kappa$ is an advected scalar, which has an initial value of
$\rho/(\chi \rho_{\rm amb})$ for cells within a distance of $r_{\rm
  c}$ from the ``edge'' of the filament, and a value of zero at
greater distances. Hence, $\kappa=1$ in the centre of the filament,
and declines outwards. The above integrations are performed only over
cells in which $\kappa$ is at least as great as the threshold value,
$\beta$. Setting $\beta = 0.5$ probes only the densest parts of the
filament and its fragments (hereafter identified with the subscript
``core''), while setting $\beta = 2/\chi$ probes the whole filament
including its low density envelope, and regions where only a small
percentage of filament material is mixed into the ambient medium
(identified with the subscript ``cloud'').

The mean density is defined as
\begin{equation}
\langle\rho\rangle = \frac{m_{\beta}}{V_{\beta}},
\end{equation}
where $V_{\beta}$ is the volume of the region with $\kappa \geq \beta$.
We also monitor the mass-weighted mean velocity of the filament in
each direction - $<v_{\rm x}>$, $<v_{\rm y}>$, and $<v_{\rm z}>$ - and the velocity dispersions in each
of the 3 directions, which are defined as 
\begin{eqnarray}
\delta v_{\rm x} = \left(\left<v_{\rm x}^{2}\right> - \langle v_{\rm
    x} \rangle^{2}\right)^{1/2}, \\
\delta v_{\rm y} = \left(\left<v_{\rm y}^{2}\right> - \langle v_{\rm y} \rangle^{2}\right)^{1/2},\\
\delta v_{\rm z} = \left(\left<v_{\rm z}^{2}\right> - \langle v_{\rm z} \rangle^{2}\right)^{1/2}.
\end{eqnarray}

The characteristic time for a spherical cloud to be crushed by the shocks
driven into it is the ``cloud crushing'' time, $t_{\rm cc} =
\chi^{1/2} r_{\rm c}/v_{\rm b}$, where $v_{\rm b}$ is the velocity of
the shock in the intercloud (ambient) medium.
\citet{Klein:1994} also introduced a ``modified cloud
crushing time'' for cylinder-shaped clouds,
\begin{equation}
\label{eq:tcc_mod_klein}
t_{\rm cc}' \equiv \frac{(\chi a_{0} c_{0})^{1/2}}{v_{\rm b}},
\end{equation}
where $a_{0}$ and $c_{0}$ are the initial radius of the cloud in the
radial and axial directions, respectively. \citet{Xu:1995} instead adopt 
\begin{equation}
\label{eq:tcc_mod_xu}
t_{\rm cs} \equiv \frac{r_{s} \chi^{1/2}}{v_{\rm b}},
\end{equation}
where $r_{s}$ is the radius of a spherical cloud of equivalent
mass. We will determine which of $t_{\rm cc}$, $t_{\rm cc}'$ and
$t_{\rm cs}$ gives the tightest comparison between models.

Several other timescales are obtained from the simulations. The time
for the average velocity of the cloud relative to that of the
postshock ambient flow to decrease by a factor of $e$ (i.e. when the
average cloud velocity $<v>_{\rm cloud} = (1 - 1/e)\,v_{\rm ps}$,
where $v_{\rm ps}$ is the speed of the postshock ambient medium as
measured in the frame of the preshock ambient gas) is defined as the
``drag time'', $t_{\rm drag}$ \citep[see][who also provide analytical
expressions for the time dependence of the mean cloud velocity for
both spherical and cylinder-shaped clouds]{Klein:1994}\footnote{Note
  that the definition of $t_{\rm drag}$ in
  \citet{Pittard:2009,Pittard:2010} is for the relative cloud velocity
  to decrease by a factor of $1/e$, which is obtained when $<v>_{\rm
    cloud} = v_{\rm ps}/e$. This definition of $t_{\rm drag}$ gives
  smaller values than the definition of \citet{Klein:1994}.}.  The
``mixing time'', $t_{\rm mix}$, is defined as the time when the mass
of the core of the filament, $m_{\rm core}$, reaches half its initial
value. Similarly, we define the ``lifetime'', $t_{\rm life}$, as the
time when $m_{\rm core}$ reaches 1\% of its initial value. The
zero-point of all time measurements occurs when the intercloud shock
is level with the centre of the filament.

\section{Results}
\label{sec:results}
As with any hydrodynamical study, it is important to investigate the
spatial resolution which is necessary to resolve key features in the
flow (e.g., the bow shock, the turbulent boundary layer,
etc.). Previous 2D studies have indicated that about 100 cells per
cloud radius is necessary \citep[e.g.,][]{Klein:1994,Nakamura:2006},
which is the point at which the turbulent boundary layer becomes
resolved \citep{Pittard:2009}. Use of a $k$-$\epsilon$ subgrid
turbulence model negates the need to resolve this layer (it is instead
handled by the subgrid physics) and can reduce the resolution
requirement to about 30 cells per cloud radius in 2D simulations
\citep{Pittard:2009}. ``Inviscid'' and $k$-$\epsilon$ 2D axisymmetric simulations can
behave significantly differently, which is likely due to the different
behaviour of vortices in 2D and 3D. The most detailed resolution study
of 3D shock-cloud interactions to date has been presented by
\citet{Pittard:2016}, where it is seen that ``inviscid'' and
$k$-$\epsilon$ simulations are in perfect agreement until the
turbulent boundary layer starts to become resolved. Resolutions of
32-64 cells per cloud radius are necessary to resolve the main flow
features and for some properties to show signs of convergence.

In Appendix~\ref{sec:restest} we carry out a similar study for
shock-filament interactions. We find that a resolution of $R_{32}$
across the semi-minor axis is the mininum necessary, and adopt this
value for the rest of this work.  Most of our calculations are for
$M=10$ and $\chi=10^{2}$ and the length and orientation of the
filament are varied. We consider values of $l$ up to 8. We adopt a
naming convention such that our reference simulation is $m10c2l8s$ -
here the $m10$ indicates the shock Mach number, the $c2$ indicates a
density contrast of $10^{2}$, the $l8$ indicates that $l=8$, and the
$s$ indicates that the filament is sideways on. Where the filament has
an oblique orientation to the shock, we replace ``$s$'' with
``$o\theta$'', where $\theta$ is the angle between the shock
\emph{surface} and the long axis of the filament (thus the
``sideways-on'' simulations could equally as well have been designated
as ``o0''). Towards the end of this section we also discuss results
from simulations with different Mach numbers and density
contrasts. Table~\ref{tab:results} summarizes the simulations
performed and some of their key parameters.  After examining the
filament morphology during the interaction, we present the evolution
of some key global quantities and then study the mixing and drag time
of the filament.

\begin{table*}
\centering
\caption[]{A summary of the simulations investigated in this work and
  some key results. $M$ is the shock Mach
  number, $\chi$ is the density contrast of the
  filament with respect to the ambient medium, and $l$ defines the length
  of the filament. The filament is either oriented at an angle
  $\theta$ between its major-axis and the shock \emph{surface}, or 
  is sideways-on to the shock ($\theta=0^{\circ}$). $t_{\rm cc}'$ is the
  modified cloud-crushing timescale of \citet{Klein:1994} (their
  Eq.~9.2) while $t_{\rm cs}$ is the cloud-crushing timescale for a
  spherical cloud of equivalent mass introduced by \citet{Xu:1995}. Key
  timescales and filament properties are also noted. Times in
  parentheses in columns 8 and 9 are in units of $t_{\rm cs}$ instead
  of $t_{\rm cc}$. Results in
  parentheses in the final three columns indicate values computed for the ``core'', while those
  without are computed for the more extended
  ``cloud''. $\langle\rho\rangle/\rho_{\rm max}$, $\langle v_{\rm
    x}\rangle/v_{\rm b}$, and $\langle x\rangle$ are evaluated at $t=t_{\rm mix}$.
}
\label{tab:results}
\begin{tabular}{lllllllllllll}
\hline
Name & $M$ & $\chi$ & $l$ & Orientation &$t_{\rm
  cc}'/t_{\rm cc}$ & $t_{\rm
  cs}/t_{\rm cc}$ & $t_{\rm drag}/t_{\rm cc/cs}$ & $t_{\rm mix}/t_{\rm
  cc/cs}$ & $t_{\rm mix}/t_{\rm drag}$ & $\langle\rho\rangle/\rho_{\rm max}$ & $\langle v_{\rm
  x}\rangle/v_{\rm b}$ & $\langle x\rangle$\\
\hline
m10c2l8s & 10 & 100 & 8 & sideways & 2.236 & 1.943 & 4.69 (2.42) & 7.06 (3.64) & 2.54 & 0.016 (0.092) & 0.589 (0.554) & (21.2)\\ 
m10c2l4s & 10 & 100 & 4 & sideways & 1.732 & 1.587 & 4.45 (2.80) & 6.32 (3.92) & 2.26 &0.017 (0.086) & 0.575 (0.543) & (17.6)\\ 
m10c2l2s & 10 & 100 & 2 & sideways & 1.414 & 1.357 & 4.25 (3.13) & 5.76 (4.20) & 2.05 & 0.017 (0.087) & 0.568 (0.524) & (14.8)\\ 

m10c2l8o85 & 10 & 100 & 8 & $85^{\circ}$ & 2.236 & 1.943 & 7.82 (4.03)& 8.40 (4.33) & 1.60 & 0.014 (0.087) & 0.514 (0.430) & (11.9)\\ 
m10c2l8o60 & 10 & 100 & 8 & $60^{\circ}$ & 2.236 & 1.943 & 8.92 (4.59)& 9.70 (5.00) & 1.62 & 0.013 (0.045) & 0.506 (0.401) & (14.9)\\ 
m10c2l8o30 & 10 & 100 & 8 & $30^{\circ}$ & 2.236 & 1.943 & 6.00 (3.09)& 8.32 (4.28) & 2.31 & 0.014 (0.071) & 0.579 (0.551) & (23.4)\\ 

m10c3l8s & 10 & $10^3$ & 8 & sideways & 2.236 & 1.943 & 6.05 (3.12) & 6.46 (3.33) & 1.49 & 0.0011 (0.011) & 0.491 (0.395) & (28.6)\\ 

m10c2l4o30 & 10 & 100 & 4 & $30^{\circ}$ & 1.732 & 1.587 & 6.07 (3.82)& 7.78 (4.83) & 2.19 & 0.013 (0.070) & 0.561 (0.510) & (19.9)\\ 
m10c2l2o30 & 10 & 100 & 2 & $30^{\circ}$ & 1.414 & 1.357 & 5.24 (3.86)& 6.54 (4.77) & 1.91 & 0.014 (0.065) & 0.585 (0.549) & (15.5)\\ 


m1.5c1l8s & 1.5 & 10 & 8 & sideways & 2.236 & 1.943 & 6.47 (3.33) & 25.4 (13.1) & 10.8 & 0.129 (0.24) & 0.317 (0.335) & (22.6)\\ 
m3c1l8s & 3 & 10 & 8 & sideways & 2.236 & 1.943 & 2.94 (1.51) & 12.8 (6.57) & 14.3 & 0.202 (0.39) & 0.526 (0.544) & (18.8)\\ 
m10c1l8s & 10 & 10 & 8 & sideways & 2.236 & 1.943 & 2.13 (1.10) & 10.1 (5.22) & 14.7 & 0.250 (0.49) & 0.603 (0.605) & (17.6)\\ 

m1.5c2l8s & 1.5 & 100 & 8 & sideways & 2.236 & 1.943 & 16.1 (8.30) &16.4 (8.45) & 2.47 & 0.0096 (0.043) & 0.266 (0.241) & (21.1) \\ 
m3c2l8s & 3 & 100 & 8 & sideways & 2.236 & 1.943 & 6.34 (3.27) & 8.21 (4.23) & 2.29 & 0.014 (0.076) & 0.492 (0.448) & (19.4)\\ 

\hline
\end{tabular}
\end{table*}

\begin{figure*}
\includegraphics[width=16.5cm]{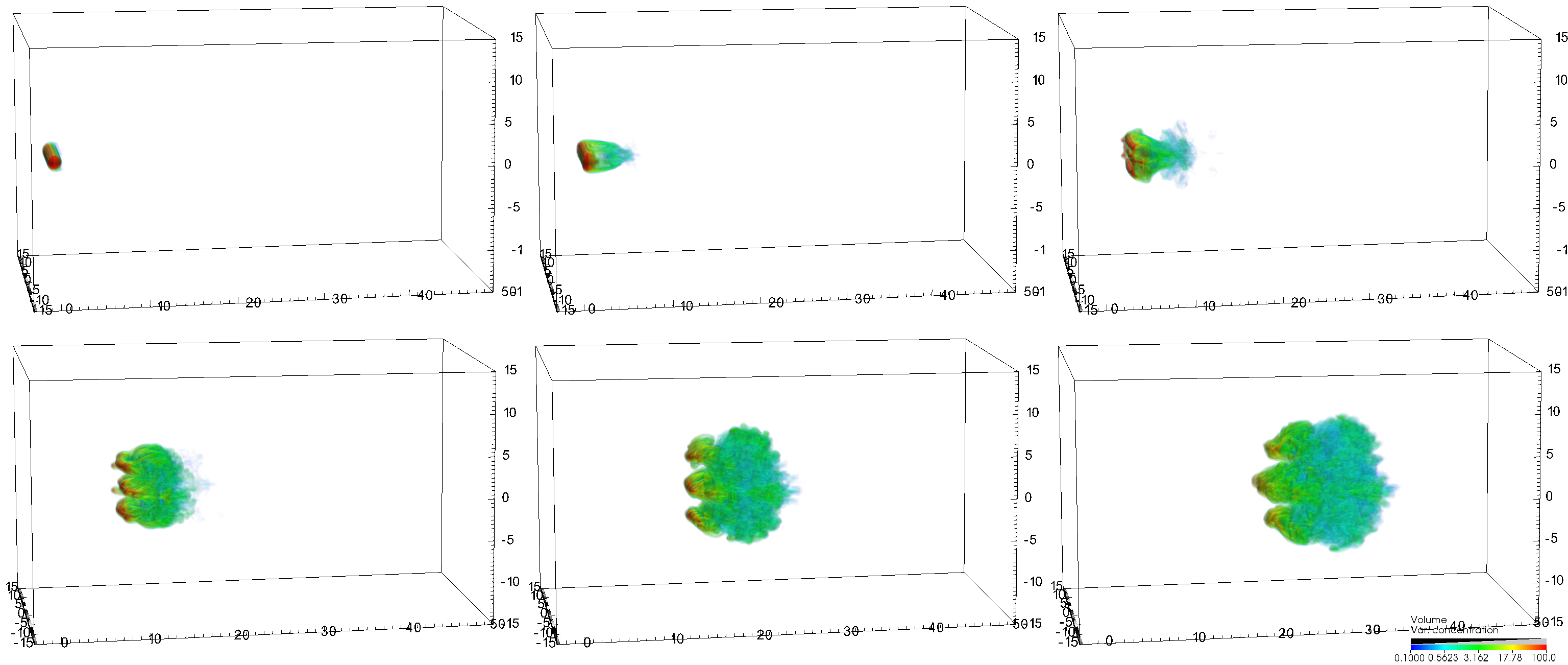}
\caption{A 3D volumetric rendering of the time evolution of simulation
  \emph{m10c2l8s}. From left to right and top to bottom the times are
  $t = 0.18, 0.88, 1.58, 2.28, 2.98$ and $3.68 \,t_{\rm cs}$ ($t = 0$ is defined
  as the time when the intercloud shock is level with the centre of
  the filament). The colour indicates the density of the filament
  material, normalized by the ambient density (i.e. the initial
  filament density is 100). Since the ambient material is not shown
  the bowshock upstream of the filament is not visible. The actual
  grid extends much further than the bounding box shown. $t_{\rm drag}
  = 2.42\,t_{\rm cs}$ and $t_{\rm mix} = 3.64\,t_{\rm cs}$.}
\label{fig:m10c2l8s_morphology}
\end{figure*}

\begin{figure*}
\includegraphics[width=16.5cm]{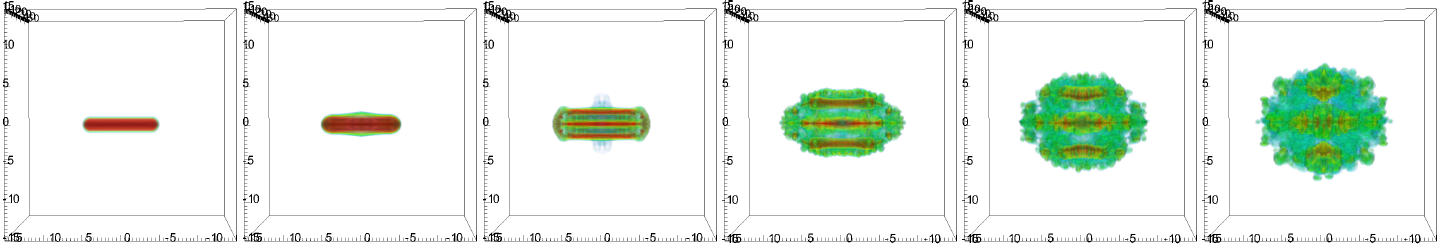}
\caption{As Fig.~\ref{fig:m10c2l8s_morphology} but viewing from the
  front. It is clear that the filament preferentially expands along
  its minor axis, parallel to the shock front.}
\label{fig:m10c2l8s_morphology_view2}
\end{figure*}

\begin{figure*}
\includegraphics[width=16.5cm]{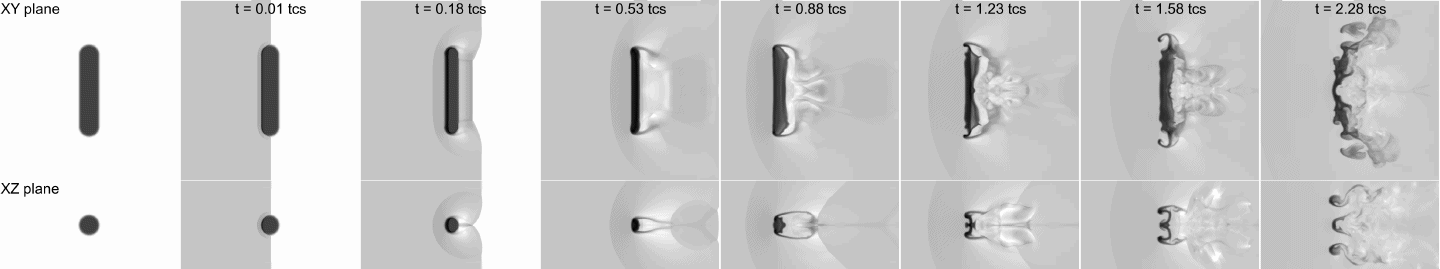}
\caption{Logarithmic density plots of the $XY$ and $XZ$ planes as a
  function of time for simulation \emph{m10c2l8s}. The grayscales shows
  the logarithm of the mass density, from $\rho_{\rm amb}$ (white) to
  $5\rho_{\rm c}$ (black). Each frame is labelled with the time. All
  frames show $-10 < Y < 10$ and $-5 < Z < 5$ (in units of $r_{\rm
    c}$). The first 4 frames show $-10 < X < 10$, while frames $5-7$ show
  $-5 < X < 15$ and frame 8 shows $0 < X < 20$. The shock is initially
at $X=-10$.}
\label{fig:m10c2l8s_XY_XZ_montage}
\end{figure*}

\begin{figure*}
\includegraphics[width=16.5cm]{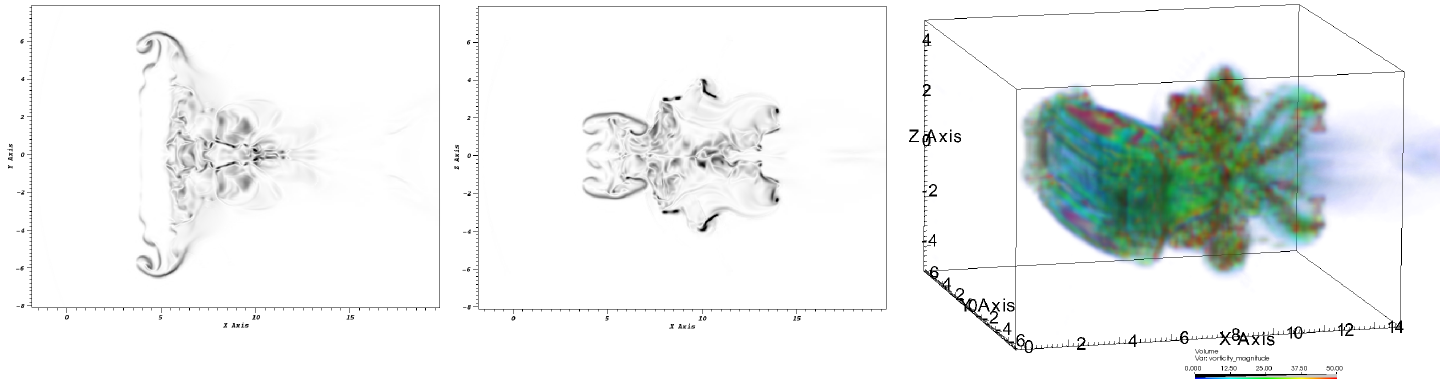}
\caption{The vorticity magnitude in simulation \emph{m10c2l8s} at
  $t=1.58\,t_{\rm cs}$. a) The $Z=0$ plane. b) The $Y=0$ plane. c) A
  3D volumetric rendering.}
\label{fig:m10fil8_vorticity_3D}
\end{figure*}

\begin{figure*}
\includegraphics[width=16.5cm]{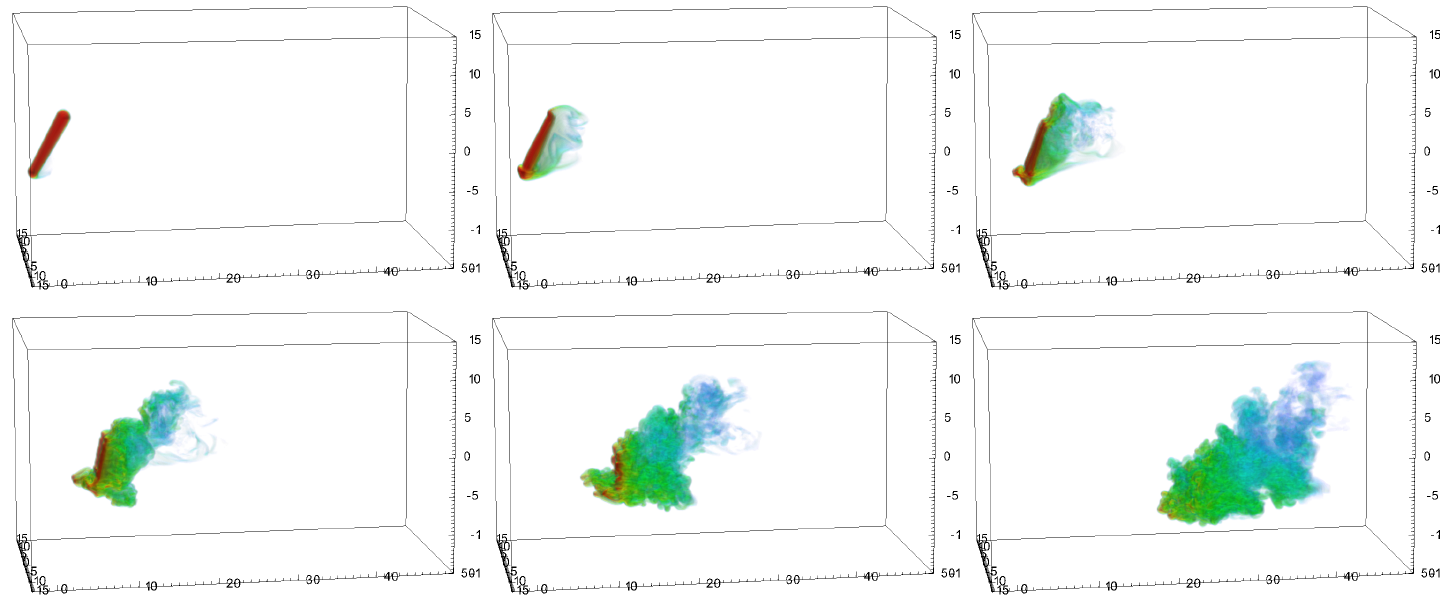}
\caption{As Fig.~\ref{fig:m10c2l8s_morphology} but for simulation
  \emph{m10c2l8o30}. Snapshots are at $t=0.18, 0.88, 1.58, 2.28, 2.98$ and
  $4.37 \,t_{\rm cs}$. $t_{\rm drag} = 3.09\,t_{\rm cs}$ and $t_{\rm mix} = 4.28\,t_{\rm cs}$.}
\label{fig:m10c2l8o30_morphology}
\end{figure*}

\begin{figure*}
\includegraphics[width=16.5cm]{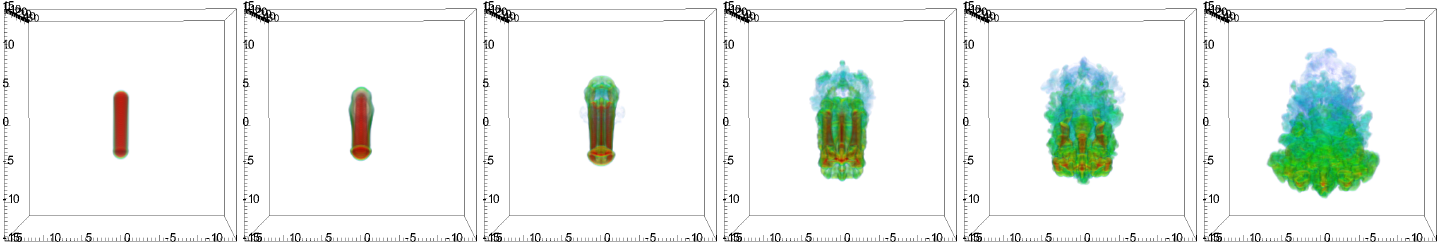}
\caption{As Fig.~\ref{fig:m10c2l8o30_morphology} but viewing from the front.}
\label{fig:m10c2l8o30_morphology_view2}
\end{figure*}

\begin{figure*}
\includegraphics[width=16.5cm]{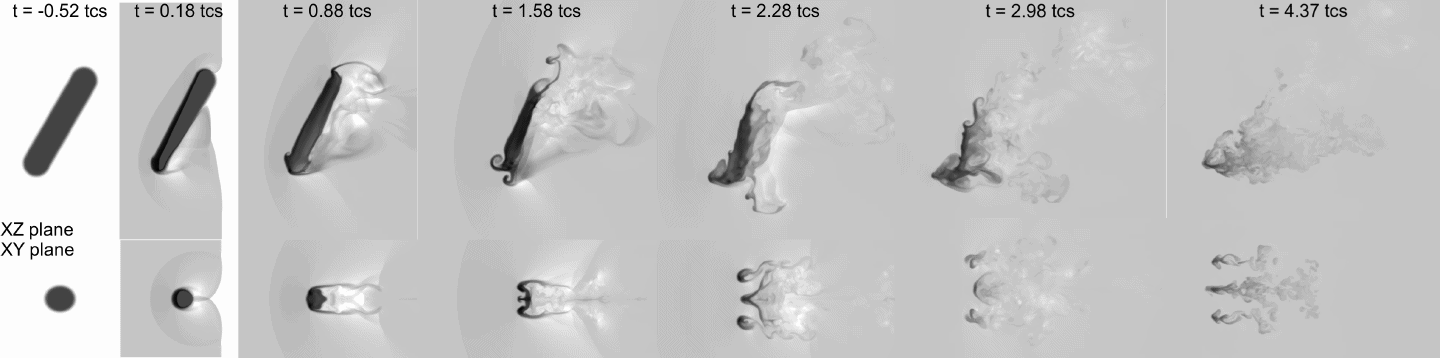}
\caption{Logarithmic density plots of the $Y=0$ and $Z=0$ (or $Z=-9$,
  for the panel at $t=4.37\,t_{\rm cs}$) planes as a function of time
  for simulation \emph{m10c2l8o30}. The grayscales shows the logarithm of
  the mass density, from $\rho_{\rm amb}$ (white) to $5\rho_{\rm c}$
  (black). Each frame is labelled with the time. The $(X,Y,Z)$ extent
  of the frames (in units of $r_{\rm c}$) as a function of time are:
  $(\pm5, \pm5, \pm10)$ at $t=-0.52$ and $0.18\,t_{\rm cs}$, $(-5$ to
  $10,\pm5,\pm10)$ at $t=0.88\,t_{\rm cs}$, $(-5$ to $15,\pm5,\pm10)$
  at $t=1.58\,t_{\rm cs}$, $(0-20,\pm5,\pm10)$ at $t=2.28\,t_{\rm
    cs}$, $(5-30,\pm6.25,\pm10)$ at $t=2.98\,t_{\rm cs}$, and
  $(15-46.25,\pm7.8125, -15$ to $10)$ at $t=4.37\,t_{\rm cs}$. The
  shock is initially at $X=-10$.}
\label{fig:m10c2l8o30_XY_XZ_montage}
\end{figure*}
 
\begin{figure}
\includegraphics[width=8.3cm]{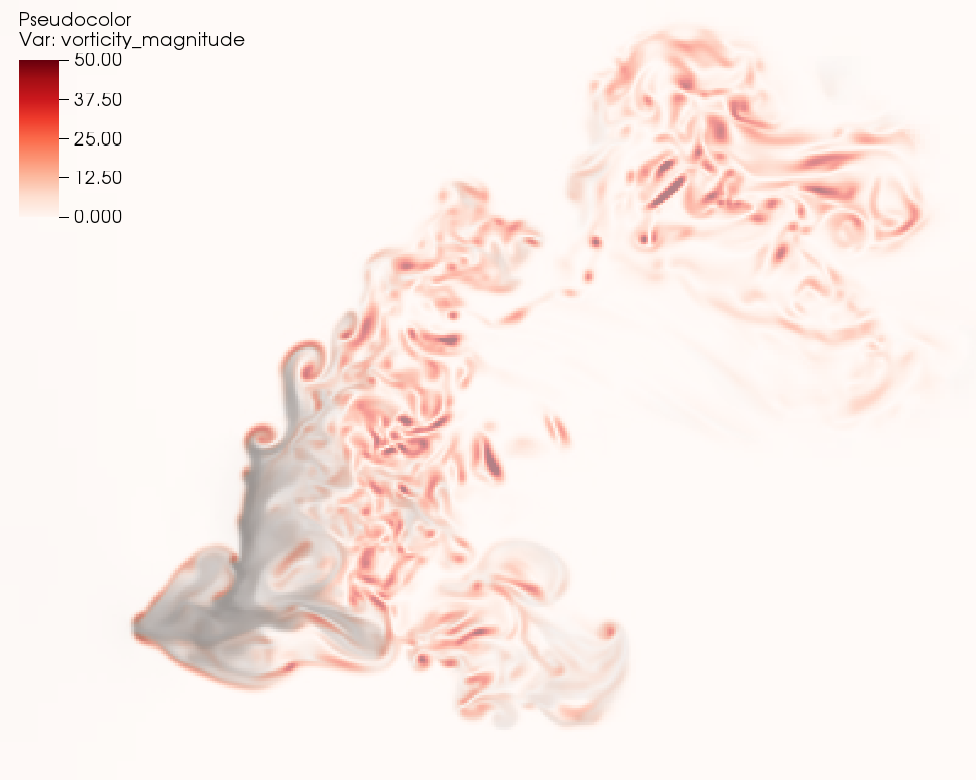}
\caption{Periodic vortex shedding in simulation \emph{m10c2l8o30}. The
  greyscale shows the logarithm of the mass density in the $XZ$ plane
  at $Y=0$ and $t=2.98\,t_{\rm cs}$. The redscale indicates the
  vorticity magnitude. The $(X,Z)$ extent
  of the frame (in units of $r_{\rm c}$) is $(5-30,\pm10)$.}
\label{fig:m10c2l8o30_XZ_vortexshedding}
\end{figure}

\subsection{Filament morphology and turbulence}
\label{sec:morphology}

\subsubsection{The interaction of an $M=10$ shock and a $\chi=10^{2}$
  sideways oriented ($\theta=0^{\circ}$) filament}

We begin by examining the morphology of the interaction for our
reference simulation, which has $M=10$, $\chi=10^{2}$, $l=8$, and has
the filament oriented sideways to the shock (simulation
\emph{m10c2l8s}). Volumetric renderings of the density of {\em
  filament} material are shown as a function of time in
Figs.~\ref{fig:m10c2l8s_morphology}
and~\ref{fig:m10c2l8s_morphology_view2}.  Because of this focus, in
these and similar figures, the bowshock and other features in the
ambient material are not visible. The snapshots shown in these figures
are at identical times. We also show the mass density on 2 orthogonal
slices through the simulation, as a function of time, in
Fig.~\ref{fig:m10c2l8s_XY_XZ_montage}. This figure shows the bowshock
and other features in the ambient material which are not visible in
Figs.~\ref{fig:m10c2l8s_morphology}
and~\ref{fig:m10c2l8s_morphology_view2}, and also adds 2 more
snapshots. Note that Fig.~\ref{fig:m10c2l8s_XY_XZ_montage} focusses on
slightly earlier times. The first panel shows the initial orientation
of the filament (the shock is located at the left edge of the
image). Shocks are easier to identify in
Fig.~\ref{fig:m10c2l8s_XY_XZ_montage}, while the 3D morphology is
better conveyed in Figs.~\ref{fig:m10c2l8s_morphology}
and~\ref{fig:m10c2l8s_morphology_view2}. Together,
Figs.~\ref{fig:m10c2l8s_morphology}-\ref{fig:m10c2l8s_XY_XZ_montage}
reveal the nature of the interaction.

Figs.~\ref{fig:m10c2l8s_morphology}-\ref{fig:m10c2l8s_XY_XZ_montage}
show the shock striking the filament from its side. A transmitted
shock is sent through the filament, while the external part of the
shock diffracts around the filament, remaining nearly normal to the
filament surface as it sweeps towards the downstream side. 
At $t=0.01\,t_{\rm cs}$ the
external shock has just passed the centre of the filament. The bowshock
upstream of the filament is visible, and the transmitted shock can be
seen entering the filament. At $t=0.18\,t_{\rm cs}$ the transmitted
shock is nearly half-way through the filament. Meanwhile the external
shock has completely swept over the filament. There is a region of
high pressure immediately downstream of the filament caused by the
focusing of the diffracted external shock and its subsequent
convergence on the $Z=0$ plane (this is best seen in the lower part of
the panel which shows the $XZ$ plane). This high pressure region
starts to drive a shock into the back of the filament, towards its
front surface. Also visible are the diffracted shocks which move from
the ends of the filament towards the $Y=0$ plane. These shocks are not
seen in spherical cloud interactions. Finally we note that some
material is already being ablated from the filament surface, from its
ends (see the $XY$ plane) and along its length (see the $XZ$ plane).

At $t=0.53\,t_{\rm cs}$ we see the filament at almost its most
compressed. The transmitted shock has just left the back of the
filament at this point, accelerating into the downstream ambient gas
(it is visible as the planar feature located just downstream of the
filament in Fig.~\ref{fig:m10c2l8s_XY_XZ_montage}).  As a consequence
of the transmitted shock reaching the back of the filament, a
rarefaction wave is launched and moves back upstream through the
filament. At this moment we see that the tips of the filament curve
upstream.  The expansion of the filament due to the rarefaction waves
passing through it is readily visible at $t=0.88\,t_{\rm cs}$. The
back of the filament has a ``fluted'' structure at this point (visible
in the $XZ$-plane in Fig.~\ref{fig:m10c2l8s_XY_XZ_montage}).
A strong tail-shock is also visible some distance downstream from the
back of the filament. 

At $t=1.23\,t_{\rm cs}$ the RT instability has given the filament a
3-pronged structure along its length, while additional RT fingers are
present at the filament ends. KH instabilities can be seen in the
$Z=0$ plane along the front of the filament at $t=1.58\,t_{\rm
  cs}$. By this time, the secondary shocks and rarefaction waves which
are moving through the filament cause it to ``reverberate''.
Vorticity deposited on the surface of the filament drives the growth
of KH instabilities which rip material from the filament and create a
complex tail. It is material in the tail which is visible above and
below the filament in the third panel of Fig.~\ref{fig:m10c2l8s_morphology_view2}.

The filament enters the late stages of its destruction by
$t=2.28\,t_{\rm cs}$, having been fragmented into 3 parallel ``rolls''
by the RT instability. The central roll displayed in the $Z=0$
$XY$-plane shows numerous small-scale features and signs of
instabilites, and is beginning to bend into an arc-like structure as
the ends of the rolls are pushed faster downstream by the external
flow.  RM fingers are visible on the front surface of the filament in
Fig.~\ref{fig:m10c2l8s_morphology}. The rolls are driven away from
each other as the postshock gas tries to flow between them. The wake
behind the filament reaches a turbulent-like state, and grows in size
as the filament material becomes increasingly mixed.

Fig.~\ref{fig:m10fil8_vorticity_3D} shows the vorticity magnitude in
simulation \emph{m10c2l8s} at $t=1.58\,t_{\rm cs}$. Vorticity grows at
the shear layers on the outside surface of the filament, and becomes
very prominent in the wake behind the filament. Vorticity is also
produced as the external shock moves around the filament ends and
interacts with the shock refracted over the top and bottom of the
filament. Later on, the vorticity field is highly intricate, and
reflects the complex nature of the interaction.

\begin{figure*}
\includegraphics[width=16.5cm]{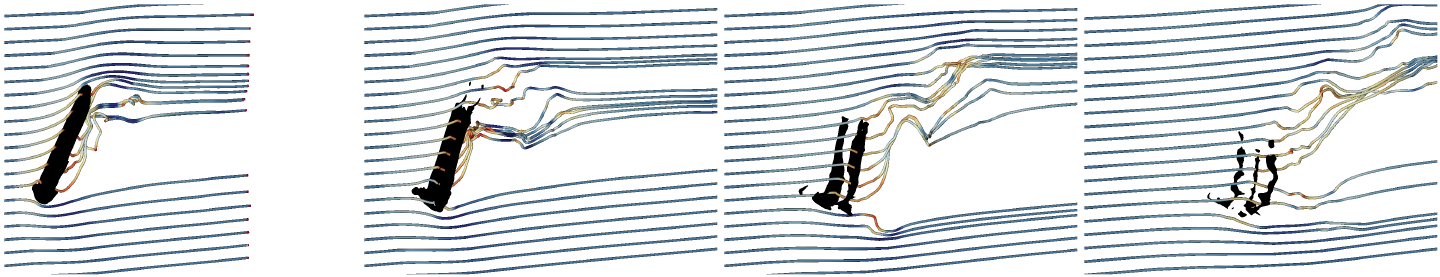}
\caption{Streamlines in simulation \emph{m10c2l8o30} at 
$t = 0.88, 1.58, 2.28$ and $2.98\,t_{\rm cs}$.
  The streamlines are in the $Y=0$ plane and
  are introduced at $X=-5$. They are colour coded with the velocity
  from dark red (low speed) to dark blue (high speed). The black
  surface is a contour at the original filament density.}
\label{fig:m10fil1_streamlines}
\end{figure*}

\begin{figure*}
\includegraphics[width=16.5cm]{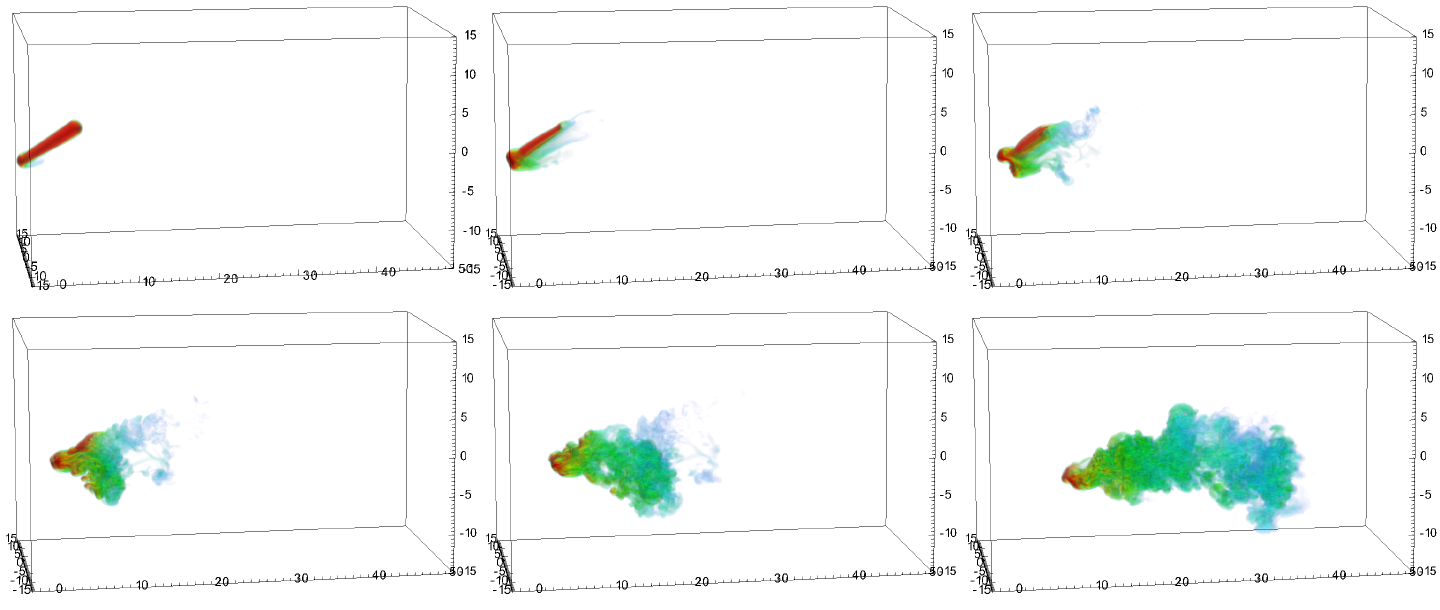}
\caption{As Fig.~\ref{fig:m10c2l8s_morphology} but for simulation
  \emph{m10c2l8o60}. Snapshots are at $t=0.18, 0.88, 1.58, 2.28, 2.98$ and
  $4.37 \,t_{\rm cs}$. $t_{\rm drag} = 4.59\,t_{\rm cs}$ and $t_{\rm mix} = 5.00\,t_{\rm cs}$.}
\label{fig:m10c2l8o60_morphology}
\end{figure*}

\begin{figure*}
\includegraphics[width=16.5cm]{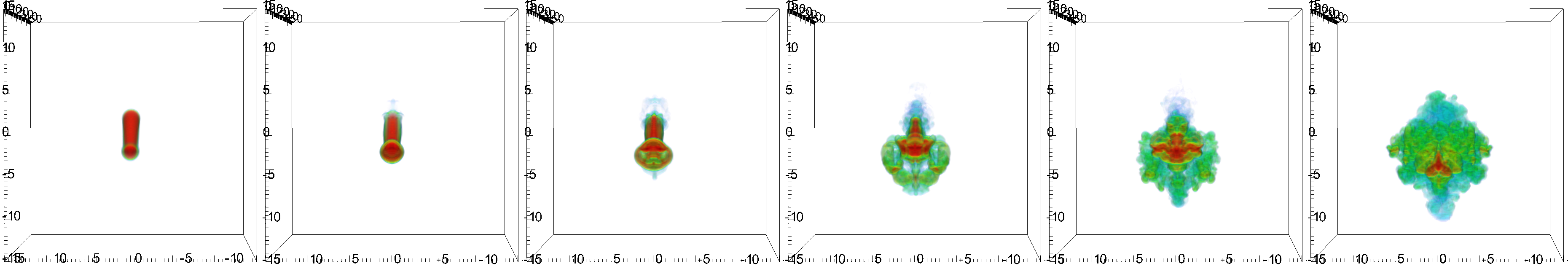}
\caption{As Fig.~\ref{fig:m10c2l8o60_morphology} but viewing from the front.}
\label{fig:m10c2l8o60_morphology_view2}
\end{figure*}

\subsubsection{Dependence on the filament orientation}
The interaction with an obliquely-oriented filament, specifically
simulation \emph{m10c2l8o30}, is shown in
Figs.~\ref{fig:m10c2l8o30_morphology}-\ref{fig:m10c2l8o30_XY_XZ_montage}. Here,
the filament's major axis is at a $30^{\circ}$ angle to the shock
surface. The shock interacts first with the upstream end of the
filament, and because of the inclination of the filament the
transmitted shock is at an angle to the long axis of the filament. At
$t=0.18\,t_{\rm cs}$, Fig.~\ref{fig:m10c2l8o30_XY_XZ_montage} shows
that the external shock which has swept over the centre of the
filament interacts with the shock which is diffracted around the
bottom of the filament. The high pressure region which forms behind
the filament when the external shock converges on the $Y=0$ plane can
be seen to drive a transmitted shock into the rear of the filament. A
vortex ring forms around the bottom end of the filament (see
Figs.~\ref{fig:m10c2l8o30_morphology} and
\ref{fig:m10c2l8o30_morphology_view2}) and at $t=0.88\,t_{\rm cs}$
regions of high vorticity spill around the bottom end of the
filament. A RT instability forms at the top of the filament and a wake
with turbulent-like characteristics is created downstream of the
filament. At $t=1.58\,t_{\rm cs}$ the most turbulent part of the wake
is near the top of the filament, while KH rolls are seen at the
bottom. A fragment breaks off the bottom of the filament by
$t=2.28\,t_{\rm cs}$. The wake extends above and downstream of the
filament, and behind the broken off fragment, and appears to become
increasingly ``turbulent'' with time. Periodic vortex shedding occurs
off the top part of the filament, due to the high pressure on the
upstream side of the filament leaking around the end of the
filament. This is highlighted in
Fig.~\ref{fig:m10c2l8o30_XZ_vortexshedding} which shows a number of
eddies forming and being advected downstream. As a result of its
initial orientation, the filament as a whole is pushed downwards by
the passage of the shock and the subsequent post-shock flow. The
face-on view displayed in Fig.~\ref{fig:m10c2l8o30_morphology_view2}
and the density plots in the $XY$ plane displayed in the part bottom
of Fig.~\ref{fig:m10c2l8o30_XY_XZ_montage} show again that the
filament breaks into 3 parallel rolls. Comparison with the reference
simulation reveals that the interaction has major differences at all
stages. At late times the wake is also noticeably longer than that in
simulation \emph{m10c2l8s}.

Fig.~\ref{fig:m10fil1_streamlines} shows streamlines in simulation
\emph{m10c2l8o30} at 4 different snapshots in time. The densest part
of the filament is highlighted by the black surface. The flow around
the filament is clearly complex. In the top left panel the streamlines
end at the shock towards the right of the plot. Moving along the
streamlines from left to right, the first change in colour and/or
direction of the streamlines indicates the position of the bowshock
upstream of the filament, with the postshock flow being slower and
typically changed in direction. On the upstream side of the filament
the streamlines bend upwards as the flow attempts to move around the
filament. The flow also accelerates around the bottom and top ends of
the filament, indicated by a darker blue colour to the streamlines at
this point. The wake behind the filament is complex, with the velocity
along some streamlines showing abrupt changes in velocity and
direction. A vortex, indicated by spiralling streamlines, is visible
downstream of the top part of the filament. There is also a clear gap
in the downstream wake where no streamlines exist. This indicates that
the flow in this region originally originated from outside the $Y=0$
plane. Similar flow behaviour is seen in the other panels. The
viewpoint is the same for all panels, so the downstream movement of
the filament can be appreciated. In the final panel we see the
filament has fragmented into 3 parts
(cf. Figs~\ref{fig:m10c2l8o30_morphology}-\ref{fig:m10c2l8o30_XY_XZ_montage}).

Figs.~\ref{fig:m10c2l8o60_morphology} and
~\ref{fig:m10c2l8o60_morphology_view2} show the nature of the
interaction when the filament is inclined yet further to the shock
(specifically simulation \emph{m10c2l8o60}). The main difference to
simulation \emph{m10c2l8o30} is that the filament does not form into 3
parallel rolls. Instead, the vortex ring which forms at the upstream
end of the filament is more dominant, and is asymmetric around the
filament because of the filament's inclination. This causes the
filament to fragment from the bottom, but the larger fragments are
from the top part of the ring. Comparing
Figs.~\ref{fig:m10c2l8s_morphology},~\ref{fig:m10c2l8o30_morphology}
and~\ref{fig:m10c2l8o60_morphology} we see a further extension in the
length of the wake at late times. This is because the filament loses
mass more slowly and has a longer lifetime than those in simulations
\emph{m10c2l8s} and \emph{m10c2l8o30} (see
Fig.~\ref{fig:mcore_var1}c).  The wake also has a smaller transverse
cross-sectional area than seen in the previous simulations.

\begin{figure*}
\includegraphics[width=16.5cm]{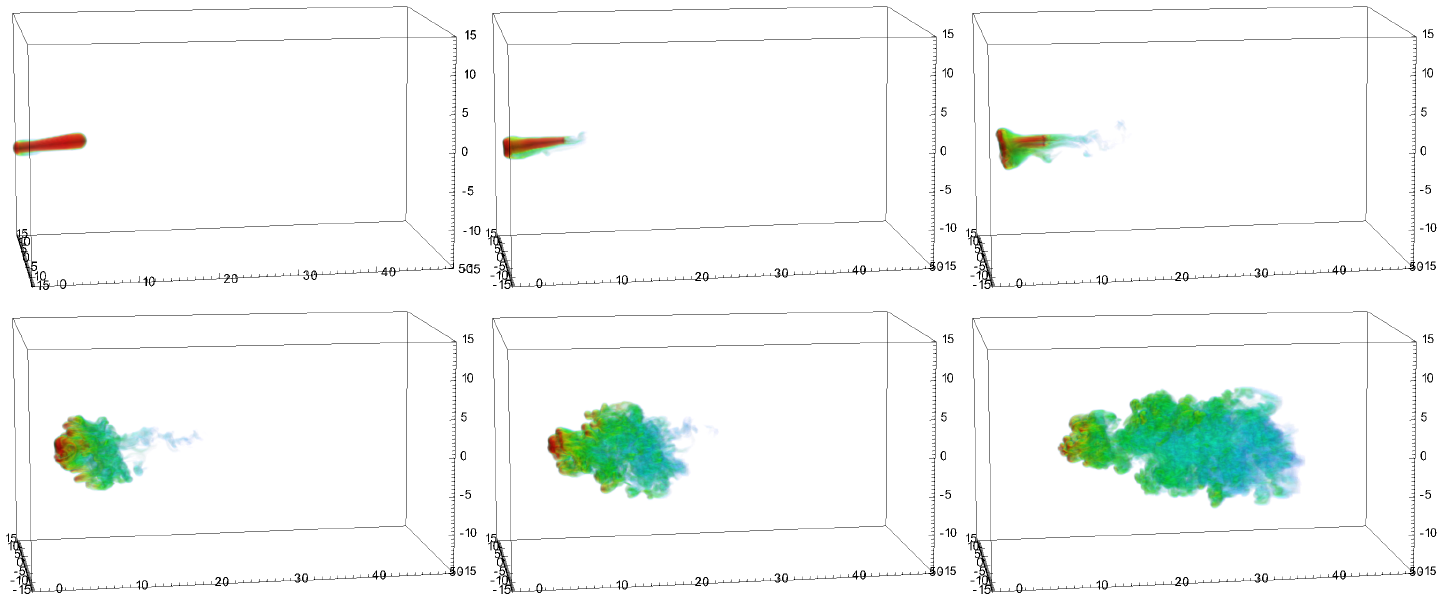}
\caption{As Fig.~\ref{fig:m10c2l8s_morphology} but for simulation
  \emph{m10c2l8o85}. Snapshots are at $t=0.18, 0.88, 1.58, 2.28, 2.98$ and
  $4.37 \,t_{\rm cs}$. $t_{\rm drag} = 4.03\,t_{\rm cs}$ and $t_{\rm mix} = 4.33\,t_{\rm cs}$.}
\label{fig:m10c2l8o85_morphology}
\end{figure*}

\begin{figure*}
\includegraphics[width=16.5cm]{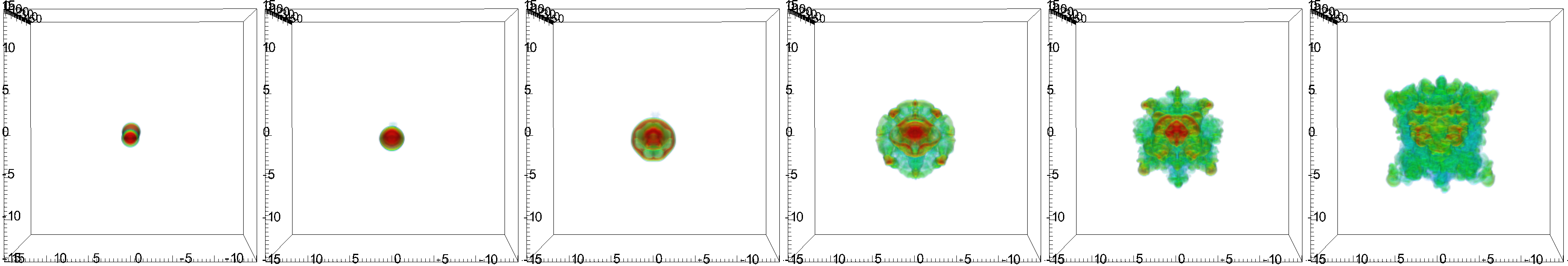}
\caption{As Fig.~\ref{fig:m10c2l8o60_morphology} but viewing from the front.}
\label{fig:m10c2l8o85_morphology_view2}
\end{figure*}

\begin{figure*}
\includegraphics[width=16.5cm]{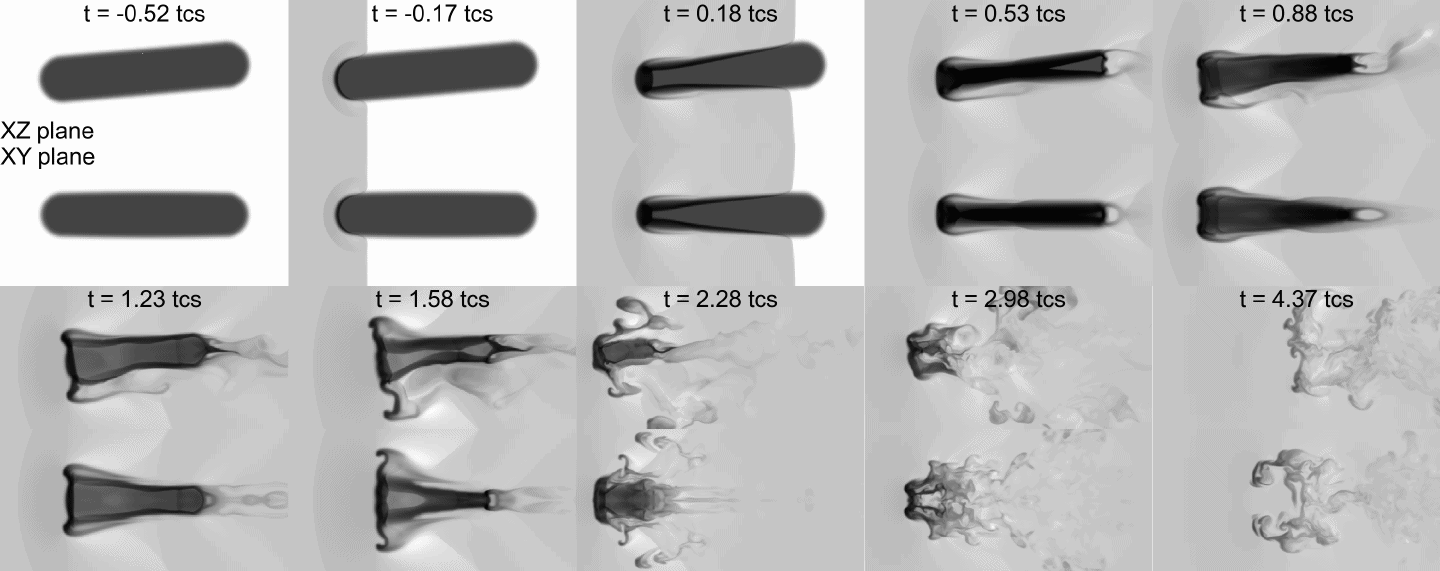}
\caption{Logarithmic density plots of the $Y=0$ and $Z=0$ planes as a
  function of time for simulation \emph{m10c2l8o85}. The grayscales shows
  the logarithm of the mass density, from $\rho_{\rm amb}$ (white) to
  $5\rho_{\rm c}$ (black). Each frame is labelled with the time. The
  $(X,Y,Z)$ extent of the frames (in units of $r_{\rm
    c}$) as a function of time are: $(\pm3.5, \pm3.5, \pm7)$ at $t=-0.52,-0.17,0.18$
  and $0.53\,t_{\rm cs}$, $(-5$ to $9,\pm3.5,\pm3.5)$ at $t=0.88,1.23$
  and $1.58\,t_{\rm
    cs}$, and $(0$ to $20,\pm5,\pm5)$ at $t=2.28,2.98$ and $4.37\,t_{\rm cs}$. The shock is initially
at $X=-10$.}
\label{fig:m10c2l8o85_XY_XZ_montage}
\end{figure*}

In Figs.~\ref{fig:m10c2l8o85_morphology},
\ref{fig:m10c2l8o85_morphology_view2}
and~\ref{fig:m10c2l8o85_XY_XZ_montage} we show the interaction of a
shock with a filament which is almost end-on to the shock (simulation
\emph{m10c2l8o85}). The transmitted shock into the filament now moves
along its length, as does the vortex ring formed at its upstream
end. However, the transmitted shock is $\approx 10\times$ slower than
the external shock, so that shocks are also driven into the filament
from its sides before the end-transmitted shock has had time to move
very far along the filament length. These ``side'' shocks move almost
perpendicular to the long filament axis. They pass through the centre
of the filament at $\sim0.5-0.6\,t_{\rm cs}$. Shortly after this they
reach the opposite surface and cause rarefaction waves which move back
towards the filament's central axis. This results in the dramatic
``hollowing'' or ``voiding'' of the filament, seen most prominently at
$t=1.23\,t_{\rm cs}$ in Fig.~\ref{fig:m10c2l8o85_XY_XZ_montage}. At
about this time, the vortex ring triggers the destruction of the
upstream end of the filament, exposing material further along the
length of the filament to the external flow. This material then
fragments as non-azimuthal instability modes grow. Note that the
filament is almost completely destroyed before the transmitted shock
has passed all the way through it, which doesn't happen until
$t\approx 5\,t_{\rm cs}$ (the transmitted shock can be seen just
downstream of the ``head'' of the filament at $t=1.23, 1.58$ and
$2.28\,t_{\rm cs}$ in Fig.~\ref{fig:m10c2l8o85_XY_XZ_montage}). Again,
the filament wake has a relatively compact cross-section.  The $XZ$
and $XY$ snapshots shown in Fig.~\ref{fig:m10c2l8o85_XY_XZ_montage}
also reveal how the initial interaction differs when the long-axis of
the filament is at a small angle to the shock normal (as in the $XZ$
plane), as compared to exactly aligned with it (as in the $XY$ plane).

\subsubsection{Dependence on filament length}
\begin{figure*}
\includegraphics[width=16.5cm]{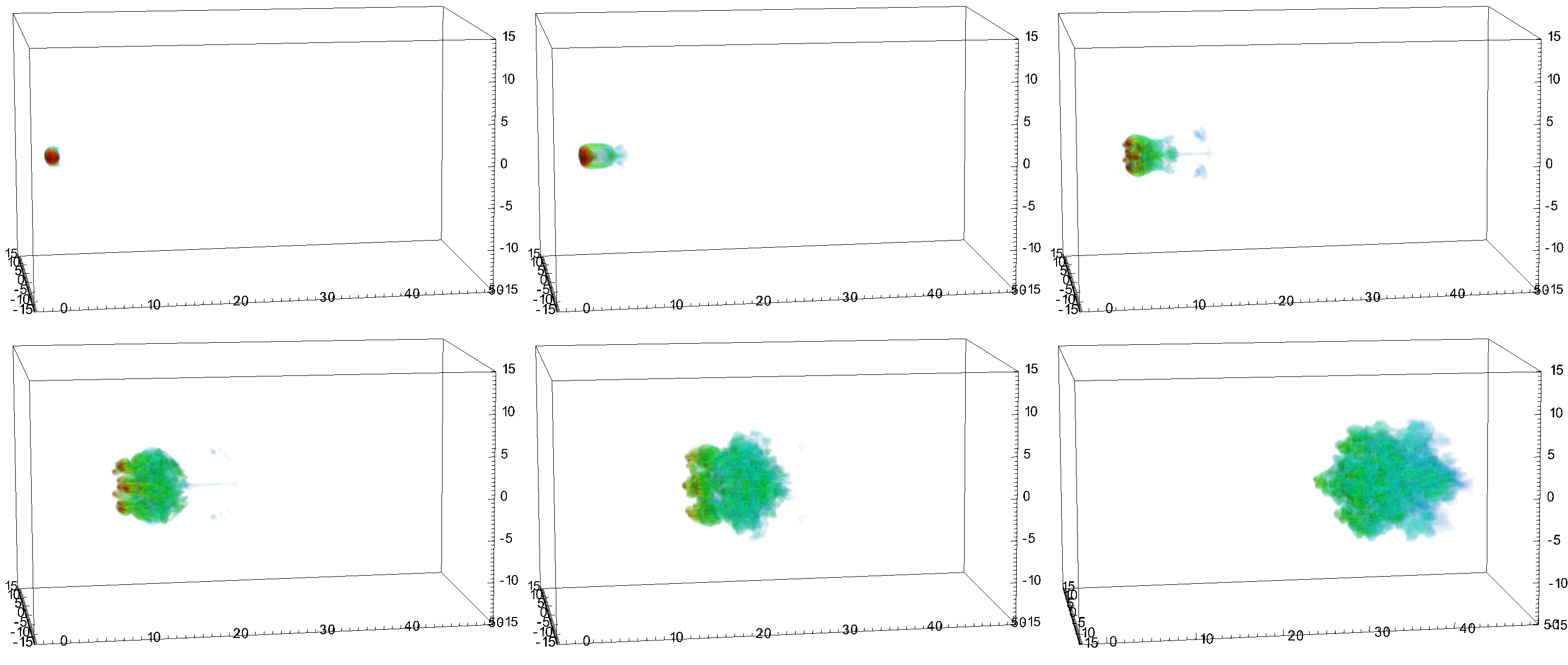}
\caption{As Fig.~\ref{fig:m10c2l8s_morphology} but for simulation
  \emph{m10c2l2s}. Snapshots are at $t=0.26. 1.26, 2.26, 3.26, 4.26$
  and $6.26 \,t_{\rm cs}$. $t_{\rm drag} = 3.13\,t_{\rm cs}$ and $t_{\rm mix} = 4.20\,t_{\rm cs}$.}
\label{fig:m10c2l2s_morphology}
\end{figure*}

\begin{figure*}
\includegraphics[width=16.5cm]{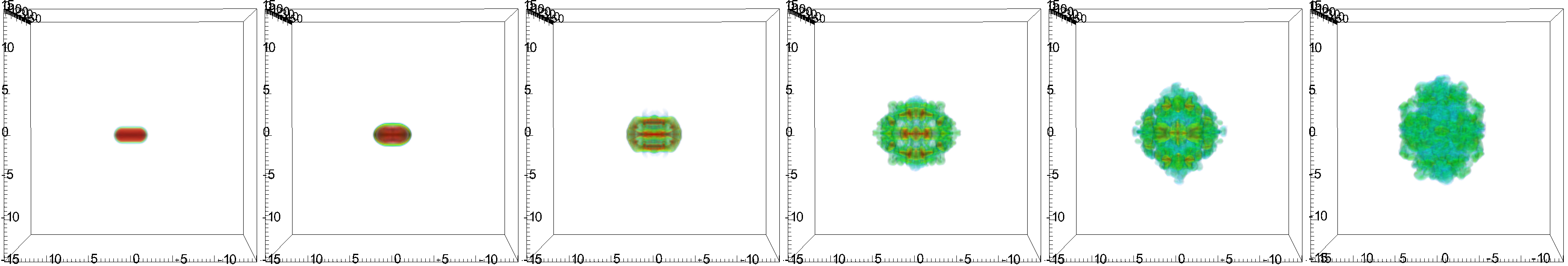}
\caption{As Fig.~\ref{fig:m10c2l2s_morphology} but viewing from the front.}
\label{fig:m10c2l2s_morphology_view2}
\end{figure*}

The 3 parallel rolls seen in simulation \emph{m10c2l8s} are a striking
feature. Since spherical clouds do not break up in this way, we wish
to determine what length of filament is needed for these rolls to
form. Figs.~\ref{fig:m10c2l2s_morphology}
and~\ref{fig:m10c2l2s_morphology_view2} show results from simulation
\emph{m10c2l2s}, where $l$ is reduced to 2.  It is clear that 3
parallel rolls still form, and the filament morphology evolves in a
broadly similar way to that in simulation \emph{m10c2l8s}.

\begin{figure*}
\includegraphics[width=16.5cm]{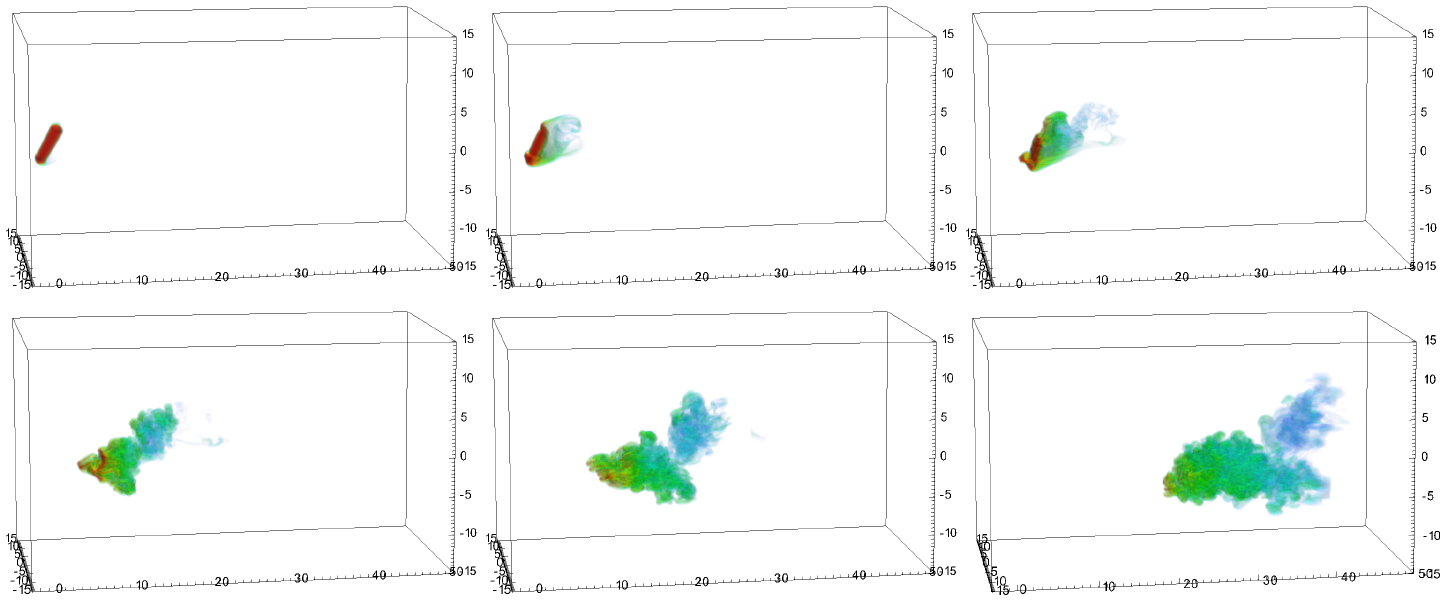}
\caption{As Fig.~\ref{fig:m10c2l8s_morphology} but for simulation
  \emph{m10c2l4o30}. Snapshots are at $t=0.23, 1.08, 1.93,
  2.79, 3.64$ and $5.35\,t_{\rm cs}$. $t_{\rm drag}
  = 3.82\,t_{\rm cs}$ and $t_{\rm mix} = 4.83\,t_{\rm cs}$.}
\label{fig:m10c2l4o30_morphology}
\end{figure*}

\begin{figure*}
\includegraphics[width=16.5cm]{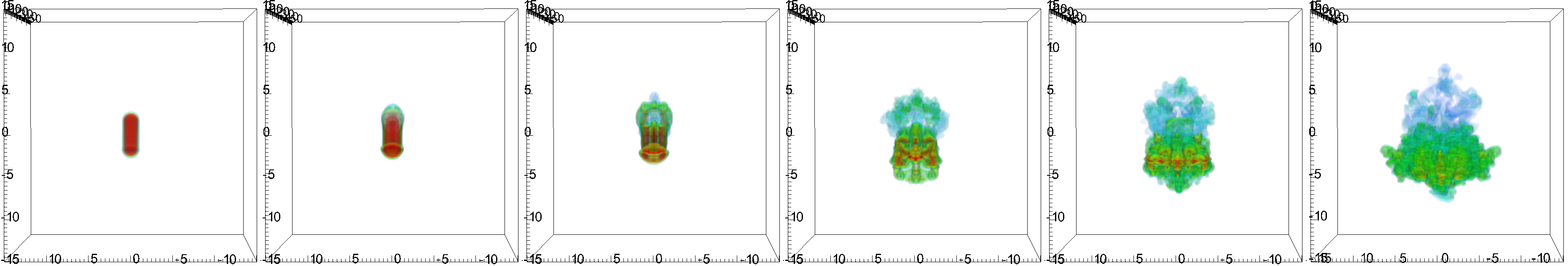}
\caption{As Fig.~\ref{fig:m10c2l4o30_morphology} but viewing from the front.}
\label{fig:m10c2l4o30_morphology_view2}
\end{figure*}

\begin{figure*}
\includegraphics[width=16.5cm]{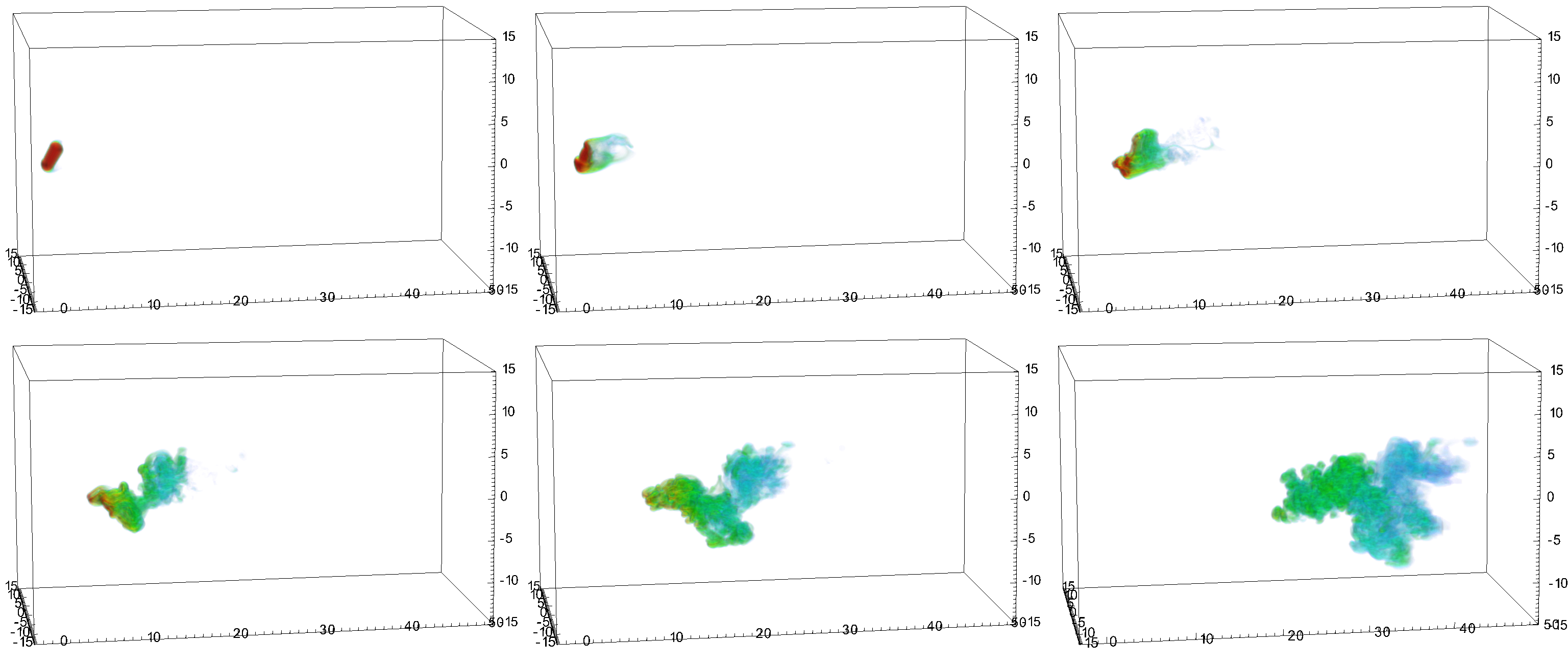}
\caption{As Fig.~\ref{fig:m10c2l8s_morphology} but for simulation
  \emph{m10c2l2o30}. Snapshots are at $t= 0.26. 1.26, 2.26,
  3.26, 4.26$ and $6.26\,t_{\rm cs}$. $t_{\rm drag} = 3.86\,t_{\rm cs}$ and $t_{\rm mix} = 4.77\,t_{\rm cs}$.}
\label{fig:m10c2l2o30_morphology}
\end{figure*}

\begin{figure*}
\includegraphics[width=16.5cm]{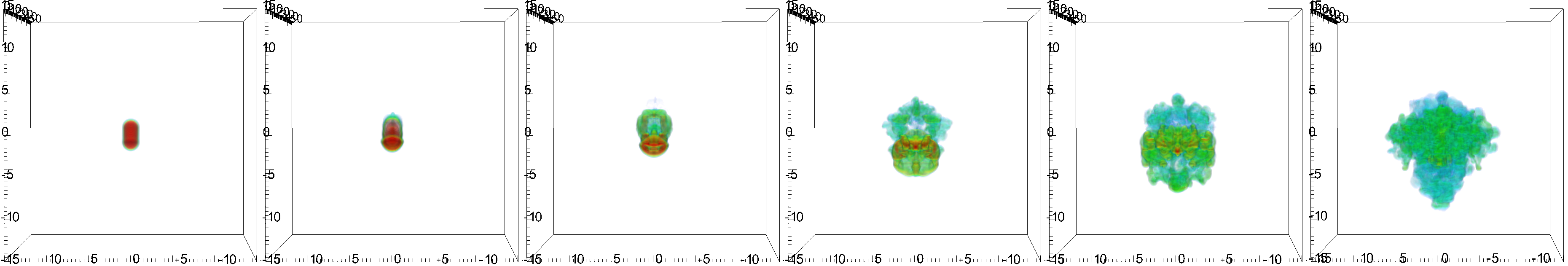}
\caption{As Fig.~\ref{fig:m10c2l2s_morphology} but viewing from the front.}
\label{fig:m10c2l2o30_morphology_view2}
\end{figure*}

It is also interesting to see how shorter filaments respond when they
have an oblique orientation to the
shock. Figs.~\ref{fig:m10c2l4o30_morphology}
and~\ref{fig:m10c2l4o30_morphology_view2} show results from simulation
\emph{m10c2l4o30}, where $l=4$ and $\theta=30^{\circ}$. This
simulation also develops 3 parallel rolls, in keeping with simulations
\emph{m10c2l4s} and \emph{m10c2l8o30}.  Reducing $l$ to 2 and keeping
$\theta=30^{\circ}$, for the first time (when $\theta < 60^{\circ}$)
we find that the interaction does not develop 3 parallel rolls - see
Figs.~\ref{fig:m10c2l2o30_morphology} and
~\ref{fig:m10c2l2o30_morphology_view2} - in contrast to simulations
\emph{m10c2l2s} and \emph{m10c2l4o30}.

\subsubsection{Mach number dependence}
All the results shown so far have involved a Mach 10 shock. Previous
work on shocks interacting with spherical clouds has revealed that the
interaction is milder, and that instabilities develop more slowly,
when the Mach number of the shock is reduced. We now wish to see how
this behaviour translates to a shock interacting with a filament.
 
\begin{figure}
\includegraphics[width=8.25cm]{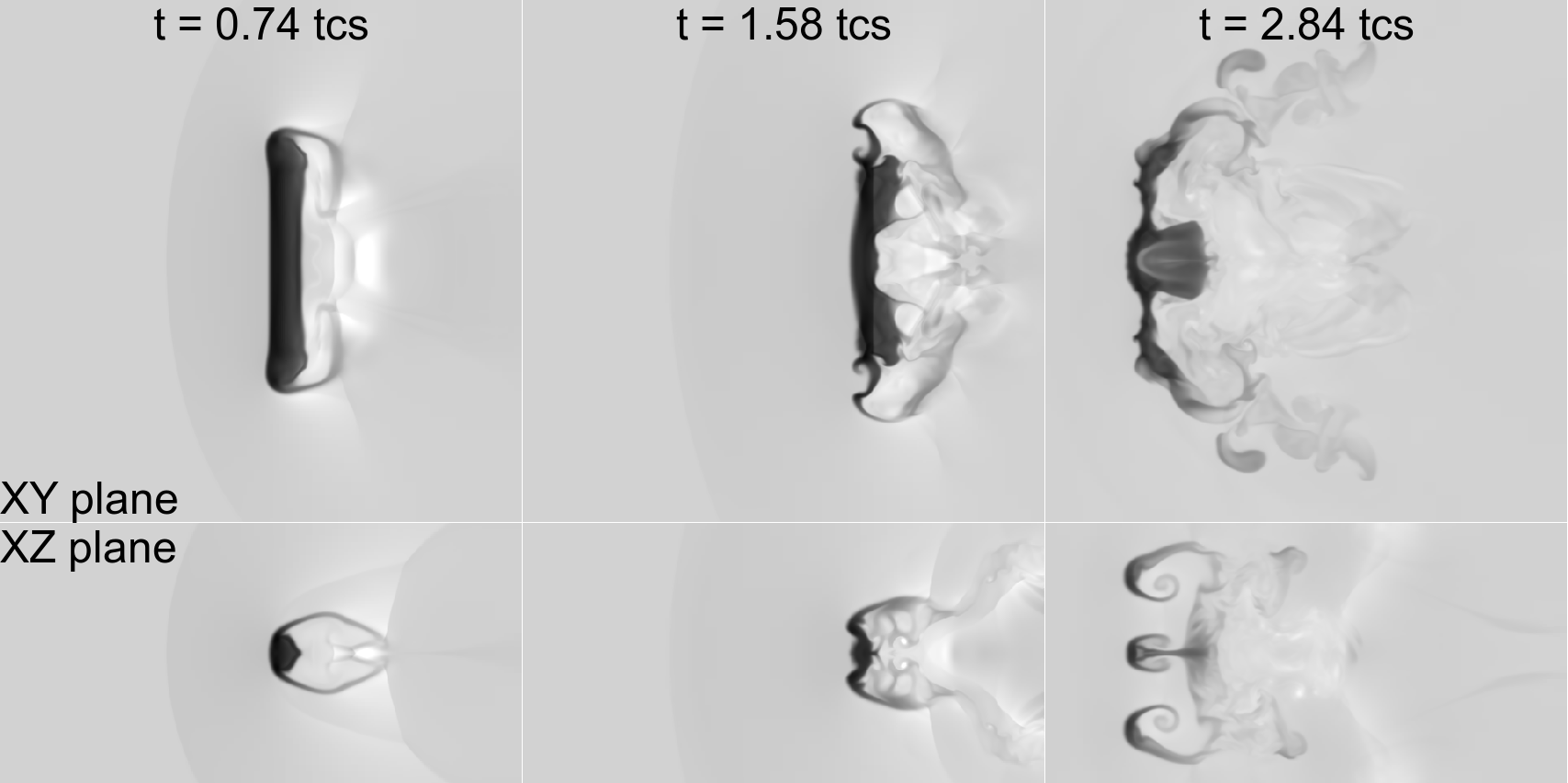}
\caption{Logarithmic density plots of the $XY$ and $XZ$ planes as a
  function of time for simulation \emph{m3c2l8s}. The grayscales shows
  the logarithm of the mass density, from $\rho_{\rm amb}$ (white) to
  $5\rho_{\rm c}$ (black). Each frame is labelled with the time.  All
  frames show $-10 < Y < 10$ and $-5 < Z < 5$ (in units of $r_{\rm
    c}$). The first two frames show $-10 < X < 10$, while the last
  shows $5 < X < 25$. The shock is initially at $X=-10$. $t_{\rm drag}
  = 3.27\,t_{\rm cs}$ and $t_{\rm mix} = 4.23\,t_{\rm cs}$.}
\label{fig:m3c2l8s_XY_XZ_montage}
\end{figure}

Fig.~\ref{fig:m3c2l8s_XY_XZ_montage} shows the interaction of a Mach 3
shock with a $\chi=10^{2}$, sideways-oriented filament (simulation
\emph{m3c2l8s}). The interaction is less ``violent'' compared to
simulations with higher Mach numbers (e.g., simulation \emph{m10c2l8s}
shown in
Figs.~\ref{fig:m10c2l8s_morphology}-\ref{fig:m10c2l8s_XY_XZ_montage}). The
transmitted shock moves into the filament relatively quicker, which
changes the timing and position of secondary shocks and rarefactions
within the filament, and its reverberations. The post-shock
compression is reduced, as is the velocity shear over the filament.
In particular, it takes longer to form the 3 parallel rolls (compare
the frames at $t=1.58 t_{\rm cs}$), though they still appear
later. Nevertheless, the interaction proceeds broadly as before.

\begin{figure*}
\includegraphics[width=16.5cm]{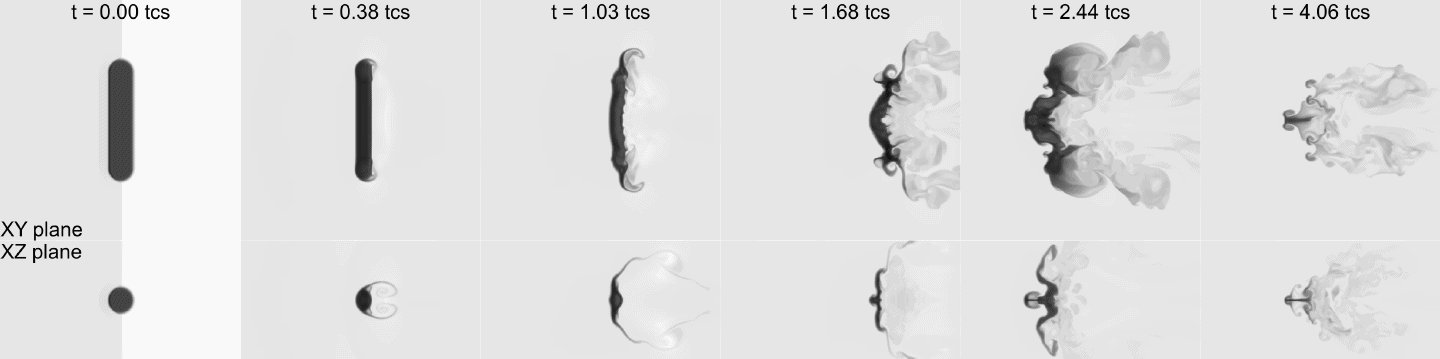}
\caption{Logarithmic density plots of the $XY$ and $XZ$ planes as a
  function of time for simulation \emph{m1.5c2l8s}. The grayscales
  shows the logarithm of the mass density, from $\rho_{\rm amb}$
  (white) to $5\rho_{\rm c}$ (black). Each frame is labelled with the
  time.  All frames show $-10 < X < 10$, $-10 < Y < 10$ and $-5 < Z <
  5$ (in units of $r_{\rm c}$). The shock is initially at
  $X=-10$. $t_{\rm drag} = 8.30\,t_{\rm cs}$ and $t_{\rm mix} =
  8.45\,t_{\rm cs}$.}
\label{fig:m1.5c2l8s_XY_XZ_montage}
\end{figure*}

A further reduction in the Mach number to 1.5 yields a still gentler
interaction (see Fig.~\ref{fig:m1.5c2l8s_XY_XZ_montage}). The external
post-shock flow is now subsonic with respect to the filament, so a
bow-wave rather than a bow-shock forms upstream. The flow around the
filament appears much more laminar, as can be seen from the large
eddies present in the filament wake in the $XZ$ plane at
$t=0.38\,t_{\rm cs}$. The drag and mixing timescales are both much
increased.

\subsubsection{Dependence on density contrast}
We now examine how the interaction changes with the density contrast
of the filament. All results so far have been for filaments with
$\chi=10^{2}$. In the interaction of a shock with a spherical cloud,
increasing the density contrast creates a more ``resistant'' obstacle,
which accelerates more slowly ($t_{\rm drag} \propto
\chi^{1/2}\,t_{\rm cc}$) and which is subject to increased KH
instabilities and hydrodynamic ablation - decreasing $\chi$ has the
opposite effect \citep[see, e.g.,][]{Pittard:2010,Pittard:2016}.

\begin{figure*}
\includegraphics[width=16.5cm]{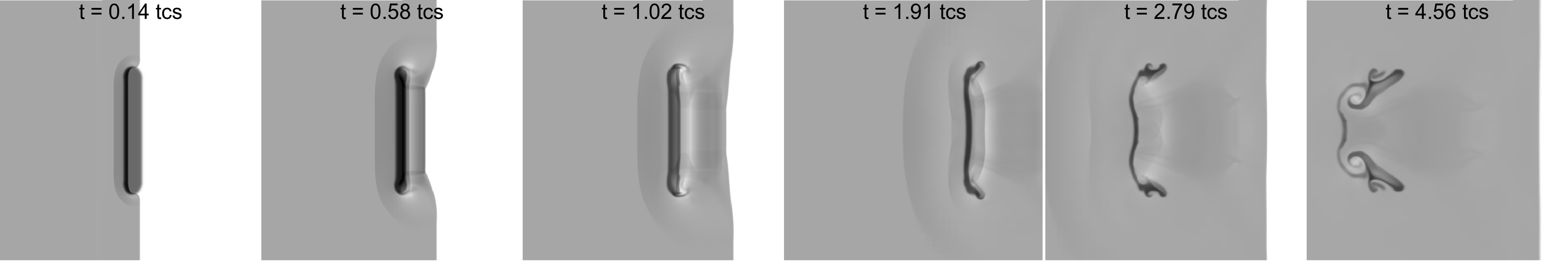}
\caption{Logarithmic density plots of the $XY$ plane as a
  function of time for simulation \emph{m10c1l8s}. The grayscales shows
  the logarithm of the mass density, from $\rho_{\rm amb}$ (white) to
  $5\rho_{\rm c}$ (black). Each frame is labelled with the time. All
  frames show $-10 < Y < 10$ (in units of $r_{\rm
    c}$). The first 4 frames show $-10 < X < 10$, while the sixth
  shows $0 < X < 20$. The final frame
  shows $10 < X < 30$. The shock is initially at $X=-10$. $t_{\rm drag}
  = 1.10\,t_{\rm cs}$ and $t_{\rm mix} = 5.22\,t_{\rm cs}$.}
\label{fig:m10c1l8s_XY_montage}
\end{figure*}

Fig.~\ref{fig:m10c1l8s_XY_montage} shows the effect of reducing
$\chi$ to 10 (simulation \emph{m10c1l8s}). The transmitted shock
moves much faster through the filament relative to the speed of the
external shock in this case (it is $\approx 3\times$ slower). The
filament experiences a much faster acceleration downstream, and
becomes quite thin (see the panel at $t=1.91\,t_{\rm cs}$ in
Fig.~\ref{fig:m10c1l8s_XY_montage}). At $t=1.91\,t_{\rm cs}$, we see
that shocks exist in the external flow either side of the filament, in
addition to the bowshock further upstream. These are caused by the
transmitted shocks exiting the filament, and accelerating into the lower
density surrounding flow.

\begin{figure*}
\includegraphics[width=16.5cm]{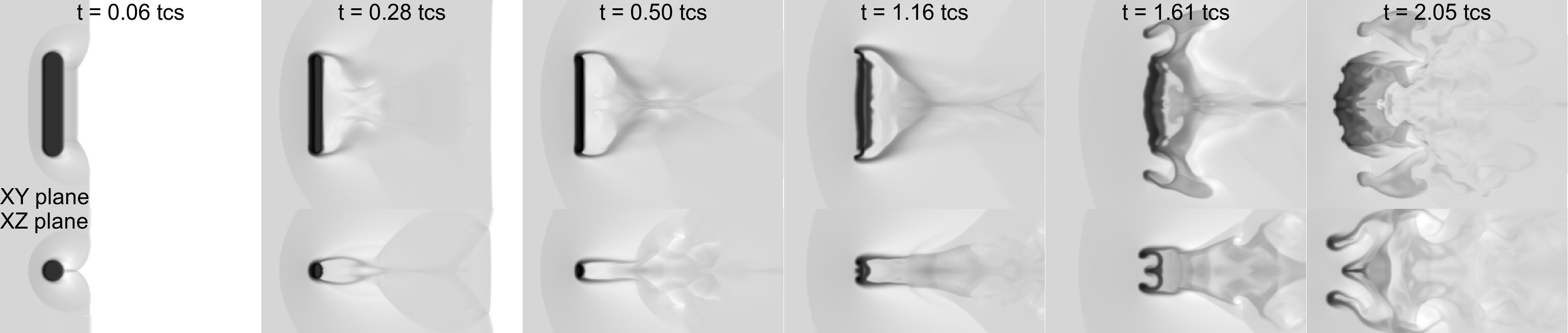}
\caption{Logarithmic density plots of the $XY$ plane as a function of
  time for simulation \emph{m10c3l8s}. The grayscales shows the
  logarithm of the mass density, from $\rho_{\rm amb}$ (white) to
  $5\rho_{\rm c}$ (black). Each frame is labelled with the time. The
  first 5 frames show $-5 < X < 20$, $-10 < Y < 10$, $-6 < Z < 6$ (in
  units of $r_{\rm c}$). The final frame shows $5 < X < 35$, $-12 < Y
  < 12$, $-7.5 < Z < 7.5$.  The shock is initially at $X=-10$.
  $t_{\rm drag} = 3.12\,t_{\rm cs}$ and $t_{\rm mix} = 3.33\,t_{\rm
    cs}$.}
\label{fig:m10c3l8s_XY_montage}
\end{figure*}

Fig.~\ref{fig:m10c3l8s_XY_montage} shows the effect of increasing
$\chi$ to $10^{3}$ (simulation \emph{m10c3l8s}). The filament now
accelerates downstream more slowly, and is subject to stronger
instabilities. Many more shocks are present in the wake. The filament
evolution is however much more akin to that in simulation
\emph{m10c2l8s} than in \emph{m10c1l8s}. Note that the resolution of
\emph{m10c3l8s} is $R_{16}$, whereas all our other simulations are at
$R_{32}$.

\begin{figure}
\includegraphics[width=8.25cm]{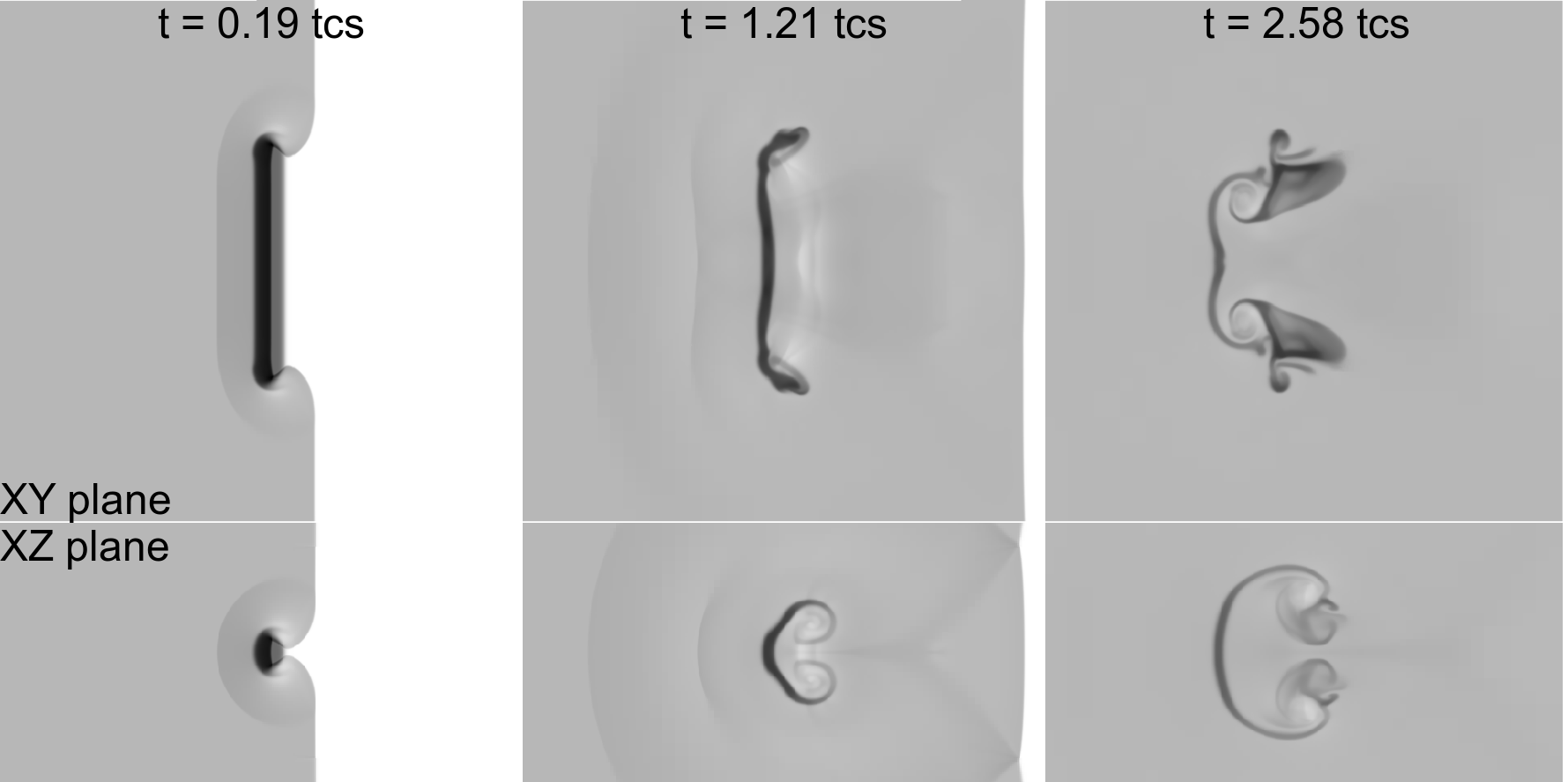}
\caption{Logarithmic density plots of the $XY$ and $XZ$ planes as a
  function of time for simulation \emph{m3c1l8s}. The grayscales shows
  the logarithm of the mass density, from $\rho_{\rm amb}$ (white) to
  $5\rho_{\rm c}$ (black). Each frame is labelled with the time. All
  frames show $-10 < Y < 10$ and $-5 < Z < 5$ (in units of $r_{\rm
    c}$). From left to right the frames show $-10 < X < 10$, $-5 < X <
  15$ and $5 < X < 25$. The shock is initially
at $X=-10$. $t_{\rm drag}
  = 1.51\,t_{\rm cs}$ and $t_{\rm mix} = 4.83\,t_{\rm cs}$.}
\label{fig:m3c1l8s_XY_XZ_montage}
\end{figure}

\begin{figure}
\includegraphics[width=8.25cm]{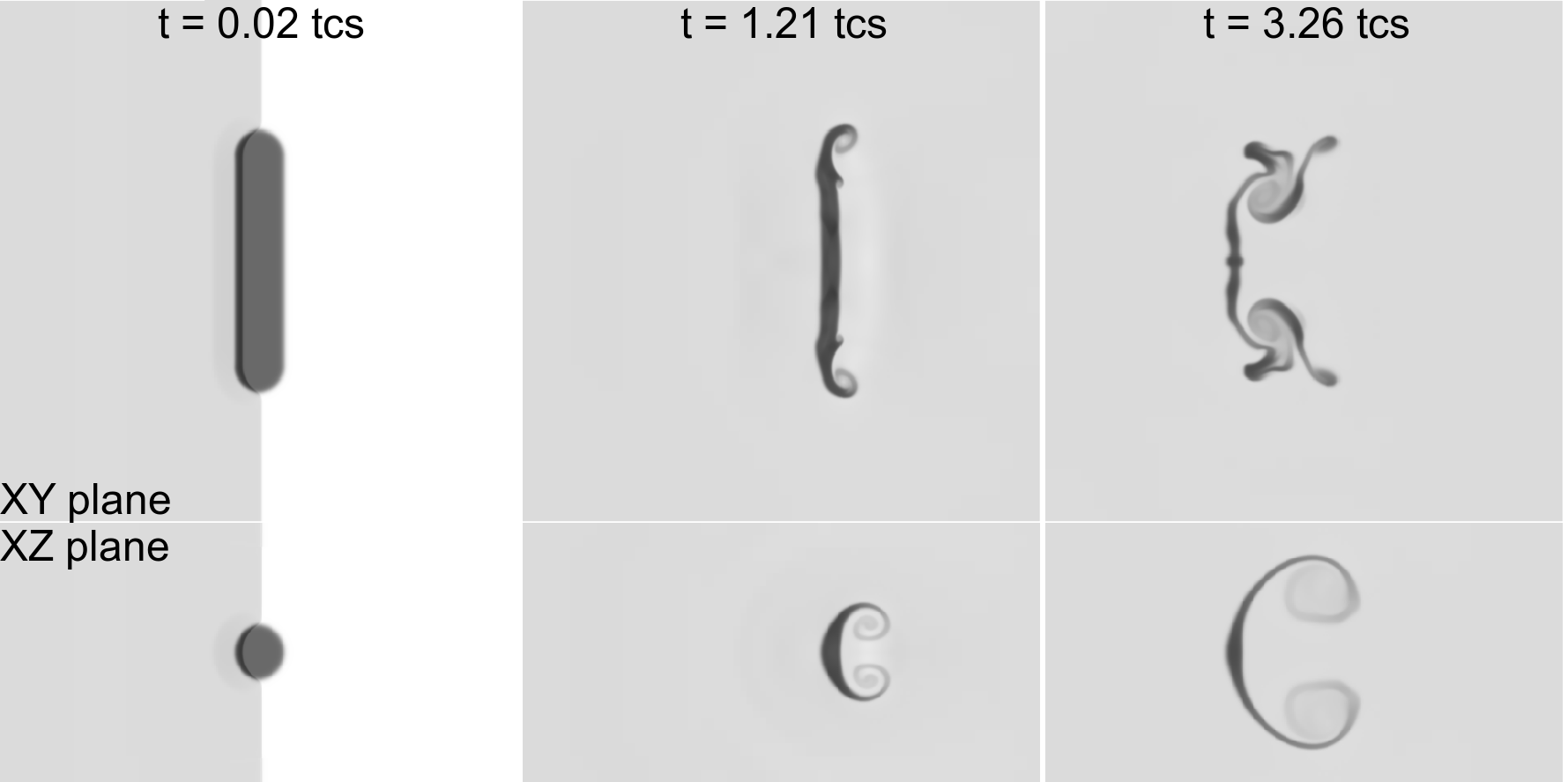}
\caption{Logarithmic density plots of the $XY$ and $XZ$ planes as a
  function of time for simulation \emph{m1.5c1l8s}. The grayscales
  shows the logarithm of the mass density, from $\rho_{\rm amb}$
  (white) to $5\rho_{\rm c}$ (black). Each frame is labelled with the
  time. All frames show $-10 < Y < 10$ and $-5 < Z < 5$ (in units of
  $r_{\rm c}$). The first two frames show $-10 < X < 10$, while the
  third frame shows $0 < X < 20$. The shock is initially at
  $X=-10$. $t_{\rm drag} = 3.33\,t_{\rm cs}$ and $t_{\rm mix} =
  13.1\,t_{\rm cs}$.}
\label{fig:m1.5c1l8s_XY_XZ_montage}
\end{figure}

Finally, we examine the interaction of low Mach number shocks with low
density contrast filaments. Fig.~\ref{fig:m3c1l8s_XY_XZ_montage} shows
results from simulation \emph{m3c1l8s}, with $M=3$ and $\chi=10$.  Key
features are the double bow-shocks seen upstream at $t=1.21\,t_{\rm
  cs}$ and the large-scale eddies seen in the $XZ$ plane. There is
general similarity with simulation \emph{m10c1l8s} (see
Fig.~\ref{fig:m10c1l8s_XY_montage}). Fig.~\ref{fig:m1.5c1l8s_XY_XZ_montage}
shows the interaction in simulation \emph{m1.5c1l8s}, with $M=1.5$ and
$\chi=10$. Again a bow-wave forms upstream. The flow around the
filament appears to be the most laminar yet, due to the gentleness of
the interaction and the rapid acceleration of the filament up to the
postshock speed.

\begin{figure*}
\includegraphics[width=16.5cm]{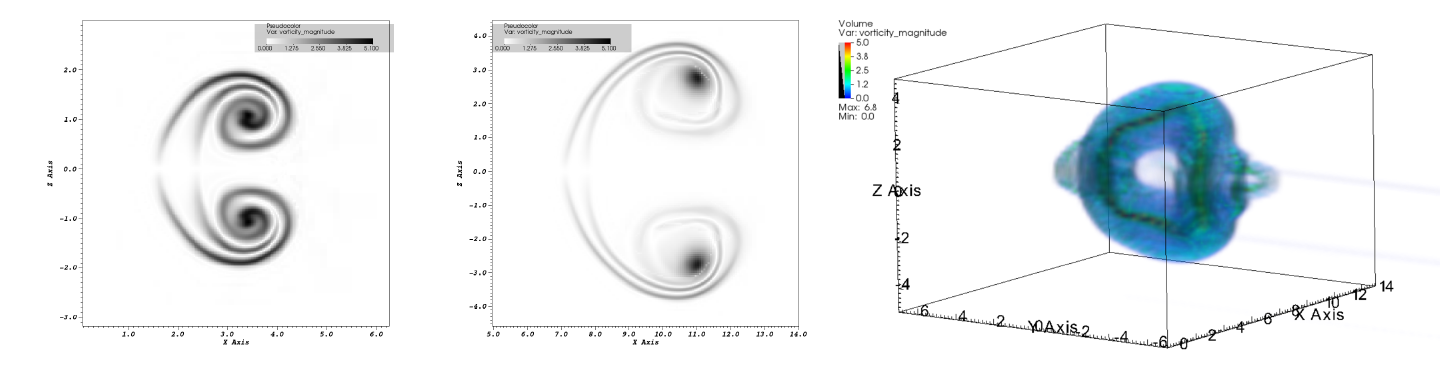}
\caption{The vorticity magnitude in simulation \emph{m1.5c1l8s} at
 $t=1.21$ (left) and $3.26\,t_{\rm cs}$ (middle and right).}
\label{fig:m1.5fil8_vorticity}
\end{figure*}

Fig.~\ref{fig:m1.5fil8_vorticity} shows the vorticity magnitude in
simulation \emph{m1.5c1l8s}. Here the vorticity magnitude highlights the main
dynamical feature of this flow which is the symmetric pair of
vortices produced in the filament wake. At this stage of the
interaction the flow in the $Y=0$ plane is mostly 2D. We can estimate
an effective Reynold's number for the 2D flow on our grid as $Re \sim
(\eta/l)^{-2}$. Here, the largest eddy has a size $l\sim R_{\rm c}$,
while the smallest eddies have a size $\eta \sim 2\,\Delta x = R_{\rm
  c}/16$. Hence we obtain $Re \sim 250$. If the filament were a rigid
structure one might expect to see vortex shedding for this value of
$Re$ (see discussion in Sec.~\ref{sec:flowPastCylinder}). The fact
that we don't see this is likely due to the non-rigidity of the
filament (which becomes significantly distorted as a result of these
vortices) and its acceleration with the flow.

\subsubsection{Comparison to Flow Past a Rigid Cylinder} 
\label{sec:flowPastCylinder}
As our work resembles flow past a cylinder of circular cross-section
it is worth examining what is known about such flows in general. Flow
around a cylinder of circular cross-section has many important
applications in engineering and is well studied by the engineering and
fluid dynamics communities. The behaviour of an incompressible fluid
flowing past an infinite cylinder is known to be a strong function of the
Reynolds number \citep[see, e.g., the review
by][]{Williamson:1996}. The flow is steady and laminar for $Re \ltsimm
50$, and the wake contains a recirculation region with two
symmetrically placed vortices.  The wake displays von Karman vortex
shedding for $100 \ltsimm Re \ltsimm 200$. 3D instabilities start to
appear at $Re \sim 200$ and low levels of turbulence start to develop
within the vortices by $Re \sim 400$. For high $Re (\sim 10^{5})$, the
turbulence spreads out of vortices and a fully turbulent wake is
formed \citep[see, e.g.,][]{Davidson:2004}. The roughness of the
cylinder and the flow quality (the turbulence levels in the oncoming
flow, as well as the character of this turbulence), also affect the
flow dynamics \citep{Williamson:1996}.

The flow past finite-length cylinders has also been studied.
Experiments by \citet{Park:2000} show that the flow past the free-end
of a finite cylinder has a complicated 3D wake structure when $Re
=2\times10^{4}$. The wake is affected by the downwash of
counter-rotating vortices separated from the free end. Other
experiments reveal that as the height-to-diameter ratio ($H/D$) of the
cylinder is reduced, the regular alternating vortex shedding is
replaced more and more by symmetrically shed vortices in the range $2
\ltsimm H/D \ltsimm 6$, and is mostly suppressed for $H/D \ltsimm 2$
\citep{Okamoto:1973,Kawamura:1984,Kappler:2002}. Large eddy
simulations by \citet{Frohlich:2004} investigated the behaviour at $Re
=4.3\times10^{4}$ of flow past a cylinder with $H/D = 2.5$. These
authors find that the relatively small height of the cylinder causes
the 2D von Karman vortices to bend and distort as they travel along
the wake. An arch-type vortex is also seen in the average (but not the
instantaneous) flow behind the cylinder.

Early theoretical and experimental research on inclined (or yawed)
cylinders attempted to associate the flow to that of a non-yawed
cylinder experiencing the flow component normal to the cylinder axis
\citep[see references in][]{Zdravkovich:2003}. The validity of this
``Independence Principle'' or ``Cosine Law'' has been assessed by
\citet{Ramberg:1983}, amongst others. Experiments by
\citet{Shirakashi:1986} and \citet{Matsumoto:1990} determined an axial
flow on the leeward surface, while \citet{Larose:2003} investigated
the effects of both $Re$ and the inclination angle. The experimental
studies are now complemented by numerical simulations of the flow
around inclined cylinders, which cover a wide range of $Re$
\citep[e.g.,][]{Yeo:2008,Vakil:2009}

Though these works are interesting and have some relevance to our
work, the differences which exist betweeen them and our work makes it
difficult to conduct a detailed comparison.  Our simulations are
different from flow past a finite circular cylinder in the following
ways: 1) our filament is not rigid (this requires very high $\chi$);
2) our interaction involves a shock and not a wind/flow; 3) the tip of
our filament is rounded and not flat; 4) our filament has a rough
surface; 5) our interaction is simulated at a lower $Re$ than many
experiments/simulations in the fluid dynamics community; 6) we
simulate compressible rather than incompressible flow (Mach numbers of
$M\ltsimm 0.1$ are needed for a compressible flow to behave similarly
to an incompressible flow, but even for our lowest Mach number
simulations with $M=1.5$ the Mach number of the post-shock flow is
0.511). For this last reason, our most relevant simulation is probably
\emph{m1.5c2l8s}. It indeed shows regular vortices on the two sides of
the cylinder. However, since it is not rigid the filament deforms in
the flow, which is probably why we do not see vortex shedding.

We can also compare the streamline plots shown in
Fig.~\ref{fig:m10fil1_streamlines} to Fig.~11 in \citet{Vakil:2009},
where streamlines around a finite inclined rigid cylinder are
shown. The flow in our simulations is clearly more complex, owing to
i) the higher effective Reynolds number of our simulation; ii) the
compressibility of our flow; iii) the dynamic behaviour and
flexibility of our object.

In Sec.~\ref{sec:timescales} we compare how $t_{\rm drag}$ for our
filament can be related to the drag coefficient measured for rigid
cylinders, and how these quantities vary as $H/D$ and the orientation of the
filament/cylinder change.

\subsection{Statistics}
\label{sec:statistics}
In this section we examine the evolution of some key global quantities of the
interaction. Our initial focus is on simulations with $M=10$ and
$\chi=10^{2}$: other Mach numbers and density contrasts are examined
later. 

We first consider the evolution of $m_{\rm core}$, the ``core'' mass
of the filament. $m_{\rm core}$ declines as material is stripped from
the filament and mixes into the surrounding flow. For spherical
clouds, the mixing time, $t_{\rm mix}$, defined as the time when
$m_{\rm core}$ reaches half of its initial value, is $\approx
6\,t_{\rm cc}$ for reasonably sharp-edged clouds at high Mach numbers,
and increases quite strongly at lower Mach numbers
\citep{Pittard:2010,Pittard:2016}.

\begin{figure*}
\resizebox{130mm}{!}{\includegraphics{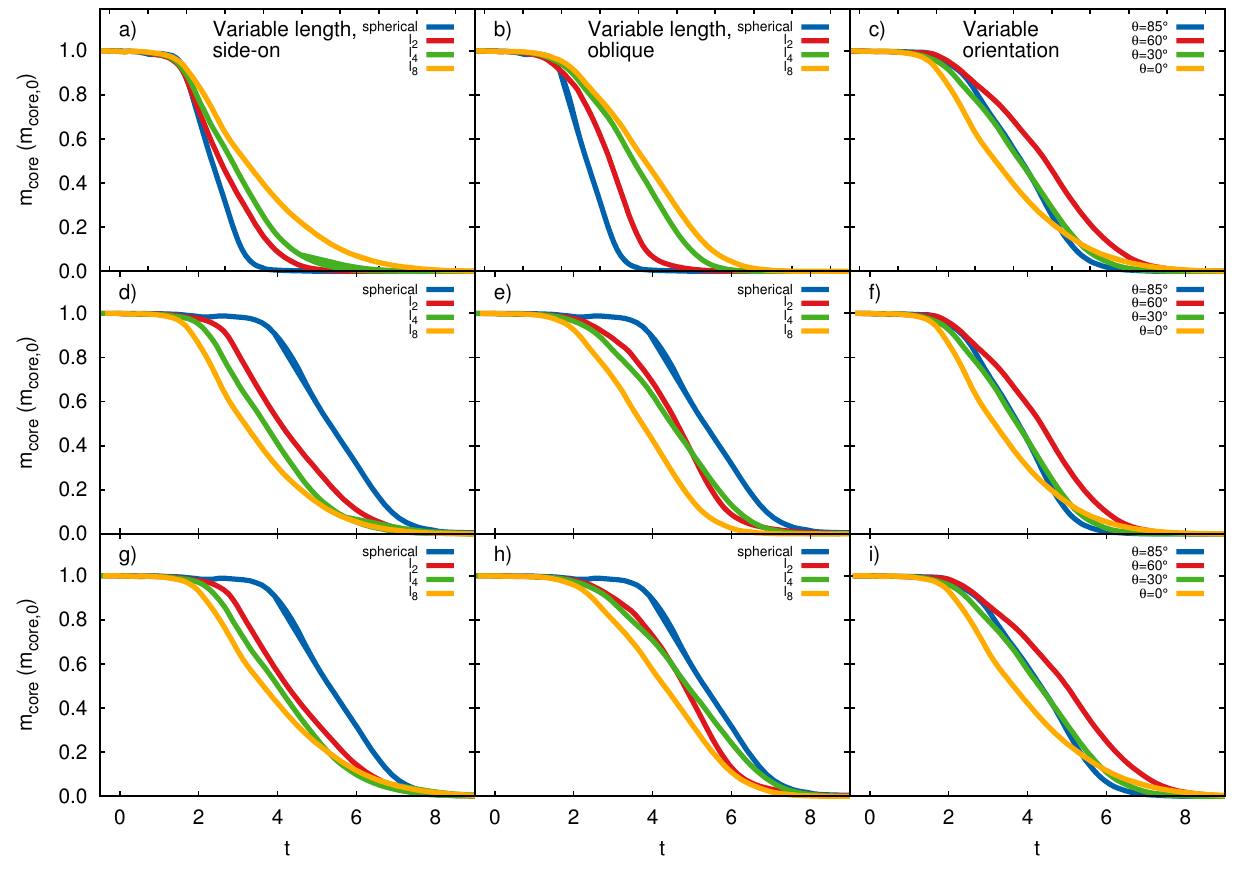}}
\vspace{-4mm}
\caption{Time evolution of the core mass, $m_{\rm core}$, normalized
  to its initial value, for various simulations with $M=10$ and
  $\chi=10^{2}$. The left panels are for ``side-on'' simulations, the
  centre panels are for simulations with $\theta=30^{\circ}$, while
  the right panels are for filaments with $l=8$. The evolution is
  normalized in terms of $t_{\rm cc}$ (top row), $t_{\rm cc}'$ (middle
  row), and $t_{\rm cs}$ (bottom row). The best convergence is
  achieved when $t_{\rm cs}$ is used to normalize the time.}
    \label{fig:mcore_var1}
\end{figure*}

Fig.~\ref{fig:mcore_var1} shows the evolution of $m_{\rm core}$ for
simulations with $M=10$ and $\chi=10^{2}$, as a function of the
filament length and initial orientation. The evolution is normalized
in terms of $t_{\rm cc}$ (top row), $t_{\rm cc}'$ (middle row), and
$t_{\rm cs}$ (bottom row). Results for a spherical cloud are also
shown. In terms of $t_{\rm cc}$, the greater mass of a filament
compared to a spherical cloud means that $m_{\rm core}$ declines more
slowly as $l$ increases (see, e.g., Figs.~\ref{fig:mcore_var1}a and
b). This is true irrespective of the filaments orientation.  On the
other hand, Fig.~\ref{fig:mcore_var1}c) shows that filaments oriented
sideways on to the shock initially lose mass most rapidly, while
mass-loss is delayed for the filaments which are most oblique
(\emph{m10c2l8o60} and \emph{m10c2l8o85}). This is likely due to the
fact that it takes longer for the shock to sweep over these filaments
(in simulation \emph{m10c2l8o85} the long axis of the filament is
almost aligned with the shock normal - see
Fig.~\ref{fig:m10c2l8o85_morphology}). However, it is interesting to
discover that the final 20\% of the mass-loss from the sideways on
filament occurs more gradually\footnote{This is also seen in the MHD
  interaction of a filament with a parallel shock
  \citep{Goldsmith:2016}.}, and that filaments oriented at an angle of
60$^{\circ}$ to the shock front lose mass more slowly than those at
other orientations. In the former instance this is likely because
these filaments are accelerated more rapidly, and thus the velocity
shear which is responsible for driving the mixing is reduced. However,
it is not obvious why a $60^{\circ}$ orientation slows the rate of
stripping.

We find that there is a general reduction in the variance between the
simulations when the evolution is measured in units of $t_{\rm
  cc}'$. In particular, the rate of decline in $m_{\rm core}$ becomes
similar for all simulations, and there is less spread in the time at
which $m_{\rm core}$ nears zero. A further tightening of the curves
occurs when the evolution is measured in units of $t_{\rm cs}$
(see panels d) and g) in Fig.~\ref{fig:mcore_var1}).  For the rest of
this work, therefore, we present our results with the timescale
normalized by $t_{\rm cs}$.

With the time normalized by either $t_{\rm cc}'$ or $t_{\rm cs}$,
Figs.~\ref{fig:mcore_var1}d) and g) reveal that filaments side-on to
the shock are destroyed \emph{more} rapidly than spherical clouds of
the same mass. This is a result of the greater surface area to volume
ratio of the filaments. The destruction timescale reduces with
increasing filament length in such cases. Note that the difference
between the spherical and \emph{m10c2l2s} simulations is greater than
that between the \emph{m10c2l2s} and \emph{m10c2l8s} simulations,
indicating that even a small change from sphericity can significantly
affect the destruction process.  Similar behaviour is observed for
oblique filaments (though the filament with $l=4$ takes slightly
longer than expected for the final 50\% of its core mass to be mixed
into the ambient flow).

\begin{figure*}
  \begin{center}
\resizebox{130mm}{!}{\includegraphics{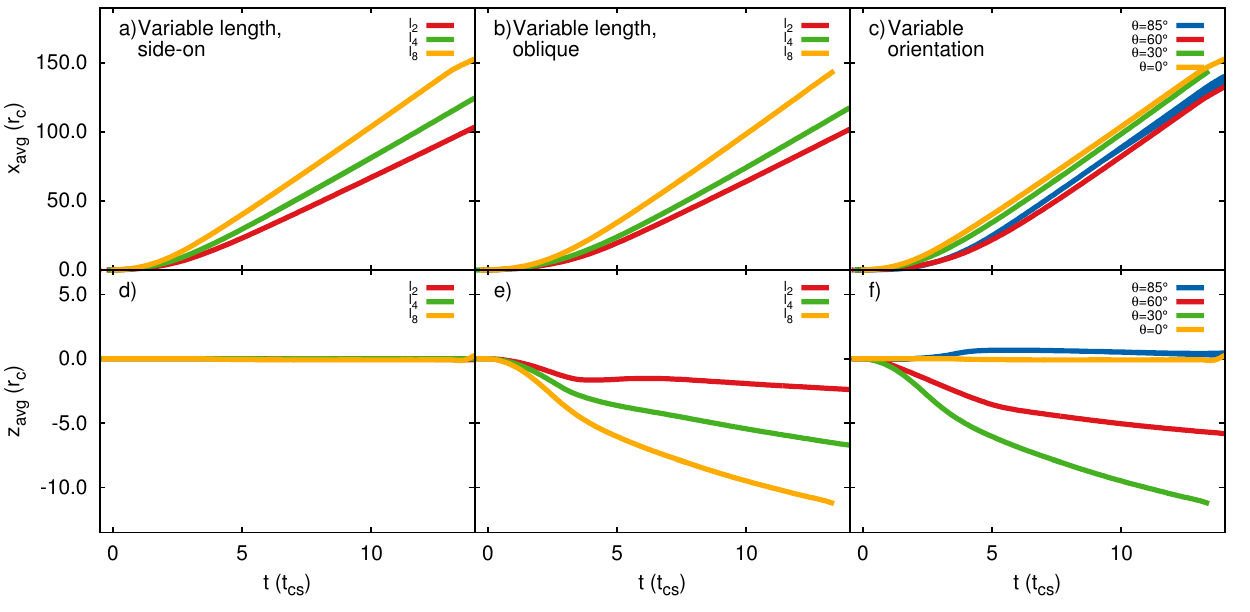}}
  \vspace{-4mm}
  \caption{The time evolution of the $X$ and $Z$ centre-of-mass
    position of the filament for various simulations (the same
    simulations as in Fig.~\ref{fig:mcore_var1}). The middle panels
    show simulations with $\theta=30^{\circ}$. The time is
    normalized to $t_{\rm cs}$.}
    \label{fig:xyzbar_cloud}
  \end{center}
\end{figure*}

The time evolution of the $X$ and $Z$ centre-of-mass position of the
filament ($\langle x\rangle_{\rm cloud}$ and $\langle z\rangle_{\rm
  cloud}$, respectively) for various simulations is shown in
Fig.~\ref{fig:xyzbar_cloud}. We see that longer filaments are
accelerated faster and move downstream more rapidly than shorter
filaments, when the time is normalized by $t_{\rm cs}$. This is again
because of their greater surface-area to volume ratio (shorter
filaments are closer to sphericity), and is also again independent of
the initial filament orientation. For filaments of fixed length,
Fig.~\ref{fig:xyzbar_cloud}c) shows that sideways on filaments are
accelerated faster downstream. Filaments with $\theta=85^{\circ}$ are
accelerated slightly faster than those with $\theta=60^{\circ}$, which
is probably due to the transmitted shock moving through the filament
in a direction very close to the normal of the incident shock.

The movement of the filament in the $Z$-direction is shown in
Fig.~\ref{fig:xyzbar_cloud}d)-f). Due to the symmetry, there is no
significant movement of the filament in the $Z$-direction for
sideways on filaments (see Fig.~\ref{fig:xyzbar_cloud}d). However,
filaments with $\theta=30^{\circ}$ are pushed downwards by the
shock and the post-shock flow. The longer filaments feel this
downwards force for longer and so experience a larger displacement in
$Z$, which can exceed $10\,r_{\rm c}$ at late
times. Fig.~\ref{fig:xyzbar_cloud}f) shows how the displacement in $Z$
depends on the initial orientation of filaments with $l=8$. The
filament which obtains the greatest displacement in $Z$ has
$\theta=30^{\circ}$ (simulation \emph{m10c2l8o30} - see
Figs.~\ref{fig:m10c2l8o30_morphology}-\ref{fig:m10c2l8o30_XY_XZ_montage}). It
is angled quite steeply to the shock normal and so experiences a good
``shove''. The downwards ``push'' is reduced when $\theta=65^{\circ}$.
Surprisingly, a slight \emph{upwards} push occurs when
$\theta=85^{\circ}$. This is noticeable when $t \gtsimm 2.5\,t_{\rm
  cs}$. It appears to be caused by the angle that the shocks
transmitted into the sides of the filament
move (for example, Fig.~\ref{fig:m10c2l8o30_XY_XZ_montage} shows that
in the $XZ$ plane at $t=0.18\,t_{\rm cs}$, the shock transmitted into
the ``bottom'' side of the filament is moving almost directly upwards
(to positive $Z$), while its counterpart on the ``top'' side of the
filament is moving at an angle to the $Z$-axis).

\begin{figure*}
  \begin{center}
\resizebox{130mm}{!}{\includegraphics{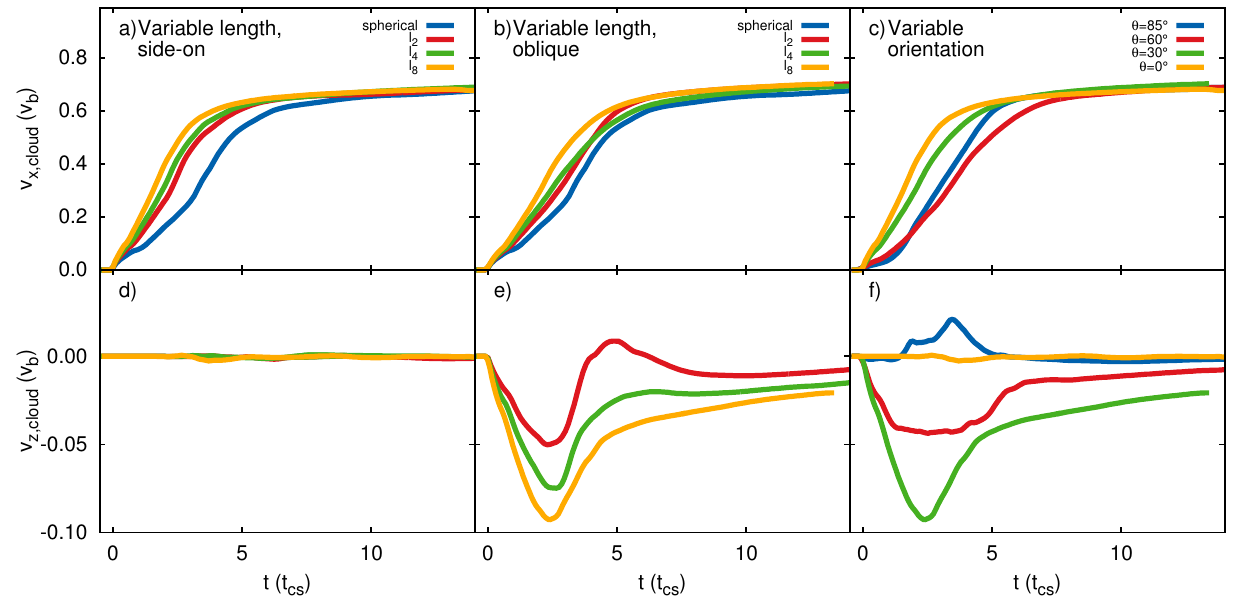}}
  \vspace{-4mm}
  \caption{As Fig.~\ref{fig:xyzbar_cloud} but showing the time
    evolution of the filament mean velocity in the direction of shock
    propagation ($\langle v_{\rm x}\rangle_{\rm cloud}$) and
    in the $Z$-direction ($\langle v_{\rm z}\rangle_{\rm cloud}$). The middle panels
    again show simulations with $\theta=30^{\circ}$.}
    \label{fig:vxz_cloud}
  \end{center}
\end{figure*}

The mean filament velocity in the direction of shock propagation is
shown in Fig.~\ref{fig:vxz_cloud}a)-c). There is essentially no
difference in the asymptotic speed reached, indicating that at late
times the filament material has been accelerated up to the speed of
the post-shock ambient flow. However, there are differences in the
acceleration behaviour of the filament. As indicated previously,
side-on filaments accelerate faster than spherical clouds, with longer
filaments accelerating the fastest (due to their high
surface-to-volume ratio). However, Fig.~\ref{fig:vxz_cloud}b) shows
that all filaments with an initial $\theta=30^{\circ}$ also accelerate
faster than spherical clouds. For filaments with $l=8$,
Fig.~\ref{fig:vxz_cloud}c) shows that side-on filaments are
accelerated downstream the fastest, while filaments with
$\theta=60^{\circ}$ experience the slowest downstream acceleration.

The obliquely-oriented filaments are able to achieve significant
velocities perpendicular to the direction of shock propagation, as
shown in Fig.~\ref{fig:vxz_cloud}e)-f). The maximum absolute velocity
gained is nearly $0.1\, v_{\rm b}$ (or about 13\% of the postshock
flow speed), which occurs in simulation
\emph{m10c2l8o30}. However, this is not sustained, and at later times
it drops back towards zero. In simulation \emph{m10c2l2o30}, the
filament first achieves an overall negative Z-velocity, before
subsequently oscillating first to a positive Z-velocity and then back
to a negative Z-velocity. This is due to the way in which the cloud
fragments (see Fig.~\ref{fig:m10c2l2o30_morphology}). Again, we note
that the filament in simulation \emph{m10c2l8o85} achieves a slight
net positive Z-velocity.

The interaction of shocks with clouds produces substantial vorticity
and velocity dispersion, which may be a key mechanism for generating
turbulent motions in the ISM
\citep[e.g.,][]{Kornreich:2000,Vazquez-Semadeni:2000,MacLow:2004,Dobbs:2007}. From
3D shock-cloud simuations, \citet{Nakamura:2006} found that the
dependence of the velocity dispersion on region size, the ``line
width-size'' relation, is time-dependent, being approximately flat at
early stages in the evolution, but showing a positive trend at later
times due to the more rapid damping of small-scale fluctuations. Their
one-dimensional velocity dispersions, $\delta v = \sqrt{(2\delta
  v_{\rm r}^{2} + \delta v_{\rm z}^2)/3} \sim 0.1\,v_{\rm b}$, are
consistent with the observed internal motions of about $1.5\,{\rm
  km\,s^{-1}}$ observed in cold-neutral-medium (CNM) clouds
\citep{Heiles:2003}.

\begin{figure*}
  \begin{center}
\resizebox{130mm}{!}{\includegraphics{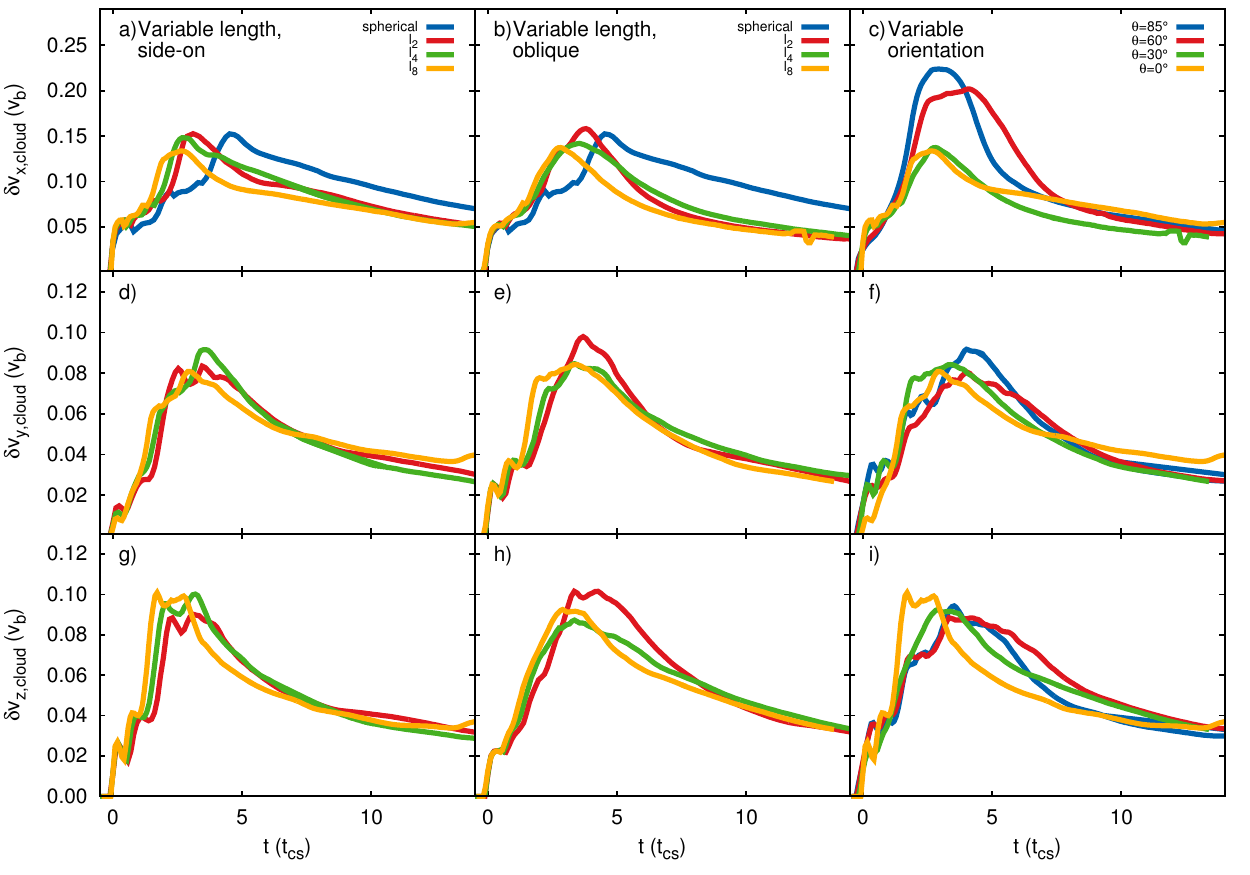}}
  \vspace{-4mm}
  \caption{As Fig.~\ref{fig:xyzbar_cloud} but showing the time
    evolution of the filament velocity dispersion in each direction. Panels
    b), e) and f) show simulations with $\theta=30^{\circ}$.}
    \label{fig:dxdydz_cloud}
  \end{center}
\end{figure*}

In Fig.~\ref{fig:dxdydz_cloud} we show the filament
velocity dispersion in each direction. The most significant finding is
that $\delta v_{\rm x,cloud}$ exceeds $0.2\,v_{\rm b}$ in some cases
(specifically simulations \emph{m10c2l8o60} and
\emph{m10c2l8o85}). This is a third higher than for spherical clouds,
and for the other filament simulations shown in this plot. It is
clearly no coincidence that these two cases correspond to filaments
with their long-axis aligned most closely to the shock normal. It
occurs because in these simulations the transmitted shock is still
travelling through the filament at the time of this peak in $\delta
v_{\rm x,cloud}$, and thus there is a large difference in $v_{\rm x}$
between different parts of the cloud at this time \citep[see also the
MHD simulations of][]{Goldsmith:2016}. Otherwise, the level of
agreement between the various filament simulations is slightly
suprisingly - $\delta v_{\rm x,cloud}$ always peaks at $\approx
0.15\,v_{\rm b}$, and $\delta v_{\rm y,cloud}$ and $\delta v_{\rm
  z,cloud}$ peak at $\approx 0.08-0.1\,v_{\rm b}$. There does not seem
to be any significant difference in $\delta v_{\rm z,cloud}$ with
filament orientation, which is unexpected given the clear differences
in $\langle v_{\rm z}\rangle_{\rm cloud}$ seen in
Fig.~\ref{fig:vxz_cloud}f).

\begin{figure}
  \begin{center}
\resizebox{75mm}{!}{\includegraphics{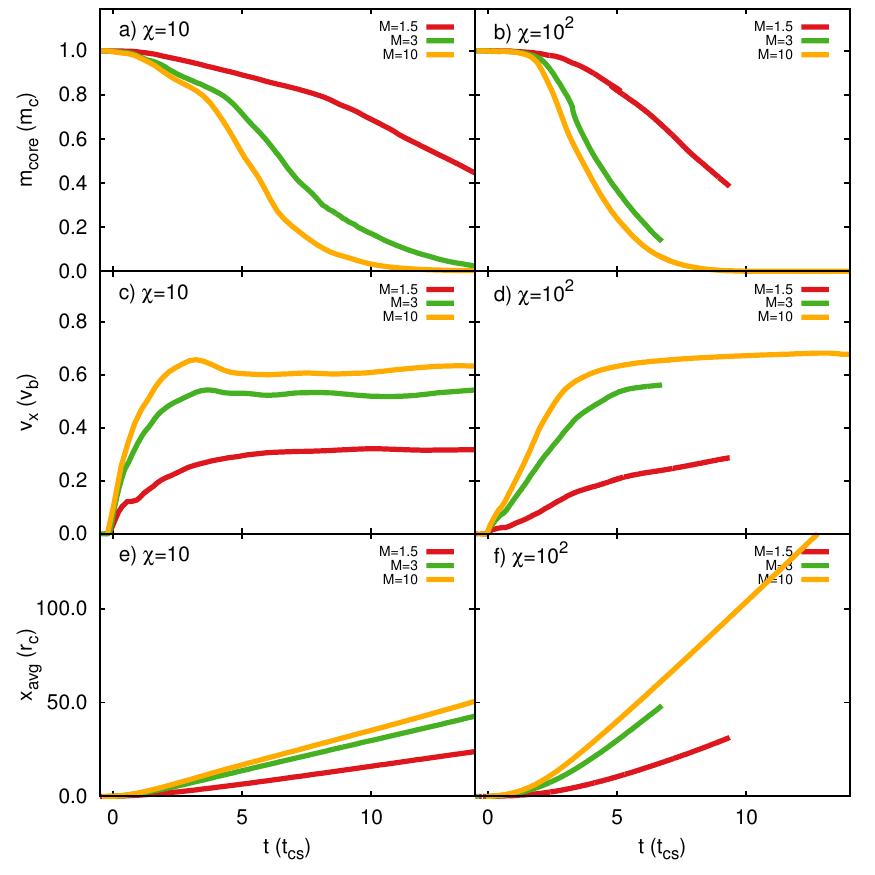}}
  \vspace{-4mm}
   \caption{The Mach number dependence of the evolution of $m_{\rm
       core}$, $\langle v_{\rm x}\rangle_{\rm cloud}$ and $x_{\rm
       avg}$, for filaments with $l=8\,r_{\rm c}$ and oriented
     sideways-on to the shock.}
    \label{fig:mach_cloud}
  \end{center}
\end{figure}

We now examine the dependence of $m_{\rm core}$, $\langle v_{\rm
  x}\rangle_{\rm cloud}$ and $\langle x\rangle_{\rm cloud}$ on the Mach number and
density contrast of a filament with $l=8$ and oriented
sideways-on to the shock. The Mach number dependence is shown in
Fig.~\ref{fig:mach_cloud}. As is the case for spherical clouds, we see
that $m_{\rm core}$ decreases more slowly, and $\langle v_{\rm
  x}\rangle_{\rm cloud}$ and $\langle x\rangle_{\rm cloud}$ both increase more slowly,
as $M$ is reduced.

\begin{figure*}
  \begin{center}
\resizebox{155mm}{!}{\includegraphics{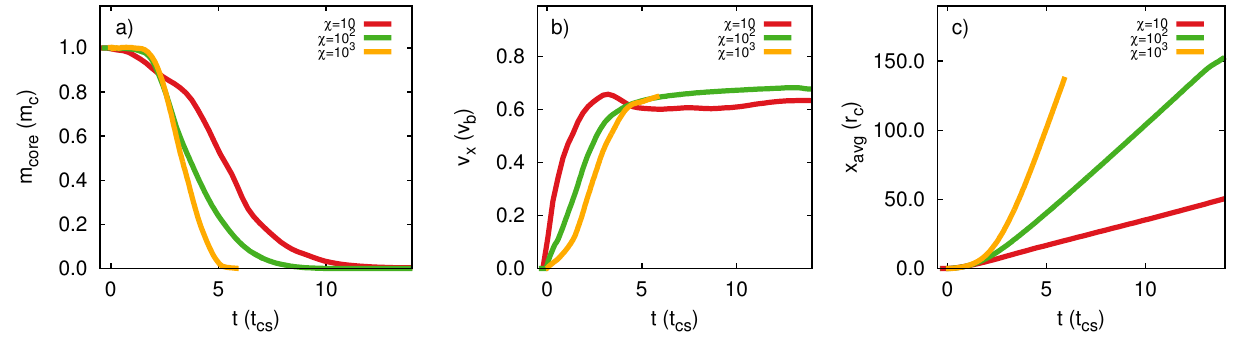}}
  \vspace{-4mm}
   \caption{$\chi$ dependence of the evolution, for simulations with
     $M=10$, and filaments of length $l=8\,r_{\rm c}$ oriented
     sideways on to the shock.}
    \label{fig:chi_cloud}
  \end{center}
\end{figure*}

Fig.~\ref{fig:chi_cloud} shows the dependence of $m_{\rm core}$,
$\langle v_{\rm x}\rangle_{\rm cloud}$ and $\langle x\rangle_{\rm
  cloud}$ as a function of the density contrast, $\chi$, for filaments
with $l=8$ oriented sideways-on to an $M=10$ shock. Denser filaments
are destroyed relatively quicker, but accelerated relatively slower,
when the time is scaled by $t_{\rm cs}$. In contrast, such montonic
behaviour is not seen for spherical clouds - Figs.~14 and~22 and
Table~4 in \citet{Pittard:2016} show that in terms of $t_{\rm cc}$,
spherical clouds with $\chi=10^{2}$ are destroyed quicker than clouds
with $\chi=10^{3}$.

\subsection{Timescales}
\label{sec:timescales}
\begin{figure}
  \begin{center}
\includegraphics[width=80mm]{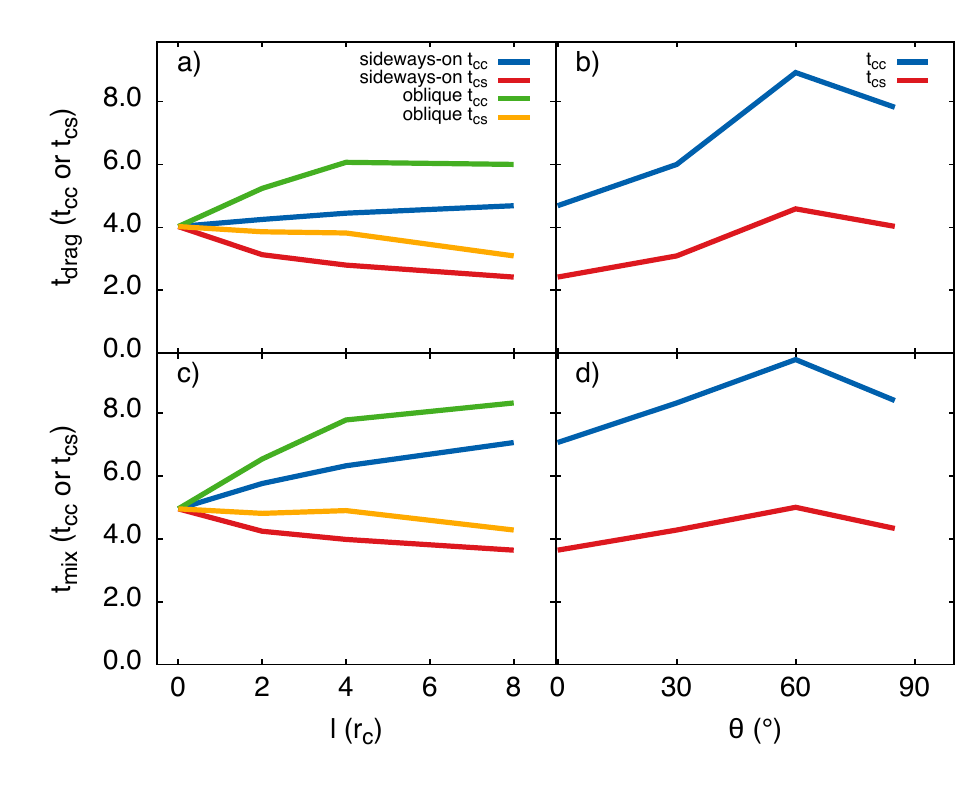}
   \caption{Top: $t_{\rm drag}$ (for the cloud); and bottom: $t_{\rm
       mix}$, as functions of the filament length (left panels) and
     orientation (right panels),
     from simulations with $M=10$ and $\chi=10^{2}$. The timescales
     are normalized by either $t_{\rm cc}$ or $t_{\rm cs}$. The
     ``oblique'' simulation results noted in the left panels are for $\theta=30^{\circ}$.}
    \label{fig:tdragtmix_cloud}
  \end{center}
\end{figure}

Values of $t_{\rm drag}$ and $t_{\rm mix}$ are noted in
Table~\ref{tab:results}. In all cases $t_{\rm drag} < t_{\rm
  mix}$. Fig.~\ref{fig:tdragtmix_cloud} shows the values of $t_{\rm
  drag}$ and $t_{\rm mix}$ from simulations with $M=10$ and
$\chi=10^{2}$, as a function of the filament length and
orientation. In Fig.~\ref{fig:tdragtmix_cloud}a) and c), values are
plotted from simulations \emph{m10c2l2s}, \emph{m10c2l4s}, and
\emph{m10c2l8s} for the ``sideways-on'' orientation, and from
simulations \emph{m10c2l2o30}, \emph{m10c2l4o30}, and
\emph{m10c2l8o30} for the ``oblique'' orientation. In
Fig.~\ref{fig:tdragtmix_cloud}b) and d), values are plotted from
simulations \emph{m10c2l8s}, \emph{m10c2l8o30}, \emph{m10c2l8o60} and
\emph{m10c2l8o85}. The drag and mixing timescales from a spherical
cloud simulation are also plotted in Fig.~\ref{fig:tdragtmix_cloud}a)
and c) as the points at $l=0$ (in this case, $t_{\rm cc} = t_{\rm
  cs}$).

Examining Fig.~\ref{fig:tdragtmix_cloud}a) we see that there is a
reduction in the spread between the sideways-on and oblique lines when
the time is scaled by $t_{\rm cs}$. $t_{\rm drag}$ also shows less
variation with length for oblique filaments when scaling to $t_{\rm cs}$
($t_{\rm drag}/t_{\rm cc}$ is surprisingly insensitive to the filament
length when the filament is sideways-on). Fig.~\ref{fig:tdragtmix_cloud}a)
also shows that sideways-on filaments have a lower value of $t_{\rm
  drag}$ (i.e. they accelerate
faster). Fig.~\ref{fig:tdragtmix_cloud}b) shows that $t_{\rm drag}$
can vary by a factor of 2 or so due to changes in the orientation of
the filament. $t_{\rm drag}$ increases as the filament orientation
moves from sideways-on towards end-on, but the variation is not
completely monotonic, with $t_{\rm drag}$ declining slightly as the
filament becomes nearly end-on. Note that $t_{\rm cs} = 1.943\,t_{\rm
  cc}$ for the results shown in Fig.~\ref{fig:tdragtmix_cloud}b) and
d).

There is also a reduction in the spread between the sideways-on and
oblique lines for $t_{\rm mix}$ when scaled by $t_{\rm cs}$ (see
Fig.~\ref{fig:tdragtmix_cloud}c). In addition, there is much less
variation with $l$ when $t_{\rm mix}$ is scaled by $t_{\rm cs}$
($t_{\rm mix}/t_{\rm cs} \approx 4$ for all cases - this further
supports our decision to present our results in terms of $t_{\rm
  cs}$). $t_{\rm mix}$ has the same behaviour
with $\theta$ as $t_{\rm drag}$.

We can compare our results against earlier work from
\citet{Nakamura:2006}, who investigated the interaction of a shock
with a soft-edged spherical cloud. These authors also considered the
3D interaction of a shock with a cylindrical cloud, which is
effectively an infinite filament hit sideways on (see their
Sec.~4.6). They note that cylindrical clouds become more flattened and
longer in the tranverse direction than spherical clouds (i.e. a larger
value of $a$). This increases the drag and leads to more rapid
acceleration of the cloud, which in turn results in a longer mixing
timescale due to the decreased velocity relative to the surrounding
flow. In interactions with $\chi=10$ and $M=10$, \citet{Nakamura:2006}
find that cylindrical clouds with very shallow density gradients
($n=2$) have $t_{\rm drag}=2.04\,t_{\rm cc}$ and $t_{\rm
  mix}=18.7\,t_{\rm cc}$, which are 0.27 and $3.95\times$ their values
for a spherical cloud. For clouds with steep density profiles the
differences are factors of 0.73 and 1.71 respectively. Our filament
has a density profile which is more akin to $n=24$ - for the closest
comparison we should therefore compare to their $n=8$ results. Our
nearest equivalent model is simulation \emph{m10c1l8s}. In terms of
$t_{\rm cc}$, we find that $t_{\rm drag}$ ($t_{\rm mix}$) is
53\% (146\%) of the values for a spherical cloud for simulation
\emph{m10c1l8s}. These factors are both smaller than the factors found by 
\citet{Nakamura:2006}, but are in reasonable agreement considering that our filaments have slightly sharper density
profiles and are finite in extent (note also that $t_{\rm mix}$
increases as the filament length increases). For interactions with filaments of
higher density contrast (specifically $\chi=10^{2}$, simulation
\emph{m10c2l8s}), we find that $t_{\rm drag}$ \emph{increases} by a factor
of 1.17 (see Fig.~\ref{fig:tdragtmix_cloud}a), while $t_{\rm mix}$
increases by a factor of 1.43 (see Fig.~\ref{fig:tdragtmix_cloud}c).

The characteristic drag time can be obtained from the equation of
motion of the cloud,
\begin{equation}
m_{\rm c} \frac{dv_{\rm c}'}{dt} = - \frac{1}{2} C_{\rm D} \rho_{i1}
v_{\rm c}'^{2} A,
\end{equation}
where $m_{\rm c}$ is the mass of the cloud, $v_{\rm c}' = |v_{\rm i1}
- v_{\rm c}|$ is the magnitude of the velocity of the cloud ($v_{\rm
  c}$) relative to the shocked intercloud medium ($v_{\rm i1}$),
$C_{\rm D}$ is the drag coefficient, $\rho_{\rm i1}$ is the density of
the shocked intercloud medium, and $A$ is the cross-sectional area of
the cloud normal to the shock normal \citep[see,
e.g.,][]{Klein:1994}. The cloud accelerates so that $v_{\rm c}' =
v_{\rm i1}/e$ at the time
\begin{equation}
t_{\rm drag} = \frac{2 m_{\rm c}}{C_{\rm
    D}\rho_{i1}A}\frac{(e-1)}{v_{\rm i1}}.
\end{equation}
For a strong shock and $\gamma=5/3$, $\rho_{i1}
\approx 4\rho_{\rm i0}$, where $\rho_{\rm i0}$ is the density of the
preshock ambient medium. Then for a spherical cloud, with $m_{\rm c} =
\frac{4}{3}\pi r_{\rm c}^{3}\rho_{i0}\chi$ and $A = \pi r_{\rm
  c}^{2}$, $v_{\rm c}' = v_{\rm i1}/e$ when $t_{\rm drag} = 1.53
\chi^{1/2} t_{\rm cc}/C_{\rm D}$. Of course, the lateral expansion of
the cloud increases $A$, so the actual drag time is considerably
smaller. Alternatively, one can keep the original value of $A$ and
instead account for this expansion with a larger value of $C_{\rm D}$.

For a filament with mass $m_{\rm c} = (l + \frac{4}{3})\pi r_{\rm
  c}^{3} \rho_{\rm i0}\chi$ and cross-sectional area $A = (lcos\theta
+ \pi)r_{\rm c}^{2}$ (where $\theta$ is the angle between the filament
long axis and the shock surface), 
\begin{equation}
\label{eq:tdrag_analytical}
t_{\rm drag} = 3.60\frac{(l + 4/3)}{(lcos\theta +
  \pi)}\frac{\chi^{1/2}}{C_{\rm D}}t_{\rm cc} = 3.96\frac{(l +
  4/3)^{2/3}}{(lcos\theta + \pi)C_{\rm D}}t_{\rm cs}.
\end{equation}
For a very long filament sideways-on to the shock ($\theta =
0^{\circ}$), $t_{\rm drag} \propto l^{-1/3}$. For very long filaments
end-on to the shock ($\theta = 90^{\circ}$), $t_{\rm drag} \propto
l^{2/3}$.

\begin{figure*}
  \begin{center}
\includegraphics[width=155mm]{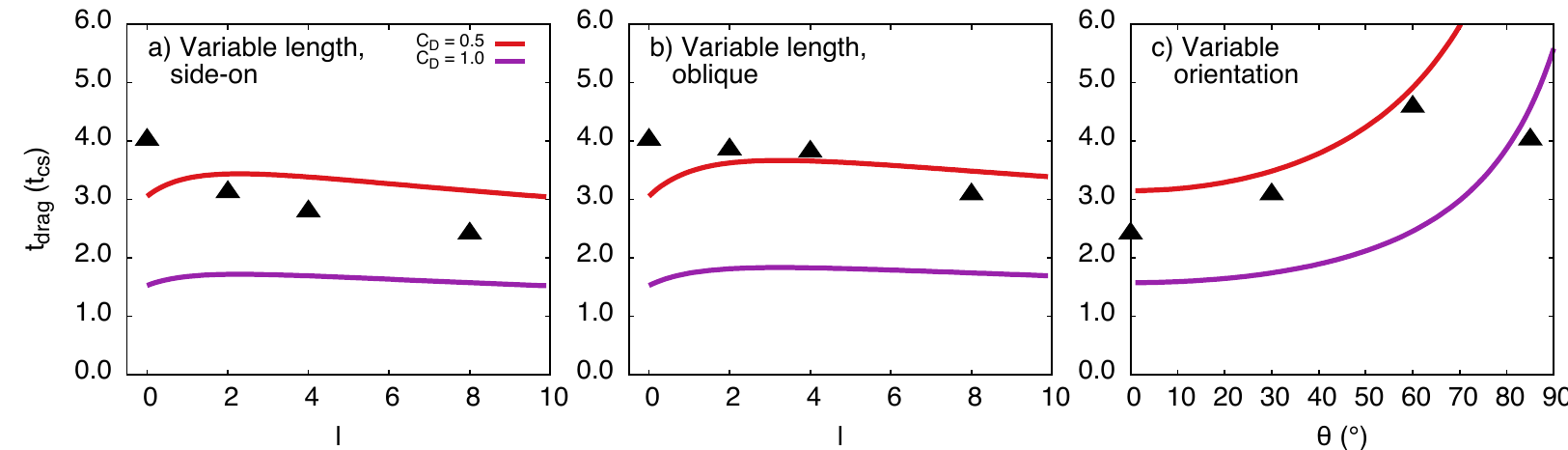}
   \caption{$t_{\rm drag}$ from the simulations compared to the
     analytical expression in Eq.~\ref{eq:tdrag_analytical}, for
     $M=10$ and $\chi=10^{2}$. a) As a function of filament length,
     for side-on filaments. b) As a function of filament length for
     oblique filaments ($\theta=30^{\circ}$). c) As a function of
     orientation for filaments of length $l=8$. The filament is
     sideways-on when $\theta=0^{\circ}$ and is end-on when $\theta=90^{\circ}$.}
    \label{fig:tdrag_theory}
  \end{center}
\end{figure*}

Fig.~\ref{fig:tdrag_theory} shows the values of $t_{\rm drag}$ from
simulations with $M=10$ and $\chi=10^{2}$ compared to the analytical
value from Eq.~\ref{eq:tdrag_analytical}. We immediately see that the
simulation values do not follow the same trend with $l$ or $\theta$ as
the simple analytical theory. In particular, we note that $t_{\rm
  drag}$ peaks at $\theta \sim 60^{\circ}$ in the simulations, whereas
the analytical theory predicts that $t_{\rm drag}$ peaks at $\theta =
90^{\circ}$ (i.e.  end-on). This indicates that the real behaviour of
the filament is more complex. For instance, its lateral expansion is
likely a function of its length and orientation, while its
acceleration is also driven by the transmitted shocks and rarefaction
waves.

For sideways-on finite \emph{rigid} cylinders, $C_{\rm D}$ decreases
as the aspect ratio ($L/D$) of the cylinder decreases \citep[see,
e.g.,][and references therein]{Baban:1991}. $C_{\rm D} \sim 1$ for
subcritical, high Reynolds numbers, but is significantly higher at
lower $Re$ \citep[e.g., at $Re=1$, $C_{\rm D} \approx 20$ for $L/D =
2$, and declines to $C_{\rm D} = 12$ at $L/D = 20$, as shown
in][]{Vakil:2009}. A decline in $C_{\rm D}$ with $L/D$ is equivalent
to $t_{\rm drag}$ increasing as $l$ decreases. This is indeed the
behaviour found in our numerical simulations as shown in
Fig.~\ref{fig:tdrag_theory}a-b).  \citet{Vakil:2009} also show that
the drag coefficient increases as the cylinder moves from end-on to
side-on, while the lift coefficient is a maximum at $\theta \approx
45^{\circ}$. The fact that $C_{\rm D}$ monotonically decreases as the
cylinder rotates to become more end-on is equivalent to $t_{\rm drag}$
monotonically increasing with $\theta$. This is of course the
behaviour obtained with Eq.~\ref{eq:tdrag_analytical}, but as already
noted our simulation results do not yield such monotonic behaviour for
$t_{\rm drag}$ (see Fig.~\ref{fig:tdrag_theory}c). The ratio of the
lift-to-drag coefficients, $C_{\rm L}/C_{\rm D}$, peaks at $\theta
\approx 50^{\circ}$. For $Re = 40$, $C_{\rm L}/C_{\rm D} = 0.3$ (0.17)
at $L/D = 5$ (2) \citep{Vakil:2009}. It is interesting to note that
the ratio of the peak values of the average cloud velocities normal to
and with the flow in simulation \emph{m10c2l8o30} is $\approx 0.15$
(see Fig.~\ref{fig:vxz_cloud}).

The values of $t_{\rm drag}$ and $t_{\rm mix}$ are very relevant to
the acceleration and survival of clouds in multi-phase galactic
winds. Recent observations indicate that the cool gas extends to large
distances from the host galaxy, up to the virial radius
\citep[e.g.,][]{Ribaudo:2011,Rudie:2012,Crighton:2013,Werk:2014}.
However, it is not clear whether clouds can be accelerated by the ram
pressure of the hot wind to the observed distances and velocities. For
instance, \citet{Scannapieco:2015} find that clouds are completely
disrupted by their interaction with a wind by the time they travel
$40\,r_{\rm c}$, and that it is therefore very difficult to explain
the 100\,kpc distances at which cool gas is seen from nearby
galaxies. They also find that clouds are accelerated to only very
small fractions ($\ltsimm 0.3$) of the wind speed. On the other hand,
\citet{McCourt:2015} find that mixing is suppressed for clouds with
tangled internal magnetic fields\footnote{This is also the case for
  magnetized shocks interacting with filaments
  \citep{Goldsmith:2016}.}, such that they \emph{can} be accelerated
up to the wind speed. Taking the lifetime of clouds as $\approx
10\,t_{\rm cc}$, in agreement with the results from shock-cloud
simulations \citep[e.g., see Fig.~19 in][]{Pittard:2010},
\citet{Zhang:2015} estimate that the KH timescale must be roughly
$3-10\times$ longer than that from purely hydrodynamic simulations in
order for cool clouds to reach the velocities seen in observations.

An obvious question is whether filaments as opposed to spherical
clouds can alleviate some of these issues. The current paper indicates
that $t_{\rm drag}/t_{\rm cc}$ varies little with filament length, and
that side-on or nearly-side on filaments accelerate at approximately
the same rate as spherical clouds ($t_{\rm drag}/t_{\rm cc}$ is
within $\pm 50$\%). For longish filaments oriented nearly
end-on ($l=8,\theta = 60^{\circ}$), $t_{\rm drag}$ can be a factor of
2 longer. But as $t_{\rm mix}/t_{\rm cc}$ is only 40\% longer, they do
not travel as far downstream by $t=t_{\rm mix}$. \citet{Nakamura:2006}
report that $t_{\rm mix}/t_{\rm drag}$ remains almost constant for
clouds with very smooth density profiles (see their Tables~1 and 2).
 
The average displacement of the filament in the downstream direction
$\langle x\rangle$ at $t=t_{\rm mix}$ is noted in
Table~\ref{tab:results}. Some trends are worth noting. Firstly,
$\langle x\rangle/r_{\rm c}$ roughly decreases with increasing $\theta$
(e.g., from a value of 21.2 for simulation \emph{m10c2l8s} to 11.9 for
simulation \emph{m10c2l8o85}). Second, $\langle x\rangle/r_{\rm c}$
increases with $l$ for sideways-on filaments (e.g., from 14.8 for
\emph{m10c2l2s} to 21.2 for \emph{m10c2l8s}), and also increases with
$\chi$ for $M=10$, $l=8$, sideways-on filaments (from 17.6 for
\emph{m10c1l8s} to 28.6 for \emph{m10c3l8s}).

In fact, Table~\ref{tab:results} indicates that simulation
\emph{m10c3l8s} has the highest value of $\langle x\rangle/r_{\rm c}$
($=28.6$) at $t=t_{\rm mix}$ ($= 6.46\,t_{\rm cc}$ for this
simulation). This filament travels nearly as far as the furthest
distance reported by \citet{Scannapieco:2015}, where $d=32\,r_{\rm c}$
for their simulation \emph{M11.4v3000} ($\chi=300$) at $t=t_{25}$
(defined as the time when only 33\% of the cloud mass is at or above
the original density, which corresponds to $t=22.2\,t_{\rm cc}$ in
this simulation). However, the velocity of the cloud material differs
markedly between our work and theirs. The core material in our
simulation has $\langle v_{\rm x}\rangle/v_{\rm b} = 0.40$ whereas in
their simulation their cloud has only reached a velocity of
$v=0.14\,v_{\rm hot}$ ($420 \,{\rm km\,s^{-1}}$). At face level our
results would seem to be more favourable compared to observations of
galactic winds. However, it is hard to make a more direct comparison
because of the different tracers used for the cloud ($m_{\rm core}$ in
our work versus $F_{1/3}$ in their work), and a dedicated study is
needed to investigate this issue further.

\section{Summary And Conclusions}
\label{sec:conclusions}
In this paper we have investigated the hydrodynamic interaction of a
shock with a filament. This extends previous investigations of a shock
striking a spherical cloud
\citep[e.g.,][]{Stone:1992,Klein:1994,Nakamura:2006,Pittard:2016} to
interactions with highly elongated clouds. Two additional parameters
are introduced: the length of the filament ($l$) and its orientation
to the shock ($\theta$). We have performed 3D simulations in which
these parameters are varied, with our main focus on a Mach 10 shock
interacting with a $\chi=10^{2}$ filament. Our main conclusions are as
follows:

\begin{enumerate}
\item We find significant differences in the interaction as $l$ and
  $\theta$ are varied. In particular, we find that filaments with $l
  \gtsimm 2\,r_{\rm c}$ which are oriented at $\theta \ltsimm
  60^{\circ}$ form 3 parallel rolls during the interaction,
  which move away from each other as the filament material mixes into
  the ambient flow. In sideways-on filaments, the filament material
  expands preferentially along its minor-axis and in the direction of
  shock propagation, and much less so along its major-axis.

\item The nature of the interaction changes significantly with the
  initial obliquity of the filament. Filaments oriented only slightly
  oblique to the shock surface (at $\theta=30^{\circ}$) spill flow
  with high vorticity around the upstream end of the filament, while
  periodic vortex shedding occurs at the downstream end. Their wakes
  are longer and less symmetrical. Increasing the obliquity further
  ($\theta=60^{\circ}$) supresses the 3-parallel rolls -
  instead the vortex ring which forms at the upstream end of the
  filament is the dominant flow feature and is the cause of the
  filament's fragmentation. Filaments which are almost end-on to the
  shock ($\theta=85^{\circ}$) suffer dramatic ``hollowing'' or
  ``voiding'', followed by strong instabilities at their upstream
  end. Destruction occurs by the time that the transmitted shock
  reaches the back of the filament.

\item Sideways-on filaments initially lose mass more quickly, but
  since they are also accelerated more quickly the final 20\% of their
  mass loss occurs more gradually. Filaments oriented at an angle of
  $60^{\circ}$ to the shock front lose mass more slowly than those at
  other orientations. Even a small elongation from spherical can
  significantly change the destruction timescale.

\item The rate at which the filaments lose mass shows
  less variance between different simulations when the time is
  normalized in terms of $t_{\rm cs}$, the ``cloud crushing time'' for
  a spherical cloud of the same density contrast and equivalent
  mass. When normalized in terms of $t_{\rm cs}$, side-on filaments
  are destroyed more quickly than spherical clouds due to their
  greater surface area to volume ratio.

\item Filaments which are oriented at an angle to the shock can obtain
  velocities of order 10\% of the shock speed in directions
  perpendicular to the shock normal, for short periods of
  time. Filaments which are closely aligned with the shock normal can
  actually be pushed in a different transverse direction than
  expected.

\item The filament velocity dispersion varies very little with length
  or orientation, except for filaments with $l=8\,r_{\rm c}$ which
  have their long-axis within $30^{\circ}$ of the shock normal. There
  does not seem to be any significant difference in $\delta v_{\rm
    z,cloud}$ with filament orientation, despite clear differences in
  $\langle v_{\rm z}\rangle_{\rm cloud}$.

\item Sideways-on filaments have a lower value of $t_{\rm drag}$
  (i.e. they accelerate faster). $t_{\rm drag}$ can vary by a factor
  of 2 or so due to changes in the orientation of the filament:
  $t_{\rm drag}$ increases as the filament orientation moves from
  sideways-on towards end-on, but the variation is not completely
  monotonic, and there is a slight reduction as the filament becomes
  nearly end-on. $t_{\rm mix}$ shows similar behaviour.

\item The nature of the interaction mirrors that of spherical cloud
  interactions as $M$ and $\chi$ are varied. The interaction is much
  less violent at lower Mach numbers, and both $t_{\rm drag}/t_{\rm cs}$
  and $t_{\rm mix}/t_{\rm cs}$ increase.  Filaments with a lower
  density contrast are less able to resist the shock and their
  immersion in the post-shock flow, and are rapidly accelerated
  towards the postshock speed. Filaments with a higher density
  contrast are accelerated more slowly and are subject to stronger
  instabilities.
\end{enumerate}

In a companion paper \citep{Goldsmith:2016} we examine
the effect of a magnetic field on shock-filament interactions, and in
future work we will explore the effects of additional physics, such as
radiative cooling, and the interaction with a wind rather than a shock.

\section*{Acknowledgements}
We would like to thank the referee for a timely and useful report.
JMP would like to thank S.\,Falle for the use of the {\sc MG}
hydrodynamics code used to calculate the simulations in this work and
S.\,van~Loo for adding SILO output to it. This work was supported by
the Science \& Technology Facilities Council [Research Grant
ST/L000628/1]. The calculations for this paper were performed on the
DiRAC Facility jointly funded by STFC, the Large Facilities Capital
Fund of BIS and the University of Leeds.





\begin{thebibliography}{99}
\bibitem[\protect\citeauthoryear{Agertz et al.}{2007}]{Agertz:2007}
Agertz O., et al., 2007, MNRAS, 380, 963
\bibitem[\protect\citeauthoryear{Andr\'{e} et al.}{2010}]{Andre:2010}
Andr\'{e} Ph., et al., 2010, A\&A, 518, L102
\bibitem[\protect\citeauthoryear{Arthur \& Henney}{1996}]{Arthur:1996} 
Arthur, S.~J., \& Henney, W.~J.\ 1996, ApJ, 457, 752
\bibitem[\protect\citeauthoryear{Arzoumanian et
    al.}{2011}]{Arzoumanian:2011}
Arzoumanian D., et al., 2011, A\&A, 529, L6
\bibitem[\protect\citeauthoryear{Baban \& So}{1991}]{Baban:1991}
Baban F., So R.~M.~C., 1991, Experiments in Fluids, 10, 313
\bibitem[\protect\citeauthoryear{Ballesteros-Paredes \& Mac
    Low}{2002}]{Ballesteros-Paredes:2002}
Ballesteros-Paredes J., Mac Low M.-M., 2002, ApJ, 570, 734
\bibitem[\protect\citeauthoryear{Bate \& Bonnell}{2005}]{Bate:2005}
Bate M.~R., Bonnell I.~A., 2005, MNRAS, 356, 1201
\bibitem[\protect\citeauthoryear{Benjamin et
    al.}{2003}]{Benjamin:2003}
Benjamin R.~A., et al., 2003, PASP, 115, 953
\bibitem[\protect\citeauthoryear{Beuther et al.}{2011}]{Beuther:2011}
Beuther H., et al., 2011, A\&A, 533, A17
\bibitem[\protect\citeauthoryear{Bolatto et al.}{2013}]{Bolatto:2013}
Bolatto A.~D., et al., 2013, Nature, 499, 450 
\bibitem[\protect\citeauthoryear{Bonnell, Dobbs \& Smith}{Bonnell et
    al.}{2013}]{Bonnell:2013}
Bonnell I.~A., Dobbs C.~L., Smith R.~J., 2013, MNRAS, 430, 1790
\bibitem[\protect\citeauthoryear{Bruhweiler et
    al.}{2010}]{Bruhweiler:2010}
Bruhweiler F.~C., Ferrero R.~F., Bourdin M.~O., Gull T.~R., 2010, ApJ,
719, 1872
\bibitem[\protect\citeauthoryear{Canning et al.}{2011}]{Canning:2011}
Canning R.~E.~A., Fabian A.~C., Johnstone R.~M., Sanders J.~S.,
Crawford C.~S., Ferland G.~J., Hatch N.~A., 2011, MNRAS, 417, 3080
\bibitem[\protect\citeauthoryear{Carey et al.}{2009}]{Carey:2009}
Carey S.~J., et al., 2009, PASP, 121, 76
\bibitem[\protect\citeauthoryear{Cecil et al.}{2001}]{Cecil:2001}
Cecil G., Bland-Hawthorn J., Veilleux S., Filippenko A.~V., 2001, ApJ,
555, 338
\bibitem[\protect\citeauthoryear{Chi\`{e}ze \& Lazareff}{1981}]{Chieze:1981}
Chi\`{e}ze, J.~P., \& Lazareff, B.\ 1981, A\&A, 95, 194
\bibitem[\protect\citeauthoryear{Churchwell et
    al.}{2009}]{Churchwell:2009}
Churchwell E., et al., 2009, PASP, 121, 213
\bibitem[\protect\citeauthoryear{Close et al.}{2013}]{Close:2013}
Close, J., Pittard, J.~M., Hartquist, T.~W., \& Falle, S.~A.~E.~G.,
2013, MNRAS, 436, 3021
\bibitem[\protect\citeauthoryear{Col\'{i}n, V\'{a}zquez-Semadeni \& G\'{o}mez}{Col\'{i}n et al.}{2013}]{Colin:2013}
Col\'{i}n P., V\'{a}zquez-Semadeni E., G\'{o}mez G.~C., 2013, MNRAS,
435, 1701
\bibitem[\protect\citeauthoryear{Conselice, Gallagher \&
    Wyse}{Conselice et al.}{2001}]{Conselice:2001}
Conselice C.~J., Gallagher III J.~S., Wyse R.~F.~G., 2001, AJ, 122, 2281
\bibitem[\protect\citeauthoryear{Cooper et al.}{2008}]{Cooper:2008}
Cooper J.~L., Bicknell G.~V., Sutherland R.~S., Bland-Hawthorn J.,
2008, ApJ, 674, 157
\bibitem[\protect\citeauthoryear{Cowie, McKee \& Ostriker}{Cowie et al.}{1981}]{Cowie:1981}
Cowie, L.~L., McKee, C.~F., \& Ostriker, J.~P.\ 1981, ApJ, 247, 908
\bibitem[\protect\citeauthoryear{Crawford et
    al.}{2005}]{Crawford:2005}
Crawford C.~S., Hatch N.~A., Fabian A.~C., Sanders J.~S., 2005, MNRAS,
363, 216
\bibitem[\protect\citeauthoryear{Crighton, Hennawi \&
    Prochaska}{2013}]{Crighton:2013}
Crighton N.~H.~M., Hennawi J.~F., Prochaska J.~X., 2013, ApJ, 776, L18
\bibitem[\protect\citeauthoryear{Dale \& Bonnell}{2011}]{Dale:2011}
Dale J.~E., Bonnell I.~A., 2011, MNRAS, 414, 321
\bibitem[\protect\citeauthoryear{Davidson}{2004}]{Davidson:2004}
Davidson P.~A., 2004, ``Turbulence. An Introduction for Scientists and
Engineers'', Oxford University Press
\bibitem[\protect\citeauthoryear{Dent et al.}{2009}]{Dent:2009}
Dent W.~R.~F., et al., 2009, MNRAS, 395, 1805
\bibitem[\protect\citeauthoryear{Dobbs \& Bonnell}{2007}]{Dobbs:2007}
Dobbs C.~L., Bonnell I.~A., 2007, MNRAS, 374, 1115
\bibitem[\protect\citeauthoryear{Dursi \& Pfrommer}{2008}]{Dursi:2008}
Dursi L.~J., Pfrommer C., 2008, ApJ, 677, 993
\bibitem[\protect\citeauthoryear{Dyson, Arthur \& Hartquist}{Dyson et al.}{2002}]{Dyson:2002}
Dyson, J.~E., Arthur, S.~J., \& Hartquist, T.~W.\ 2002, A\&A, 390, 1063
\bibitem[\protect\citeauthoryear{Dyson \& Hartquist}{1987}]{Dyson:1987}
Dyson, J.~E., \& Hartquist, T.~W.\ 1987, MNRAS, 228, 453
\bibitem[\protect\citeauthoryear{Elmegreen \& Scalo}{2004}]{Elmegreen:2004}
Elmegreen B.~G., Scalo J., 2004, ARA\&A, 42, 211
\bibitem[\protect\citeauthoryear{Engelbracht et
    al.}{2006}]{Engelbracht:2006}
Engelbracht C.~W., et al., 2006, ApJ, 642, L127
\bibitem[\protect\citeauthoryear{Falle}{1991}]{Falle:1991}
Falle S.~A.~E.~G., 1991, MNRAS, 250, 581
\bibitem[\protect\citeauthoryear{Farris \& Russell}{1994}]{Farris:1994}
Farris M.~H., Russell C.~T., 1994, J. Geophys. Research, 99, 17681
\bibitem[\protect\citeauthoryear{Federrath et
    al.}{2010}]{Federrath:2010}
Federrath C., Roman-Duval J., Klessen R.~S., Schmidt W., Mac Low
M.-M., 2010, A\&A, 512, A81
\bibitem[\protect\citeauthoryear{Fern\'{a}ndez-L\'{o}pez et al.}{2014}]{Fernandez-Lopez:2014}
Fern\'{a}ndez-L\'{o}pez M., et al., 2014, ApJL, 790, L19
\bibitem[\protect\citeauthoryear{Forman et al.}{2007}]{Forman:2007}
Forman W., et al., 2007, ApJ, 665, 1057
\bibitem[\protect\citeauthoryear{Fr\"{o}hlich \&
    Rodi}{2004}]{Frohlich:2004}
Fr\"{o}hlich J., Rodi W., 2004, Int. Journ. Heat and Fluid Flow, 25, 537
\bibitem[\protect\citeauthoryear{Fujita et al.}{2009}]{Fujita:2009}
Fujita A., Martin C.~L., Mac Low M.-M., New K.~C.~B., Weaver R., 2009,
ApJ, 698, 693
\bibitem[\protect\citeauthoryear{Gaspari, Ruszkowski \&
    Sharma}{Gaspari et al.}{2012}]{Gaspari:2012}
Gaspari M., Ruszkowski M., Sharma P., 2012, ApJ, 746, 94
\bibitem[\protect\citeauthoryear{Georgievskiy, Levin \& Sutyrin}{Georgievskiy et
    al.}{2015}]{Georgievskiy:2015}
Georgievskiy P.~Yu., Levin V.~A., Sutyrin O.~G., 2015, Shock Waves,
25, 357
\bibitem[\protect\citeauthoryear{Glover \& Clark}{2012}]{Glover:2012}
Glover S.~C.~O., Clark P.~C., 2012, MNRAS, 426, 377
\bibitem[\protect\citeauthoryear{Goldsmith \&
    Pittard}{2016}]{Goldsmith:2016}
Goldsmith K.~J.~A., Pittard J.~M., 2016, MNRAS, submitted
\bibitem[\protect\citeauthoryear{Goodman et al.}{2014}]{Goodman:2014}
Goodman A.~A., et al., 2014, ApJ, 797, 53
\bibitem[\protect\citeauthoryear{G\'{o}mez \&
    V\'{a}zquez-Semadeni}{2014}]{Gomez:2014}
G\'{o}mez G.~C., V\'{a}zquez-Semadeni E., 2014, ApJ, 791, 124
\bibitem[\protect\citeauthoryear{Gregori et al.}{2000}]{Gregori:2000}
Gregori G., Miniati F., Ryu D., Jones T.~W., 2000, ApJ, 543, 775
\bibitem[\protect\citeauthoryear{Hacar et al.}{2013}]{Hacar:2013}
Hacar A., Tafalla M., Kauffmann J., Kov\'{a}cs A., 2013, A\&A, 554,
A55
\bibitem[\protect\citeauthoryear{Heeson et al.}{2011}]{Heeson:2011}
Heeson V., Beck R., Krause M., Dettmar R.-J., 2011, A\&A, 535, A79
\bibitem[\protect\citeauthoryear{Heiles \&
    Troland}{2003}]{Heiles:2003}
Heiles C., Troland T.~H., 2003, ApJ, 586, 1067
\bibitem[\protect\citeauthoryear{Heitsch \&
    Hartmann}{2008}]{Heitsch:2008}
Heitsch F., Hartmann L., 2008, ApJ, 689, 290
\bibitem[\protect\citeauthoryear{Hennebelle}{2013}]{Hennebelle:2013}
Hennebelle P., 2013, A\&A, 556, A153
\bibitem[\protect\citeauthoryear{Hennemann et
    al.}{2012}]{Hennemann:2012}
Hennemann M., et al., 2012, A\&A, 543, L3
\bibitem[\protect\citeauthoryear{Henning et al.}{2010}]{Henning:2010}
Henning Th., et al., 2010, A\&A, 518, L95
\bibitem[\protect\citeauthoryear{Henshaw et al.}{2014}]{Henshaw:2014}
Henshaw J.~D., Caselli P., Fontani F., Jim\'{e}nez-Serra I., Tan J.~C.,
2014, MNRAS, 440, 2860
\bibitem[\protect\citeauthoryear{Hoopes et al.}{2005}]{Hoopes:2005}
Hoopes C.~G., et al., 2005, ApJ, 619, L99
\bibitem[\protect\citeauthoryear{Jackson et al.}{2010}]{Jackson:2010}
Jackson J.~M., et al., 2010, ApJ, 719, L185
\bibitem[\protect\citeauthoryear{Johansson \& Ziegler}{2013}]{Johansson:2013}
Johansson E.~P.~G., Ziegler U., 2013, ApJ, 766, 45
\bibitem[\protect\citeauthoryear{Juvela et al.}{2012}]{Juvela:2012}
Juvela M., et al., 2012, A\&A, 541, A12
\bibitem[\protect\citeauthoryear{Kappler}{2002}]{Kappler:2002}
Kappler M., 2002, Ph.D. thesis, Institute for Hydromechanics,
University of Karlsruhe
\bibitem[\protect\citeauthoryear{Kawamura et
    al.}{1984}]{Kawamura:1984}
Kawamura T., Hiwada M., Hibino T., Mabuchi I., Kamuda M., 1984, ``Flow
around a finite circular cylinder on a flat plate'', Bull. JSME 27
(232), 2142
\bibitem[\protect\citeauthoryear{Kirk et al.}{2015}]{Kirk:2015}
Kirk H., Klassen M., Pudritz R., Pillsworth S., 2015, ApJ, 802, 75
\bibitem[\protect\citeauthoryear{Klein, McKee \& Colella}{Klein et al.}{1994}]{Klein:1994}
Klein R.~I., McKee C.~F., Colella P., 1994, ApJ, 420, 213
\bibitem[\protect\citeauthoryear{Klein et al.}{2003}]{Klein:2003}
Klein R.~I., Budil K.~S., Perry T.~S., Bach D.~R., 2003, ApJ, 583, 245
\bibitem[\protect\citeauthoryear{Klessen \&
    Burkert}{2000}]{Klessen:2000}
Klessen R.~S., Burkert A., 2000, ApJS, 128, 827
\bibitem[\protect\citeauthoryear{Kornreich \& Scalo}{2000}]{Kornreich:2000}
Kornreich P., Scalo J., 2000, ApJ, 531, 366
\bibitem[\protect\citeauthoryear{Krumholz, Klein \& McKee}{Krumholz et
    al.}{2011}]{Krumholz:2011}
Krumholz M.~R., Klein R.~I., McKee C.~F., 2011, ApJ, 740, 74
\bibitem[\protect\citeauthoryear{Larose, Savage \& Jakobsen}{Larose et
    al.}{2003}]{Larose:2003}
Larose G.~L., Savage M.~G., Jakobsen J.~B., 2003, Wind tunnel
experiments on an inclined and yawed circular cylinder in the critical
Reynolds number range, in Proceedings of the Eleventh International
Conference on Wind Engineering, Lubbock, TX, p.~1705
\bibitem[\protect\citeauthoryear{Leao et al.}{2009}]{Leao:2009}
Lea\~{o} M.~R.~M., de Gouveia Dal Pino E.~M., Falceta-Gon\c{c}alves D., Melioli
C., Geraissate F.~G., 2009, MNRAS, 394, 157
\bibitem[\protect\citeauthoryear{Li, Frank \& Blackman}{Li et al.}{2013}]{Li:2013}
Li S., Frank A., Blackman E.~G., 2013, ApJ, 774, 133
\bibitem[\protect\citeauthoryear{Li et al.}{2010}]{Li:2010}
Li Z.-Y., Wang P., Abel T., Nakamura F., 2010, ApJ, 720, L26
\bibitem[\protect\citeauthoryear{Mac Low \& Klessen}{2004}]{MacLow:2004}
Mac Low M.-M., Klessen R., 2004, Rev. Mod. Phys., 76, 125
\bibitem[\protect\citeauthoryear{Martin, Kobulnicky \& Heckman}{Martin
    et al.}{2002}]{Martin:2002}
Martin C.~L., Kobulnicky H.~A., Heckman T.~M., 2002, ApJ, 574, 663
\bibitem[\protect\citeauthoryear{Matsumoto et
    al.}{1990}]{Matsumoto:1990}
Matsumoto M., Shiraishi N., Kitazawa M., Knisely C., Shirato H., Kim
Y., Tsujii M., 1990, J. Wind Eng. Ind. Aerodyn., 33, 63
\bibitem[\protect\citeauthoryear{McCourt et al.}{2015}]{McCourt:2015}
McCourt M., O'Leary R.~M., Madigan A.-M., Quataert E., 2015, MNRAS,
449, 2
\bibitem[\protect\citeauthoryear{McDonald et
    al.}{2010}]{McDonald:2010}
McDonald M., Veilleux S., Rupke D.~S.~N., Mushotzky R., 2010, ApJ,
721, 1262
\bibitem[\protect\citeauthoryear{McKee \& Ostriker}{1977}]{McKee:1977} 
McKee, C.~F., \& Ostriker, J.~P.\ 1977, ApJ, 218, 148
\bibitem[\protect\citeauthoryear{Men'shchikov et
    al.}{2010}]{Menshchikov:2010}
Menshchikov A., et al., 2010, A\&A, 518, L103
\bibitem[\protect\citeauthoryear{Moeckel \&
    Burkert}{2015}]{Moeckel:2015}
Moeckel N., Burkert A., 2015, ApJ, 807, 67
\bibitem[\protect\citeauthoryear{Molinari et
    al.}{2010}]{Molinari:2010}
Molinari S., et al., 2010, A\&A, 518, L100
\bibitem[\protect\citeauthoryear{Motte et al.}{2010}]{Motte:2010}
Motte F., et al., 2010, A\&A, 518, L77
\bibitem[\protect\citeauthoryear{Nakamura et al.}{2006}]{Nakamura:2006}
Nakamura F., McKee C.~F., Klein R.~I., Fisher R.~T., 2006, ApJSS, 164, 477
\bibitem[\protect\citeauthoryear{Niederhaus}{2007}]{Niederhaus:2007} 
Niederhaus J.~H.~J., 2007, PhD thesis, University of Wisconsin - Madison
\bibitem[\protect\citeauthoryear{Niederhaus et
    al.}{2008}]{Niederhaus:2008} 
Niederhaus J.~H.~J., Greenough J.~A., Oakley J.~G., Ranjan D.,
Anderson M.~H., Bonazza R., 2008, J. Fluid Mech., 594, 85
\bibitem[\protect\citeauthoryear{Ntormousi et
    al.}{2011}]{Ntormousi:2011}
Ntormousi E., Burkert A., Fierlinger K., Heitsch F., 2011, ApJ, 731,
13
\bibitem[\protect\citeauthoryear{Ohyama et al.}{2002}]{Ohyama:2002}
Ohyama Y., et al., 2002, PASJ, 54, 891
\bibitem[\protect\citeauthoryear{Okamoto \&
    Yagita}{1973}]{Okamoto:1973}
Okamoto T., Yagita M., 1973, Bull. JSME, 95, 805
\bibitem[\protect\citeauthoryear{Orlando et al.}{2005}]{Orlando:2005}
Orlando S., Peres G., Reale F., Bocchino F., Rosner R., Plewa T., Siegel A.,
2005, A\&A, 444, 505
\bibitem[\protect\citeauthoryear{Padoan et al.}{2001}]{Padoan:2001}
Padoan P., et al., 2001, ApJ, 553, 227
\bibitem[\protect\citeauthoryear{Padoan et al.}{2006}]{Padoan:2006}
Padoan P., Cambr\'{e}sy L., Juvela M., Kritsuk A., Langer W.~D.,
Norman M.~L., 2006, ApJ, 649, 807
\bibitem[\protect\citeauthoryear{Palmeirim et
    al.}{2013}]{Palmeirim:2013}
Palmeirim P., et al., 2013, A\&A, 550, A38
\bibitem[\protect\citeauthoryear{Park \& Lee}{2000}]{Park:2000}
Park C.-W., Lee S.-J., 2000, Journal Wind Engineering and Industrial
Aerodynamics, 88, 231
\bibitem[\protect\citeauthoryear{Peretto et al.}{2012}]{Peretto:2012}
Peretto N., et al., 2012, A\&A, 541, A63
\bibitem[\protect\citeauthoryear{Pittard et al.}{2003}]{Pittard:2003}
Pittard, J.~M., Arthur, S.~J., Dyson, J.~E., Falle, S.~A.~E.~G., Hartquist,
T.~W., Knight M.~I., \& Pexton M.\ 2003, A\&A, 401, 1027
\bibitem[\protect\citeauthoryear{Pittard et al.}{2009}]{Pittard:2009}
Pittard J.~M., Falle S.~A.~E.~G., Hartquist T.~W., Dyson J.~E., 2009, MNRAS, 394, 1351
\bibitem[\protect\citeauthoryear{Pittard et al.}{2010}]{Pittard:2010}
Pittard J.~M., Hartquist T.~W., Falle S.~A.~E.~G., 2010, MNRAS, 405,
821
\bibitem[\protect\citeauthoryear{Pittard \& Parkin}{2016}]{Pittard:2016}
Pittard J.~M., Parkin E.~R., 2016, MNRAS, accepted
\bibitem[\protect\citeauthoryear{Porter et al.}{1994}]{Porter:1994}
Porter D., et al., 1994, Phys. Fluids, 6, 2133
\bibitem[\protect\citeauthoryear{Raga et al.}{2007}]{Raga:2007}
Raga A.~C., Esquivel A., Riera A., Vel\'{a}zquez P.~F., 2007, ApJ,
668, 310
\bibitem[\protect\citeauthoryear{Ramberg}{1983}]{Ramberg:1983}
Ramberg S.~E., 1983, J. Fluid Mech., 128, 81
\bibitem[\protect\citeauthoryear{Ribaudo et al.}{2011}]{Ribaudo:2011}
Ribaudo J., Lehner N., Howk J.~C., Werk J.~K., Tripp T.~M., Prochaska
J.~X., Meiring J.~D., Tumlinson J., 2011, ApJ, 743, 207
\bibitem[\protect\citeauthoryear{Rich et al.}{2010}]{Rich:2010}
Rich J.~A., Dopita M.~A., Kewley L.~J., Rupke D.~S.~N., 2010, ApJ,
721, 505
\bibitem[\protect\citeauthoryear{Rodr\'{i}guez-Gonz\'{a}lez et
    al.}{2008}]{Rodriguez-Gonzalez:2008}
Rodr\'{i}guez-Gonz\'{a}lez A., Esquivel A., Raga A.~C., Cant\'{o} J.,
2008, ApJ, 684, 1384
\bibitem[\protect\citeauthoryear{Rogers \&
    Pittard}{2013}]{Rogers:2013}
Rogers H., Pittard J.~M., 2013, MNRAS, 431, 1337
\bibitem[\protect\citeauthoryear{Rudie et al.}{2012}]{Rudie:2012}
Rudie G.~C., et al., 2012, ApJ, 750, 67
\bibitem[\protect\citeauthoryear{Sales et al.}{2010}]{Sales:2010}
Sales L.~V., Navarro J.~F., Schaye J., Dalla Vecchia C., Springel V.,
Booth C.~M., 2010, MNRAS, 409, 1541
\bibitem[\protect\citeauthoryear{Scalo \& Elmegreen}{2004}]{Scalo:2004}
Scalo J., Elmegreen B.~G., 2004, ARA\&A, 42, 275
\bibitem[\protect\citeauthoryear{Scannapieco \&
    Br\"{u}ggen}{2015}]{Scannapieco:2015}
Scannapieco E., Br\"{u}ggen M., 2015, ApJ, 805, 158
\bibitem[\protect\citeauthoryear{Schneider et al.}{2006}]{Schneider:2006}
Schneider N., et al., 2006, A\&A, 458, 855
\bibitem[\protect\citeauthoryear{Schneider \&
    Elmegreen}{1979}]{Schneider:1979}
Schneider S., Elmegreen B.~G., 1979, ApJS, 41, 87
\bibitem[\protect\citeauthoryear{Seifried \&
    Walch}{2015}]{Seifried:2015}
Seifried D., Walch S., 2015, MNRAS, 452, 2410
\bibitem[\protect\citeauthoryear{Sharp \&
    Bland-Hawthorn}{2010}]{Sharp:2010}
Sharp R.~G., Bland-Hawthorn J., 2010, ApJ, 711, 818
\bibitem[\protect\citeauthoryear{Shin, Stone \& Snyder}{Shin et al.}{2008}]{Shin:2008}
Shin M.-S., Stone J.~M., Snyder G.~F., 2008, ApJ, 680, 336
\bibitem[\protect\citeauthoryear{Shirakashi, Hasegawa \&
    Wakiya}{Shirakashi et al.}{1986}]{Shirakashi:1986}
Shirakashi M., Hasegawa A., Wakiya S., 1986, Bull. Jpn. Soc. Mech. Eng. 29, 250, 1124
\bibitem[\protect\citeauthoryear{Shopbell \&
    Bland-Hawthorn}{1998}]{Shopbell:1998}
Shopbell P.~L., Bland-Hawthorn J., 1998, ApJ, 493, 129
\bibitem[\protect\citeauthoryear{Smith, Glover \&
    Klessen}{Smith et al.}{2014}]{Smith:2014}
Smith R.~J., Glover S.~C.~O., Klessen R.~S., 2014, MNRAS, 445, 2900
\bibitem[\protect\citeauthoryear{Stone \& Norman}{1992}]{Stone:1992}
Stone J.~M., Norman M.~L., 1992, ApJ, 390, L17
\bibitem[\protect\citeauthoryear{Strickland \&
    Stevens}{2000}]{Strickland:2000}
Strickland D.~K, Stevens I.~R., 2000, MNRAS, 314, 511
\bibitem[\protect\citeauthoryear{Strickland et
    al.}{2004}]{Strickland:2004}
Strickland D.~K., Heckman T.~M., Colbert E.~J.~M., Hoopes C.~G.,
Weaver K.~A., 2004, ApJSS, 151, 193
\bibitem[\protect\citeauthoryear{Vaidya, Hartquist \& Falle}{Vaidya et al.}{2013}]{Vaidya:2013}
Vaidya B., Hartquist T.~W., Falle S.~A.~E.~G., 2013, MNRAS, 433, 1258
\bibitem[\protect\citeauthoryear{Vakil \& Green}{2009}]{Vakil:2009}
Vakil A., Green S.~I., 2009, Computers \& Fluids, 38, 1771
\bibitem[\protect\citeauthoryear{Van Loo, Falle \& Hartquist}{Van Loo et al.}{2010}]{VanLoo:2010}
Van~Loo S., Falle S.~A.~E.~G., Hartquist T.~W., 2010, MNRAS, 406, 1260
\bibitem[\protect\citeauthoryear{V\'{a}zquez-Semadeni}{1994}]{Vazquez-Semadeni:1994}
V\'{a}zquez-Semadeni E., 1994, ApJ, 423, 681
\bibitem[\protect\citeauthoryear{V\'{a}zquez-Semadeni et al.}{2000}]{Vazquez-Semadeni:2000}
V\'{a}zquez-Semadeni E., Ostriker E.~C., Passot T., Gammie C.~F.,
Stone J.~M., 2000, in Protostars and Planets IV, ed. V.~Mannings,
A.~P.~Boss \& S.~S.~Russell (Tucson: Univ. Arizona Press), 3 
\bibitem[\protect\citeauthoryear{V\'{a}zquez-Semadeni}{2006}]{Vazquez-Semadeni:2006}
V\'{a}zquez-Semadeni E., Ryu D., Passot T., Gonz\'{a}lez R.~F., Gazol A., 2006, ApJ, 643, 245
\bibitem[\protect\citeauthoryear{V\'{a}zquez-Semadeni}{2010}]{Vazquez-Semadeni:2010}
V\'{a}zquez-Semadeni E., et al., 2010, ApJ, 715, 1302
\bibitem[\protect\citeauthoryear{Veilleux \&
    Rupke}{2002}]{Veilleux:2002}
Veilleux S., Ruple D.~S.~N., 2002, ApJ, 565, L63
\bibitem[\protect\citeauthoryear{Veilleux, Cecil \&
    Bland-Hawthorn}{Veilleux et al.}{2005}]{Veilleux:2005}
Veilleux S., Cecil G., Bland-Hawthorn J., 2005, ARA\&A, 43, 769
\bibitem[\protect\citeauthoryear{Vig et al.}{2007}]{Vig:2007}
Vig S., Testi L., Walmsley M., Molinari S., Carey S., Noriega-Crespo
A., 2007, A\&A, 470, 977
\bibitem[\protect\citeauthoryear{Werk et al.}{2014}]{Werk:2014}
Werk J.~K., et al., 2014, ApJ, 792, 8
\bibitem[\protect\citeauthoryear{Westmoquette, Smith \&
    Gallagher}{Westmoquette et al.}{2011}]{Westmoquette:2011}
Westmoquette M.~S., Smith L.~J, Gallagher III J.~S., 2011, MNRAS, 414, 3719
\bibitem[\protect\citeauthoryear{White \& Long}{1991}]{White:1991}
White, R.~L., \& Long, K.~S.\ 1991, ApJ, 373, 543
\bibitem[\protect\citeauthoryear{Williamson}{1996}]{Williamson:1996}
Williamson C.~H.~K., 1996, Annu. Rev. Fluid Mech., 28, 477
\bibitem[\protect\citeauthoryear{Wilson et al.}{2005}]{Wilson:2005}
Wilson B.~A., Dame T.~M., Masheder M.~R.~W., Thaddeus P., 2005, A\&A,
430, 523
\bibitem[\protect\citeauthoryear{Xu \& Stone}{1995}]{Xu:1995}
Xu J., Stone J.~M., 1995, ApJ, 454, 172
\bibitem[\protect\citeauthoryear{Yeo \& Jones}{2008}]{Yeo:2008}
Yeo D., Jones N.~P., 2008, J. Wind Eng. Ind. Aerodyn., 96, 1947
\bibitem[\protect\citeauthoryear{Yirak, Frank \& Cunningham}{Yirak et al.}{2010}]{Yirak:2010}
Yirak K., Frank A., Cunningham A., 2010, ApJ, 722, 412 
\bibitem[\protect\citeauthoryear{Zdravkovich}{2003}]{Zdravkovich:2003}
Zdravkovich M.~M., 2003, Flow Around Circular Cylinders, Vol. 2:
Applications, Oxford University Press
\bibitem[\protect\citeauthoryear{Zhang et al.}{2015}]{Zhang:2015}
Zhang D., Thompson T.~A., Quataert E., Murray N., 2015,
arXiv:1507.01951
\end{thebibliography}




\appendix

\section{Resolution Test}
\label{sec:restest}
The damping of hydromagnetic waves, through either particle collisions
or wave-particle interactions, sets the length scale, $\eta$, of the
smallest instabilities in the actual interaction of a shock with an
obstacle \citep[see the discussion in][]{Pittard:2009}. The Reynolds
number of the interaction is given by ${\rm Re} = (l/\eta)^{4/3}$,
where $l$ is the characteristic size of the largest eddies (typically
the size of the obstacle).  In astrophysical settings, Re can easily
exceed $10^{5}-10^{6}$. At such high values, the flow will develop
turbulent-like characteristics (i.e., rapid variations in the fluid
properties in time and space). Resolving the smallest eddies in a
numerical simulation can be very challenging, which is why some
studies make use of subgrid turbulent viscosity models (e.g., the
$k$-$\epsilon$ model) which add turbulent-specific viscosity and
diffusion terms to the Euler equations
\citep[e.g.,][]{Pittard:2009,Pittard:2010,Pittard:2016}.

For simulations which simply solve the Euler equations for inviscid
fluid flow, new unstable scales will be added as the resolution of the
simulation is increased, since there is no prescription for
small-scale dissipative physics. In such ``inviscid'' simulations, the
nature of the interaction of a shock with an obstacle will depend on
the resolution used. At higher resolution, smaller instabilities can
develop and features in the flow (e.g., shocks and interfaces) become
sharper. The effective or grid-scale Reynolds number of the flow is
$Re = (l/\eta)^{4/3}$, where $\eta$ is now the minimum eddy size in
the simulation ($\eta \approx 2 \Delta x$, where $\Delta x$ is the
grid cell size). Hence, calculations at lower resolution effectively
simulate a more viscous flow. In shock-cloud simulations, this can
affect the rate at which material is stripped from the cloud and mixed
into the post-shock flow, and the acceleration of the cloud. Important
features of the flow may not be present at very low resolution, and
the simulated interaction will compare poorly to reality.

Simulations of problems for which there is no analytical solution
typically rely on a demonstration of self-convergence. However, care
must be taken to avoid ``false convergence''
\citep[e.g.,][]{Niederhaus:2007}. Formal convergence may be impossible
in ``inviscid'' simulations - where convergence is demonstrated, it
may only apply to specific global quantities, and at the earlier
stages of the interaction. Rather than attempting to demonstrate
formal convergence, some previous studies have instead focussed on
resolving key features in the flow. In shock-cloud simulations this
could include the stand-off distance of the bowshock
\citep[e.g.,][]{Farris:1994}, the thickness of the turbulent boundary
layer \citep[e.g.,][]{Pittard:2009}, the cooling layer behind shocks
\citep{Yirak:2010}, and the magnetic draping layer in magnetized
interactions \citep{Dursi:2008}.

A resolution test for 3D adiabatic shock-cloud simulations was
performed by \citet{Pittard:2016}. It was found that $32-64$ cells per
cloud radius were needed to capture the main flow features and for
reasonable convergence of some key global quantities. Here we examine how
shock-filament interactions are affected by the resolution used.
 
\subsection{Filament morphology}

\begin{figure*}
\includegraphics[width=16.0cm]{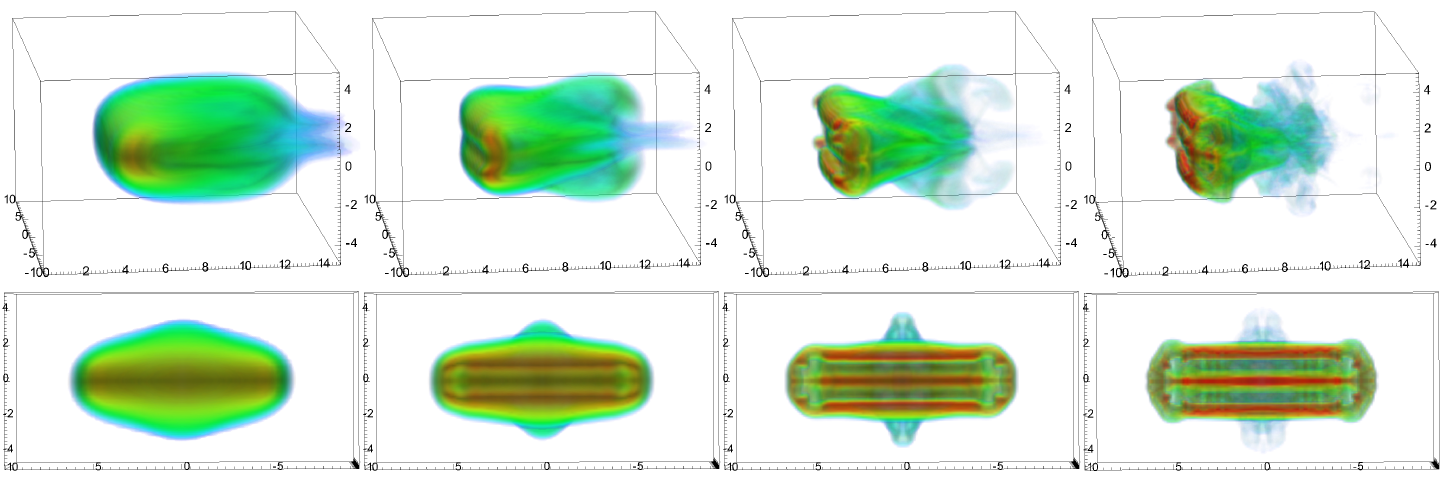}
\caption{The effect of the grid resolution on simulation
  \emph{m10c2l8s} at $t=1.58 \,t_{\rm cs}$. From left to right the
  resolution is $R_4$, $R_8$, $R_{16}$ and $R_{32}$. Side-views are
  shown in the top row and face-on views in the bottom row.}
\label{fig:m10c2l8s_morphology_restest}
\end{figure*}

\begin{figure*}
\includegraphics[width=16.0cm]{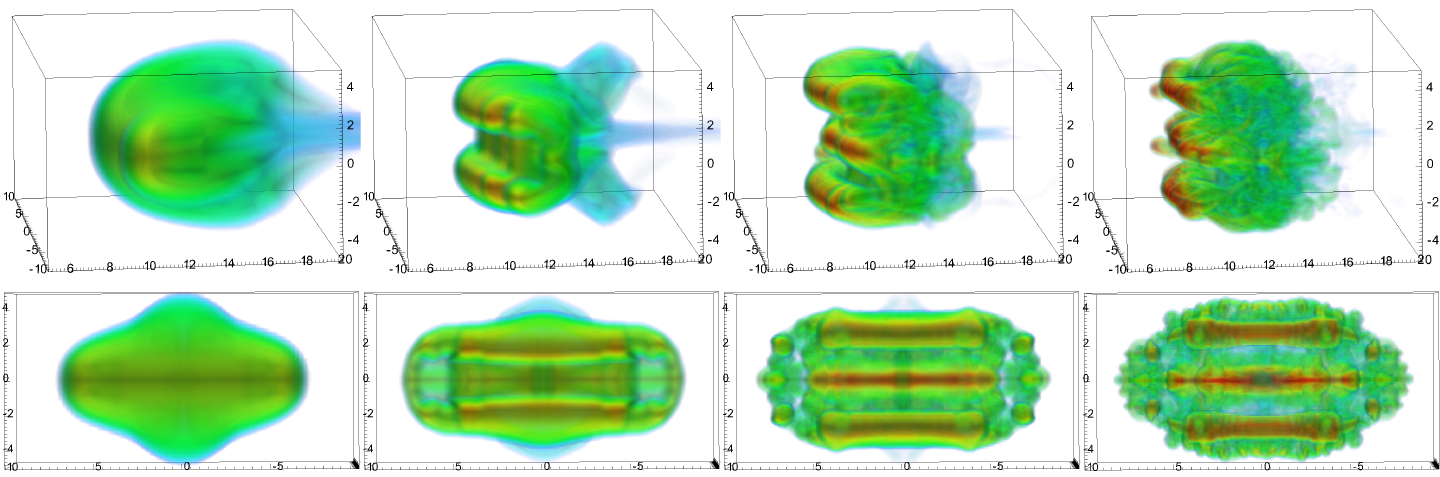}
\caption{As Fig.~\ref{fig:m10c2l8s_morphology_restest} but at $t=2.28 \,t_{\rm cs}$.}
\label{fig:m10c2l8s_morphology_restest2}
\end{figure*}

Fig.~\ref{fig:m10c2l8s_morphology_restest} shows the effect of the
resolution on the interaction in simulation \emph{m10c2l8s}, at $t=1.58
\,t_{\rm cs}$. The filament is oriented side-on to the shock, and we
view it from one end in the top row of panels, and face-on in the
bottom row of panels. As the resolution increases the shape of the
cloud changes, from smooth and relatively featureless, to displaying 2
and then 3 parallel rolls at the two highest resolutions. The
downstream wake shows characteristics of turbulence at resolution
$R_{32}$. Fig.~\ref{fig:m10c2l8s_morphology_restest2} shows equivalent
snapshots at the slightly later time of $t=2.28 \,t_{\rm cs}$. The
filament in the lowest resolution simulation has moved slightly
further downstream than the others. The shape of the rolls is
slightly different between simulations at $R_{16}$ and $R_{32}$, but
generally speaking the $R_{16}$ simulation appears to have captured
the main features of the interaction. Additional smaller-scale
structure is present at $R_{32}$.

\begin{figure*}
\includegraphics[width=16.0cm]{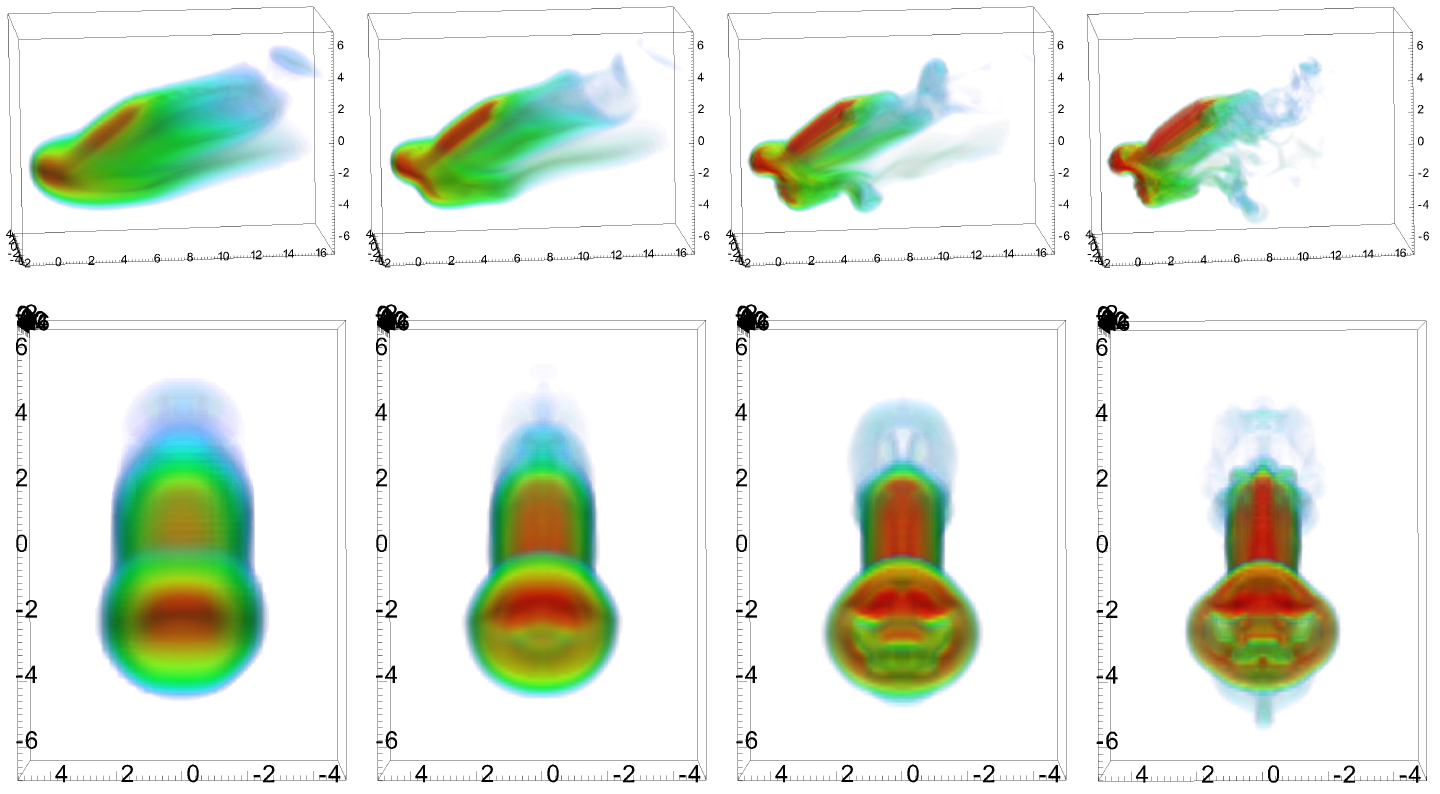}
\caption{The effect of the grid resolution on simulation
  \emph{m10c2l8o60} at $t=1.58 \,t_{\rm cs}$. From left to right the
  resolution is $R_4$, $R_8$, $R_{16}$ and $R_{32}$. Side-views are
  shown in the top row and face-on views in the bottom row.}
\label{fig:m10c2l8o60_morphology_restest}
\end{figure*}

Fig.~\ref{fig:m10c2l8o60_morphology_restest} shows the effect of the
grid resolution on simulation \emph{m10c2l8o60}. Here the filament is
oriented at an angle of $30^{\circ}$ to the shock normal. Here the
broad features of the flow are present even at $R_{4}$, but increasing
resolution shows the up-stream tip of the filament becoming more and
more hollow as an asymmetric vortex roll forms around it. Material at
the up-stream tip is thus ripped off the filament in this way.

\begin{figure*}
\includegraphics[width=16.0cm]{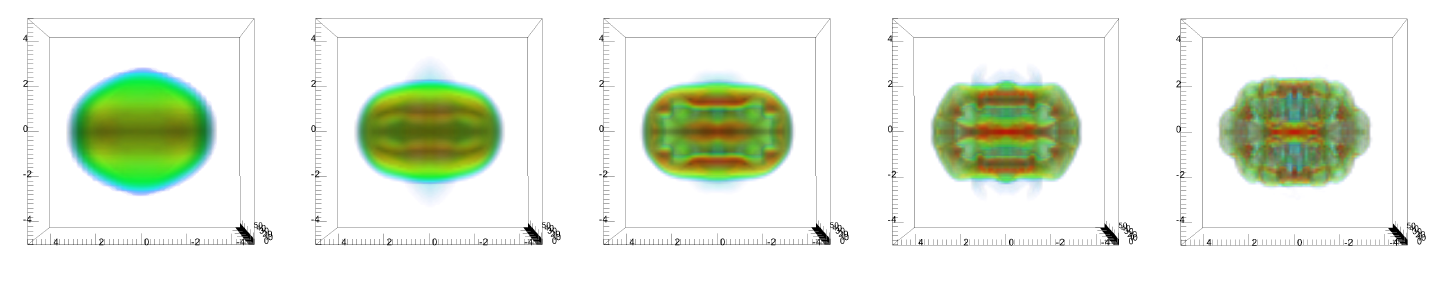}
\caption{The effect of the grid resolution on simulation
  \emph{m10c2l2s} at $t=2.26 \,t_{\rm cs}$. From left to right the
  resolution is $R_4$, $R_8$, $R_{16}$, $R_{32}$ and $R_{64}$.}
\label{fig:m10c2l2s_morphology_restest}
\end{figure*}

\begin{figure*}
\includegraphics[width=16.0cm]{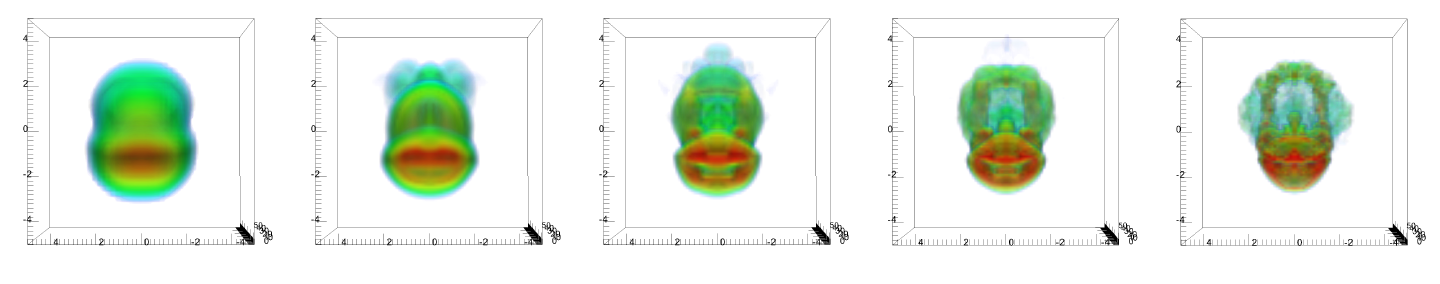}
\caption{The effect of the grid resolution on simulation
  \emph{m10c2l2o30} at $t=2.26 \,t_{\rm cs}$. From left to right the
  resolution is $R_4$, $R_8$, $R_{16}$, $R_{32}$ and $R_{64}$.}
\label{fig:m10c2l2o30_morphology_restest}
\end{figure*}

Figs.~\ref{fig:m10c2l2s_morphology_restest}
and~\ref{fig:m10c2l2o30_morphology_restest} show the effect of the
grid resolution on two simulations where the filament is significantly
shorter (models \emph{m10c2l2s} and \emph{m10c2l2o30},
respectively). Fig.~\ref{fig:m10c2l2s_morphology_restest} also shows
that the cloud forms first 2, and then 3 parallel rolls as the
resolution increases from $R_{8}$ to $R_{16}$ and $R_{32}$. However,
increasing the resolution further to $R_{64}$ shows that the
prominence of the rolls reduces slightly as additional
non-axisymmetric structure forms. In
Fig.~\ref{fig:m10c2l2o30_morphology_restest} we see the same hollowing
of the filament tip seen for longer filaments (cf.
Fig.~\ref{fig:m10c2l8o60_morphology_restest}). In this case the
$R_{64}$ simulation shows additional structure but also largely maintains the
general features seen at $R_{32}$.

In summary, one obtains the general impression that $R_{16}$ often
captures the main morphological features of the interaction, but that
$R_{32}$ is really the minimum needed for a more accurate description
of the flow. This requirement is maybe a little below that for spherical clouds,
where $\sim R_{64}$ is identified as the minimum necessary resolution
in purely hydrodynamical interactions \citep{Xu:1995,Pittard:2016}.
However, the lower resolution requirement for filaments compared to spherical
clouds is perhaps not too unexpected given that we measure the resolution
across the short axis of the filament. To compare more fairly, we
could instead consider the respective volumes of the filament and the
spherical cloud. The volume of the filament in model \emph{m10c2l8s} is
$7\times$ that of a spherical cloud of equivalent cross-section. A
spherical cloud with the same volume as this filament would thus have
a radius $1.91\times$ larger. Hence the \emph{effective} resolution of
an $R_{32}$ simulation of model \emph{m10c2l8s} could be argued to be
$R_{61}$.

\subsection{Time Evolution}

\begin{figure*}
  \begin{center}
\resizebox{155mm}{!}{\includegraphics{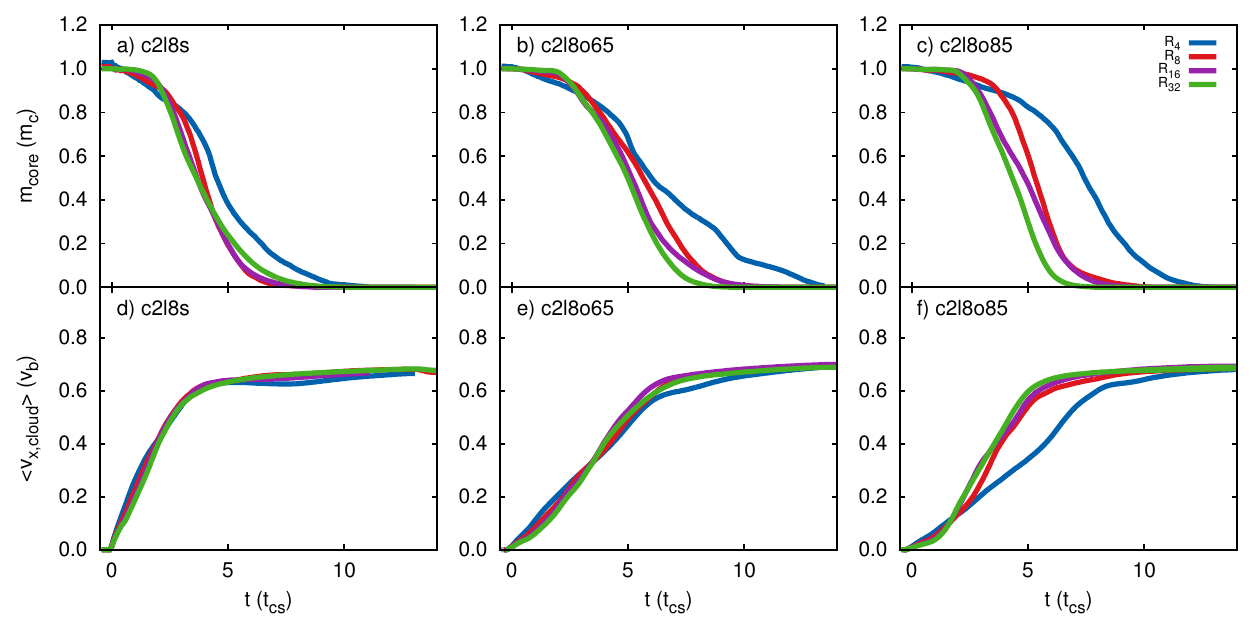}}
   \caption{Time evolution of the core mass, $m_{\rm core}$, and the
     mean filament speed, $<v_{\rm x,cloud}>$, for
     simulations 
     (a) \emph{m10c2l8s}, (b) \emph{m10c2l8o60} and (c) \emph{m10c2l8o85}.}
    \label{fig:mcore_restest_fil}
  \end{center}
\end{figure*}
 
Fig.~\ref{fig:mcore_restest_fil} shows the time evolution of the core
mass, $m_{\rm core}$, and the mean filament speed, $<v_{\rm
  x,cloud}>$, for a number of simulations where the filament length
$l=8\,r_{\rm c}$. In all cases we observe good convergence, with only
the $R_{4}$ resolution simulations showing significant divergence from
the others. Closer examination reveals that there is less difference
between the $R_{16}$ and $R_{32}$ simulations, than between the
$R_{8}$ and $R_{16}$ simulations, consistent with convergence.

The only other resolution test for non-spherical clouds which we are
aware of in the literature is in \citet{Xu:1995}, where prolate clouds
with an axial ratio of 2:1 and $\chi=10$ are struck by a Mach 10
shock. Their Fig.~8 shows the resolution dependence of the time evolution of
various global properties of the cloud when it is initially oriented
with its long-axis perpendicular to the shock normal (i.e. sideways
on). The mean cloud speed is very insensitive to the range of
resolutions considered (corresponding to $R_{7}$, $R_{16}$, and
$R_{32}$ in our terminology), though the mixing fraction, $f_{\rm
  mix}$ is significantly different at $R_{7}$. Our results for longer
and denser filaments are roughly consistent with their findings.

In their Fig.~10, \citet{Xu:1995} show similar plots for an inclined
cloud (with the semi-major axis at $45^{\circ}$ to the shock
normal). Only $R_{7}$ and $R_{14}$ calculations are compared - again
the mean cloud velocity is very insensitive to the resolutions
examined, but there are some more significant differences for $f_{\rm
  mix}$. In Fig.~\ref{fig:mcore_restest_fil} we find that the
simulations are less converged as the filament is oriented more end-on
to the shock.

To conclude, it seems that the time evolution of $m_{\rm core}$ and
$<v_{\rm x,cloud}>$ are reasonably converged by
$R_{16}-R_{32}$. However, we caution that other quantities are
likely to show greater variation with resolution.

\subsection{Convergence Tests}
\label{sec:restest_convergence}
To gain further insight into the effect of the grid resolution on our
simulations we examine the variation of some integral quantities
computed from the datasets at a particular moment in time. Formal
convergence demands that there is an asymptotic levelling off with
increasing resolution of a particular quantity.

\begin{figure*}
\includegraphics{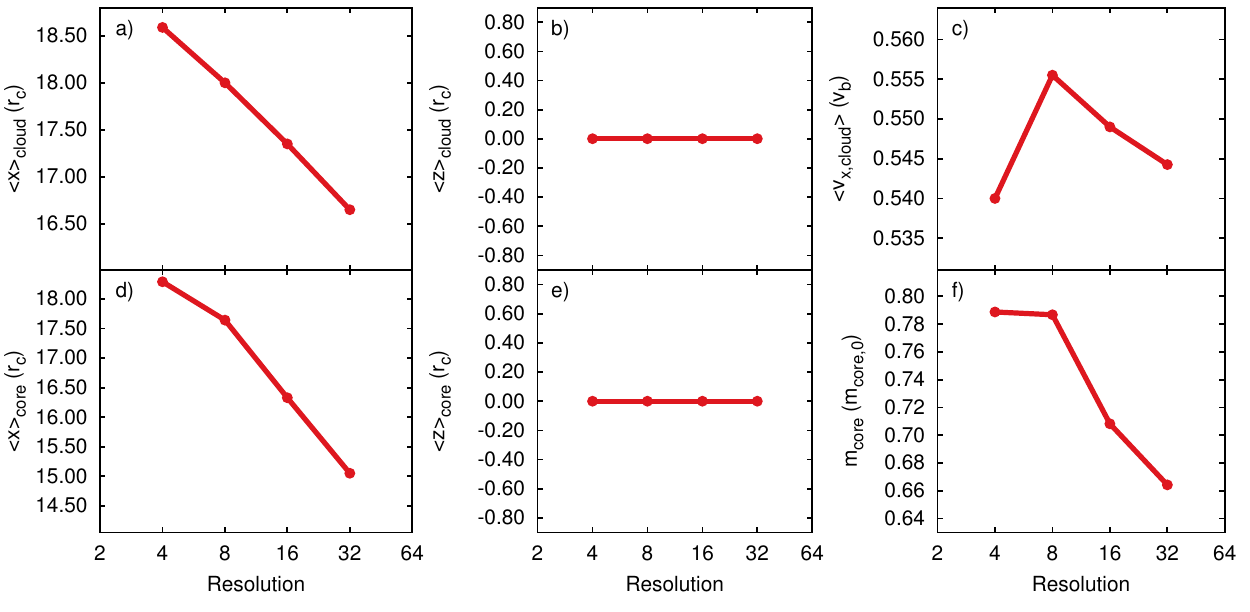}
   \caption{Integral quantities from simulation \emph{m10c2l8s} at $t=3.0\,t_{\rm cs}$, plotted as a function of the
     grid resolution.}
    \label{fig:m10fil8_convergence}
\end{figure*}

The variation in $\langle x\rangle_{\rm cloud}$, $\langle
z\rangle_{\rm cloud}$, $<v_{\rm x,cloud}>$, $\langle x\rangle_{\rm
  core}$, $\langle z\rangle_{\rm core}$ and $m_{\rm core}$ with the
spatial resolution for simulation \emph{m10c2l8s} is shown in
Fig.~\ref{fig:m10fil8_convergence}. It is clear that $\langle
x\rangle_{\rm cloud}$ and $\langle x\rangle_{\rm core}$ are not yet
showing signs of convergence. However, $<v_{\rm x,cloud}>$ and $m_{\rm
  core}$ show perhaps the early signs of convergence between $R_{8}$
and $R_{32}$ (the simulation at $R_{4}$ is clearly completely
unresolved and is ignored). 

\begin{figure*}
\includegraphics{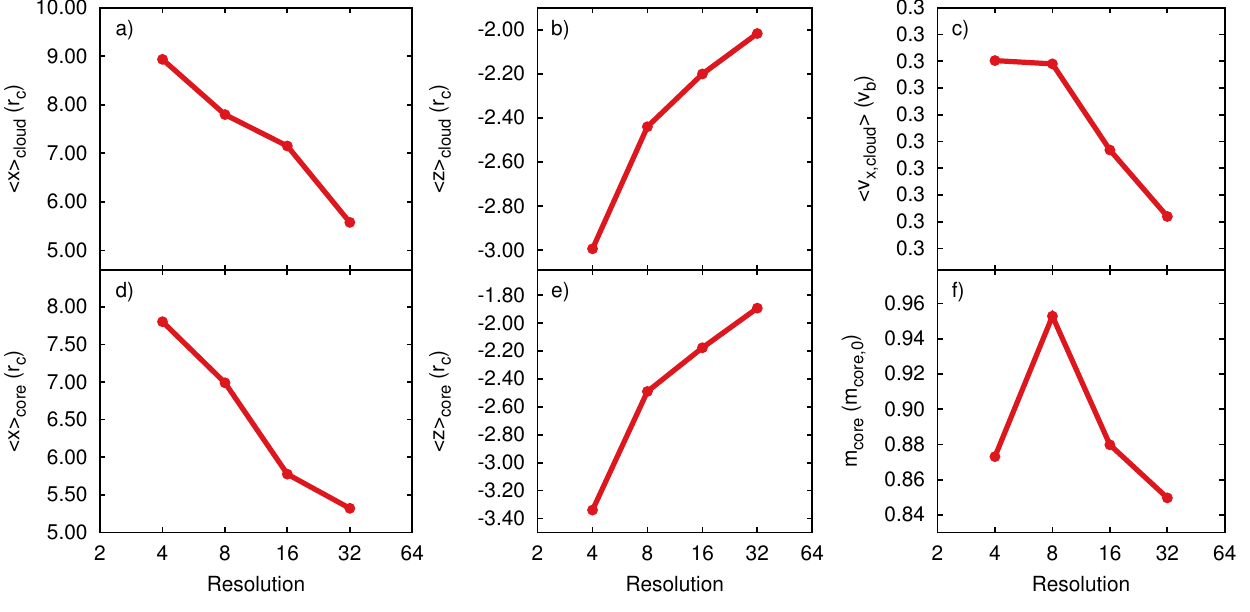}
   \caption{As Fig.~\ref{fig:m10fil8_convergence} but for simulation \emph{m10c2l8o60}.}
    \label{fig:m10fil2_convergence}
\end{figure*}

\begin{figure*}
\includegraphics{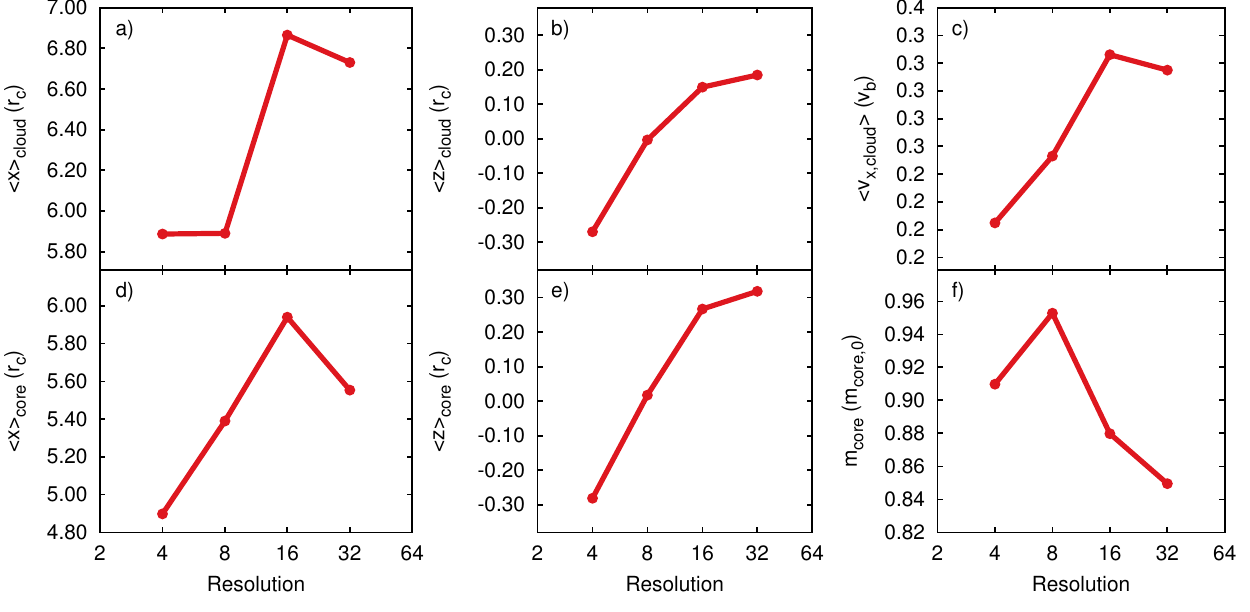}
   \caption{As Fig.~\ref{fig:m10fil8_convergence} but for simulation \emph{m10c2l8o85}.}
    \label{fig:m10fil3_convergence}
\end{figure*}

Figs.~\ref{fig:m10fil2_convergence} and~\ref{fig:m10fil3_convergence}
examine the convergence properties for simulations \emph{m10c2l8o60}
and \emph{m10c2l8o85}. Broadly similar behaviour is seen. We conclude
that our simulations are not formally converged, but that the highest
resolution simulations have sufficient resolution that some of the
integral quantities are showing signs of convergence at $t=3.0\,t_{\rm
  cs}$. Higher resolution simulations remain desirable to investigate
this further.

\subsection{Timescales}

\begin{figure*}
\includegraphics[width=12.7cm]{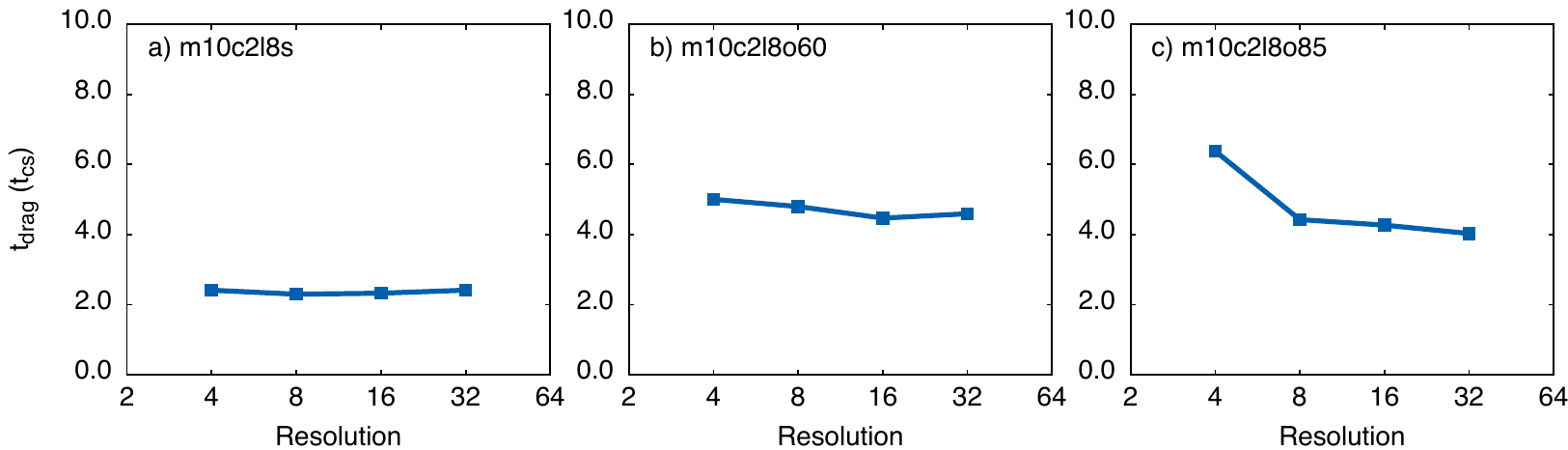}
   \caption{Resolution dependence of $t_{\rm drag}$ (for the
      cloud) for simulations a) \emph{m10c2l8s}; b) \emph{m10c2l8o60};
    c) \emph{m10c2l8o85}.}
    \label{fig:tdragrestest}
\end{figure*}

\begin{figure*}
\includegraphics{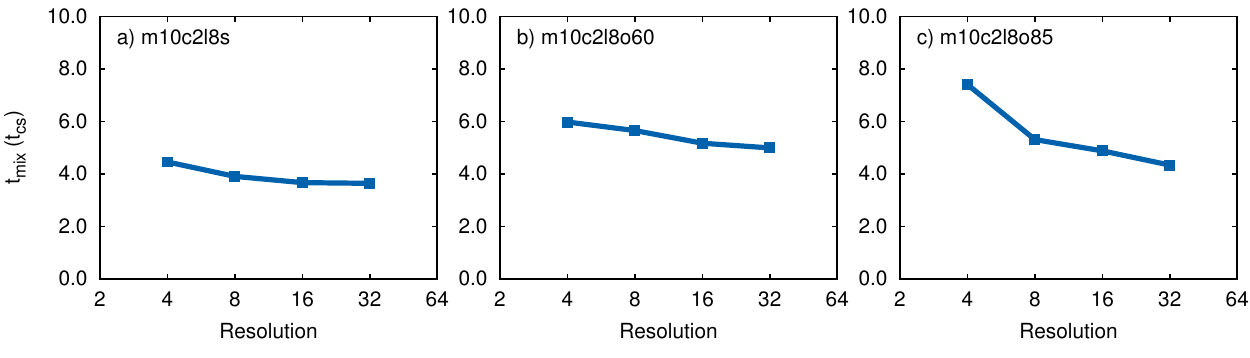}
   \caption{Resolution dependence of $t_{\rm mix}$ (for the
      cloud) for simulations a) \emph{m10c2l8s}; b) \emph{m10c2l8o60};
    c) \emph{m10c2l8o85}.}
    \label{fig:tmixrestest}
\end{figure*}

Figs.~\ref{fig:tdragrestest} and~\ref{fig:tmixrestest} examine the
resolution dependence of $t_{\rm drag}$ and $t_{\rm mix}$. In general,
$t_{\rm drag}$ is broadly stable with resolution.
However, $t_{\rm mix}$ appears to decrease with increasing resolution,
which is different to the behaviour for spherical clouds. For
simulation \emph{m10c2l8s} we see convergence towards an asymptote,
for $t_{\rm mix}$.  So this simulation appears
to be formally converged for this particular quantitie. However, we
do not see such behaviour for the other simulations, so formally they
are non-convergent. Having said this, most of them show some signs of
levelling off in $t_{\rm drag}$ and $t_{\rm mix}$, indicating that
their values may be reasonably close to the ``true'' value. In
comparison, we note that at resolution $R_{32}$ values of $t_{\rm
  drag}$ and $t_{\rm mix}$ are on average within a few percent of
their values at higher resolutions in simulations of shocks striking
spherical clouds \citep{Pittard:2016}. We might expect this to be true
for shock-filament interactions also, but in order to check for false
convergence higher resolution simulations are needed.

\bsp	
\label{lastpage}
\end{document}